\documentclass[12pt, oneside, a4paper]{article}
\usepackage{amsmath,amssymb,mathtools}
\usepackage{natbib}
\usepackage{natbib, lscape}
\usepackage{float}

\usepackage{booktabs}
\usepackage{tabularx}
\usepackage{setspace}
\usepackage{authblk}

\title{Horseshoe Priors for Time-Varying AR and GARCH Processes}
\author[1]{John W. G. Addy}
\author[2]{Chloe Maclaren}
\author[1]{Kirsty Hassall}
\affil[1]{Intelligent Data Ecosystems, Rothamsted Research, UK}
\affil[2]{Department of Crop Production Ecology, Swedish University of Agricultural Science, Sweden}
\date{\today}

\begin{document}

\maketitle
\begin{abstract}
Grassland ecosystems support a wide range of species and provide key services including food production, carbon storage, biodiversity support, and flood mitigation. However, yield stability in these grassland systems is not yet well understood, with recent evidence suggesting water stress throughout summer and warmer temperatures in late summer reduce yield stability. In this study we investigate how grassland yield stability of the Park Grass Experiment, UK, has changed over time by developing a Bayesian time-varying Autoregressive and time-varying Generalised Autoregressive Conditional Heterogeneity model using the variance-parameterised Gamma likelihood function.
\end{abstract}

\section{Introduction}
Climate change is pushing Earth’s biosphere into conditions that have not been experienced for millennia and that are novel to present-day ecosystems \citep{bova2021seasonal}. To forecast how ecosystems will respond, a robust understanding of their stability is needed. Less stable ecosystems are more prone to disruptions in or losses of function, posing a greater risk to the biodiversity and human communities that depend on those ecosystems for life-supporting goods and services \citep{willcock2023earlier}. Grasslands are a globally important ecosystem that support a wide range of species and provide key services to human society including food production, carbon storage, biodiversity support, and flood mitigation \citep{bengtsson2019grasslands}. In the UK, grasslands cover nearly 40\% of the land area, and may be highly vulnerable to climate change. Grassland productivity in the south-east of England forecast to decline by as much as 48-50\% by 2080 if global temperature increases exceed 1.5$^\circ$C \citep{addy2022heteroskedastic}.\\
\\
Previous studies have found that lower grassland yields in the south-east of England are associated with warmer and drier years, which are predicted to increase under climate change \citep{addy2021changes, cashen1947influence}. However, the yield stability (or variance in yield over time) in these grassland systems is not yet well understood. Recent empirical evidence suggests that water stress throughout summer (between March and October) and warmer temperatures in late summer reduce yield stability \citep{macholdt2023long}, but these trends have not yet been formally modelled through time-series processes.\\
\\
Yields from grasslands ecosystems are known to follow an autoregressive lag-1 process and are highly correlated with yield in the previous year \citep{jenkinson1994trends}. This is likely due to high-yielding years resulting in higher seed production or higher allocation of resources to energy-storage organs (such as rhizomes), allowing the grassland to make better use of the conditions and available resources in the following year \citep{ma2021soil}. Such processes may also lead to autoregression in yield stability. Seeds produced or energy stored in good years could provide a buffer to unfavourable conditions in subsequent years, while low seed production and scarce energy storage in bad years could mean grasslands are subsequently more susceptible to the effects of favourable vs poor conditions. Similarly, shifts in grassland community composition and diversity in response to changing conditions may affect the overall productivity and/or resilience of the ecosystem, leading to patterns in stability over time \citep{zhang2022intensification, zhang2018climate}.\\
\\
In this study, we seek greater insight into how grassland yield stability has responded to change over time by using Autoregressive Conditional Heterogeneity (ARCH) and Generalised Autoregressive Conditional Heterogeneity (GARCH) processes. These are further extensions of time-series models enabling a formal approach to modelling yield stability. An ARCH process occurs when the squared residual error $\varepsilon^2$ is autoregressive \citep{engle1982autoregressive}. In addition to an ARCH process, a GARCH process also considers an additional moving average process of the error variance $\sigma^2$ \citep{bollerslev1986generalized}. Time-Varying Autoregression (TV-AR) and Time-Varying Generalised Autoregressive Conditional Heterogeneity (TV-GARCH) models estimate time-dependent coefficients similar to time-varying regression models \citep{zhang2012inference}. Although there are methods which consider the yield stability of cropping systems \citep{piepho1998methods}, these methods do not consider the time-varying GARCH process of grassland yields. Incorporating a TV-AR and a TV-GARCH process can provide insights into how sensitive mean and variance estimates, and therefore yield stability of grassland yields, are over time.\\
\\
With TV-AR and TV-GARCH processes being powerful functions which assess how AR and GARCH terms change over time, they are typically estimated by smoothed non-parametric functions \citep{hoover1998nonparametric} which may need regulating to prevent over-fitting of the data \citep{tibshirani1996regression}. If smoothed non-parametric functions are not regulated this can lead to unrealistic estimates and the misinterpretation of the smoothed relationship. These smoothed non-parametric functions can be considered as a linear combination of basis functions and basis coefficients, penalised-splines (P-splines) have introduced the global shrinkage of basis coefficients as a method to avoid over-fitting \citep{eilers1996flexible}. Bayesian approaches consider priors for P-splines to assume a zero-centred multivariate-Normal distribution on spline basis coefficients \citep{wood2017generalized} with a user defined penalty matrix and global smoothing parameter representing the covariance structure \citep{lang2004bayesian}. However, the method of P-splines and global shrinkage can dominate the smoothing of a non-parametric functions. Horseshoe priors \citep{carvalho2010horseshoe} have been developed to allow more localised shrinkage for each coefficient through a series of zero-centred Normal distributions. Although they are more robust in dealing with over-fitting, posterior estimates from Horseshoe priors are often correlated with no formalised prior structure and little understanding of the influence of covariate shrinkage introduced by localised shrinkage parameters. We compare how different prior specifications regulate smoothed non-parametric functions for TV-AR and TV-GARCH processes.\\
\\
To compare how different prior specifications regulate smoothed non-parametric functions, we first generalise the global shrinkage parameter as a covariance matrix by introducing an Inverse-Wishart prior \citep{kass2006default} on the variance component of a zero-centred multivariate-Normal distribution. Additionally, we develop and apply multivariate Horseshoe priors for smooth non-parametric time-varying functions by introducing the LKJ-prior \citep{lewandowski2009generating} on the correlation structure of a decomposed covariance matrix \citep{barnard2000modeling}. The influence on the effect of global and localised shrinkage is compared for these distinct prior specifications. Two simulation studies are presented addressing how different prior specifications for basis coefficients of the non-parametric TV-AR(1) and TV-GARCH(1, 1) processes influence model prediction and regulate the estimated time-varying function. \\
\\
With TV-AR and TV-GARCH processes used in economics and the medical sciences \citep{giudici2023bayesian}, we use these processes to better understand and formalise how autoregressive and conditional heterogeneous terms change over time for grassland yield using data from the Park Grass Experiment, Hertfordshire. To ensure mean estimates of yield are strictly non-negative, we develop a Bayesian TV-AR and TV-GARCH model using the variance-parameterised Gamma likelihood function \citep{addy2022heteroskedastic} along with a mean-variance relationship across various treatments of the Park Grass Experiment.

\section{Data}
\subsection{Park Grass Experiment}
The herbage yield data analysed in this work comes from the Park Grass Experiment at Rothamsted Research. Different fertilizer inputs affect the herbage yield from each plot (Figure 1) along with the species composition of the original meadowland, with different species composition across plots \citep{macdonald2018guide}. Herbage yield data were taken from five treatments, two with no fertilizer inputs (plots 2.2 and 3), nitrogen inputs at 96 kg N ha $^{-1}$ plus minerals (plot 14.2), 48 kg N ha $^{-1}$ plus minerals (plot 16) and mineral only plot (plot 7). Each treatment has four liming sections with a maintained pH of 7, 6, 5 and unlimed. There are 20 sections of Park Grass used in this analysis (five treatments and 4 liming sections) with data from 1966 to 2018. Data from 2003 was missing for plots 14.2, 16, 2.2, 7 that were maintained at pH levels 6 and unlimed. Also from 2003 data was missing for plot 3 unlimed. Missing data was imputed using mean estimates from regression terms and TV-AR(1) estimates \citep{little2019statistical}. \\
\\
One of the main objectives of this study is to apply time-varying time-series approaches to herbage yield data to better understand, integrate and formally model how AR and GARCH processes have changed over time and if these processes have been influenced by climate or environmental change. For example, yields from the late-1990s to early-2000s were slightly more variable compared to any other time-period (Figure 1). We also develop a TV-GARCH model including a heteroskedastic process fitted as a mean-variance relationship with a Gamma model. It is known that climatic conditions affect grassland production and with global temperatures likely to exceed 1.5$^{\circ}$C by 2100 \citep{IPCC2023}, understanding how AR and GARCH processes change with time will better inform how grassland production will change over time. Seasonal mean temperature ($^{\circ}$C) and total rainfall (mm) were provided by the Rothamsted Meteorological Station from 1966 to 2018.

\section{Smoothed Functions and Priors}
\subsection{Non-Parametric Smoothed Functions}
Consider $f(x)$ to be a smooth non-parametric function between predictor $x$ and response $y$. Estimates of the smoothed function $f(x)$ can be considered as a linear combination of basis functions $B_p$ and basis coefficients $\theta_p$, over $p$ basis dimensions,
\begin{equation*}
f(x_i| \theta) = \sum_{p = 1}^{m}\theta_{p}B_{p}(x_{i})
\end{equation*}
\citep{wood2017generalized}. In this study we only consider $p$ equidistant Gaussian functions over a total of $m$ dimensions. Although the method of non-parametric smoothing is considered flexible, when $\theta_p$ is un-penalised this can result in the smoothed function $f(x)$ over-fitting the data. Consequences of over-fitting can result in unrealistic estimates of the smooth relationship which influences the interpretation of the model and the overall conclusions of the analysis.

\subsection{P-Splines Priors and Global Shrinkage}
Shrinking of the basis coefficient's sample space compensates for model over-fitting by narrowing the plausible sample region for each coefficient. This results in penalised basis coefficients $\theta_p$ and Penalised-splines (P-splines) \citep{tibshirani1996regression}. To obtain P-splines in a Bayesian context a prior distribution for $\theta_p$ and a global smoothing parameter ${\lambda^G}$ are introduced. Priors for Basis coefficients $\theta_p$ are given as zero-centred multivariate Normal distribution, $\boldsymbol{\theta} | \lambda \sim \text{MVN}(0, {\lambda^G}^2 K^{-1})$ with $\boldsymbol{\theta}$ a vector of $\theta_p$ terms \citep{wood2017generalized}, and $K$ a $m$-by-$m$ penalty matrix defined as the squared difference on adjacent basis coefficients \citep{eilers1996flexible}. Although the zero-centred multivariate-Normal prior for $\theta_p$ is proper, $K$ is rank deficient and the prior for the basis coefficients is now given as an improper multivariate-Normal distribution, $\theta_p | \lambda^G \propto \text{exp}\left(-\frac{1}{2{\lambda^G}^2}\boldsymbol{\theta}'K\boldsymbol{\theta} \right)$ \citep{lang2004bayesian}. Prior densities for the global shrinkage parameter have been given as $\lambda^G \sim \text{Inverse-Gamma}(1, 0.005)$, where a half-Cauchy, $\lambda^G \sim \text{C}^{+}(0, 1)$ is preferred to prevent the prior dominating posterior estimates of global shrinkage \citep{gelman2006prior}. The properties of the additional smoothing parameter is as $\lambda^G \to 0$ the relationship becomes a straight line, and as $\lambda^G \to \infty$ the relationship tends to an un-penalised relationship.

\subsection{Horseshoe Priors and Local Shrinkage}
Although global shrinkage may resolve the issue of over-fitting of smoothed functions, individual basis coefficients $\theta_p$ may become too localised around zero and become dominated by global shrinkage \citep{faulkner2018locally}. Horseshoe priors introduce localised shrinkage parameter ($\lambda_{\theta_p}$) by decomposing the variance structure for each basis coefficient prior into local and global shrinkage (${\lambda^G_\theta}$) parameters. Where priors for $\theta_p$ and $\lambda_{\theta_p}$ are given as $\theta_p | \lambda_{\theta_p}, \lambda^G_\theta \sim N(0, \lambda^2_{\theta_p}{\lambda^G_\theta}^2)$ and $\lambda_{\theta_p} \sim C^{+} (0, 1)$ \citep{carvalho2010horseshoe}. The introduction of Horseshoe priors allows for $\theta_p$ coefficients to assume larger values which may otherwise be smoothed out with a single global shrinkage parameter. Estimating the shrinkage coefficient $\kappa_{\theta_p}$ for each $\lambda_{\theta_p}$ is given as\[
\kappa_{\theta_p} = \frac{1}{1 + {\lambda^G_\theta}^2 \lambda^2_{\theta_p}}\] \citep{carvalho2010horseshoe}. The global smoothing parameter $\lambda^G_\theta$ is assumed to be 1 for examples presented in this study, and given $\lambda_{\theta_p} \sim C^{+} (0, 1)$ the prior of $\kappa_{\theta_p}$ can be shown to be proportional to a $\text{Beta}(0.5, 0.5)$. As $\kappa_{\theta_p} \to 0$ there is no shrinkage of coefficient $\theta_p$, and as $\kappa_{\theta_p} \to 1$ there is total shrinkage of coefficient $\theta_p$ to zero (Supplementary Figure 1).

\subsection{Covariance Structures for Global Shrinkage}
With the introduction of local shrinkage priors to $\theta_p$, the underlying prior for $\theta_p$ has also changed from a zero-centred multivariate-Normal distribution for all $m$ terms to $m$ zero-centred Normal distributions. A series of $m$ zero-centred Normal distributions do provide adequate local and global shrinkage for smoothed function \citep{faulkner2020horseshoe}. However, posterior estimates of $m$ total basis coefficients are often correlated and covariate shrinkage of basis coefficients is not identified or regulated with a series of $m$ univariate Horseshoe priors.\\
\\
Although a covariance structure for a zero-centred multivariate-Normal prior on basis coefficients is pre-specified for a global shrinkage parameter $\lambda^G$ given by penalty matrix $K$ \citep{lang2004bayesian}, $K$ is rank deficient and often specified as the squared difference on adjacent basis coefficients. An Inverse-Wishart prior on the covariance matrix of the zero-centred multivariate-Normal prior is a proper prior for global shrinkage with a covariance structure of basis coefficients \citep{kass2006default}. Now $\boldsymbol{\theta} | \Sigma_\theta, \psi_\theta \sim \text{MVN}(0, \Sigma_\theta)$ and $\Sigma_\theta | \psi_\theta \sim \text{Inv-Wishart} (I_m, \psi_\theta)$, with $I_m$ the identity scale matrix for $m$ coefficients and $\psi_\theta$ defined the degrees of freedom. As $\psi_\theta \to \infty$ the Inverse-Wishart prior contributes to more shrinkage than the Horseshoe prior (Supplementary Figure 1), this results in a shrinking of the prior sample space for $\theta_p$ (Supplementary Figure 2). Therefore, there is a lack of independence between the number of $\theta_p$ and the total shrinkage of basis coefficients, as the degrees of freedom for the Inverse-Wishart distribution needs to be larger than the number of total $m$ basis functions. In this study we set $\psi_\theta$ to be sufficiently large at 20 degrees of freedom. $\Sigma_\theta$ can be considered as the generalised covariance form of global shrinkage for basis coefficients without the need of prior specification of a penalty matrix $K$. With the introduction of a covariance structure to the shrinkage parameter the generalised covariate shrinkage coefficient is given as\[\kappa_{\theta_{pq}} = \frac{1}{1 + \Sigma^*_{\theta_{pq}}}.
\] $\Sigma^*$ is given as the absolute value of covariance matrix $\Sigma$ as the prior sampling space for $\theta$ is symmetric. However, $\Sigma_\theta$ is not specified with local shrinkage parameters and further considerations of the decomposition of the covariance structure should be considered to allow for localised shrinkage and co-shrinkage.

\subsection{Correlation Structures for Local Shrinkage}
For a multivariate Horseshoe prior, consider the vector of basis coefficients and local shrinkage parameters as $\boldsymbol{\theta}$ and $\boldsymbol{\lambda_\theta}$. The Horseshoe prior for incorporating a covariance matrix for spline basis coefficients decomposes $\Sigma_\theta$ into local and global shrinkage terms along with a correlation structure $\boldsymbol{\theta} | \boldsymbol{\lambda_\theta}, \lambda^G_\theta, \Omega_\theta \sim \text{MVN}(0, {\lambda^G_\theta}^2\boldsymbol{\lambda_\theta}\boldsymbol{\lambda}^T_{\boldsymbol{\theta}} \Omega_\theta)$ \citep{barnard2000modeling}. With $\Omega_\theta$ given as the basis coefficient correlation matrix with the prior covariance matrix of $\boldsymbol{\theta}$ given as ${\lambda^G_\theta}^2\boldsymbol{\lambda_\theta}\boldsymbol{\lambda}^T_{\boldsymbol{\theta}} \Omega_\theta$. The prior covariance matrix can be further decomposed into $\Sigma_\theta = {\lambda^G_\theta}^2\boldsymbol{\lambda_\theta}\boldsymbol{\lambda}^T_{\boldsymbol{\theta}} L_\theta L^{'}_\theta$. $L_\theta$ is the Cholesky decomposition of the correlation matrix $\Omega_\theta$. The prior for $\Omega_\theta$ is given as an LKJ($\phi_\theta$) distribution \citep{lewandowski2009generating} and shape parameter $\phi_\theta > 0$. With $\phi_\theta = 1$ the prior is uniform over all possible correlation values, $0 < \phi_\theta < 1$ the prior has higher densities around -1 and 1 with a wider ridge along bivariate parameter estimates, and as $\phi_\theta \to \infty$ the prior is localised around zero where there is a narrower ridge along bivariate parameter estimates (Figure 2). To regulate $\phi_\theta$ estimates, a half-Cauchy$(0, 1)$ prior is assumed. As the correlation between basis functions increases the effect of covariate shrinkage decreases, whereas as the correlation between coefficients decreases this introduces more covariate shrinkage (Figure 2). By decomposing the multivariate-Normal covariance matrix into local shrinkage parameters with a correlation structure there is more capacity to regulate covariate shrinkage compared to assuming a Inverse-Wishart prior on the covariance matrix with 20 degrees of freedom (Figure 2 \& Supplementary Figure 2).

\section{Gamma Model}
\subsection{Mean and Variance Parameterisation}
Here we specify the reparameterisation of the Gamma model to include TV-AR and TV-GARCH terms. Consider a non-negative continuous response variable $Y \sim \text{Gamma}(\alpha, \nu)$ with shape $\alpha$ and rate $\nu$ parameters. Mean parameterisation of Gamma likelihood is given as $\mu_i = \alpha/ \nu_i$. With mean estimates of the Gamma likelihood given by a linear combination of predictors $x_{ij}$ and $j$ model parameters $\beta_j$ via a link function and residual error ($\varepsilon_i$), such that \[
\eta_i = \sum_{j} x_{ij}\beta_j + \varepsilon_i.
\]The link function for the Gamma distribution is $\eta_i = \sum_{j} x_{ij}\beta_j$ such that $\mu_i = e^{\eta_i}$ to ensure positive estimates of the response \citep{dobson1990introduction}. Estimates of $\nu$ are given as $\nu_i = \alpha/\mu_i$ and the likelihood function is reparameterised by mean estimates $\mu$ and shape parameter $\alpha$, the likelihood is given as $p(y|\mu, \alpha)$.\\
\\
The variance from a Gamma process is given as $\sigma^2_i = \mu^2_i / \alpha$ \citep{GLM}, and $1/\alpha$ is the measure of overall dispersion. Variance structures can be modelled by a smoothed non-parametric function $g(x_i|\tau)$ and basis coefficients for the variance function $\tau$ \citep{rice1991estimating, silverman1985some}. With $\sigma^2_i = g(x_i|\tau)$, the shape parameter $\alpha$ is given by $\alpha_i = \sigma^2_i / \mu^2_i$ \citep{addy2022heteroskedastic}. The model likelihood can now be given in terms of mean and variance estimates $p(y|\mu, \sigma^2)$.

\subsection{Time-Varying Autoregression}
For an AR(1) process predictions at time $y_t$ become $y_t = a y_{t-1} + \varepsilon_t$ with predictions based upon the previous year. TV-AR processes allows for the investigation of how lag-$1$ processes vary over time. In this manuscript we focus on a TV-AR(1) processes as grassland herbage yields are know to have a lag-one autoregressive effect. TV-AR processes can be represented using spline basis functions \citep{zhang2012inference} modelling the autoregressive terms as a smoothed non-parametric relationship through time. Mean estimates from the link function for Gamma model with regression terms and a TV-AR(1) process becomes \[
\eta_{it} = \sum_{j}x_{ij}\beta_j + a_{t} y_{t-1} + \varepsilon_{it},
\]
with $a_{t} = \sum_{p}a_p B_p (t)$, representing the smooth relationship between the autoregressive lag-one term and time via TV-AR(1) basis coefficients $a_p$.

\subsection{Time-Varying GARCH Process}
An ARCH process of the $1$st order occurs when the variance of the model through its squared residual error $\varepsilon_t^2$ is autoregressive at lag $1$ \citep{engle1982autoregressive}. A $1$st order GARCH process also considers the moving average process of the error variance $\sigma^2$ along the time-series at lag $1$ \citep{bollerslev1986generalized}. A time-varying GARCH($1$, $1$) process can be estimated via smoothed non-parametric functions \citep{karmakar2021bayesian} to identify time-varying processes on the models error variance. In this study we only consider a TV-GARCH(1, 1) process formally represented as, \[
\sigma^2_t = e^{\tau_0 + b_t\varepsilon^2_{t-1} + c_t\sigma^2_{t-1}}.
\] TV-GARCH terms are given as smooth functions, $b_t = \sum_p b_p B_p (t)$ and $c_t = \sum_p c_p B_p (t)$, and $\tau_0$ is the variance intercept. The GARCH process presented in this study considers heterogeneity in model residuals conditional on mean estimates.

\section{Simulation Study}
For all simulations within this paper, each non-parametric function was estimated from 15 basis coefficients from 15 equally spaced Gaussian dimensions. Standardised residual error at time $t$ was used to model the TV-ARCH process due to the shifting of the model variance over time, with $\varepsilon^{s}_t = \varepsilon_t/ \sqrt{\sigma^2_t}$. A two-step model fitting process was used, first the TV-AR(1) processes was modelled followed by the TV-GARCH(1, 1) process. This follows the procedure to model GARCH processes from the RStan package \citep{RStan}.

\subsection{Individual TV-AR(1), TV-ARCH(1) and TV-GARCH(0, 1) Processes}
Three individual TV-AR(1), TV-ARCH(1) and TV-GARCH(0, 1) processes were simulated from 1000 time points with time-varying function
\[f(t) = 0.45 \text{sin}(1.25 \times 10^{-2}t) + (5 \times 10^{-4}t),\]
and $Y\sim \text{Gamma}(\mu, \sigma^2)$.
Although no other time-varying process was used for each example, we still estimated TV-AR(1) and TV-GARCH(1, 1) processes. This was to assess the stability in the the model approach and how different prior specifications would estimate a time-varying process even if one was not present. The fitted model across all prior specifications on basis functions are given in Supplementary Figures 3-5. \\
\\
Across all three simulations all three prior specifications for basis coefficients (Inverse-Wishart prior on a zero-centred multivariate-Normal covariance matrix, Horseshoe and multivariate Horseshoe) gave good estimates for the non-parametric time-varying processes. The Inverse-Wishart prior on the covariance matrix generally gave slightly wigglier estimates, with Horseshoe and multivariate Horseshoe priors on basis coefficients resulting in similar smoothed relationships, where the multivariate Horseshoe prior allowed for regulation of basis coefficients via the LKJ-distribution. Although the same function was used to simulate all time-varying processes, there was more variation associated with the time-varying ARCH(1) and GARCH(0, 1) processes compared to the TV-AR(1) process. Therefore, there is more uncertainty on variance estimates and the need for better specification on basis function regulation.\\
\\
All prior specifications on basis coefficients managed to adequately identify no relationship for the TV-AR(1) process for the TV-ARCH(1) and TV-GARCH(1) simulation. However, the use of the Inverse-Wishart prior provided wiggiler estimates for the TV-ARCH(1) and TV-GARCH(1) when no processes were simulated compared to the Horseshoe or multivariate Horseshoe prior specifications. This shows how strict co-regulation can dominate smooth non-parametric relationships on variance terms and how decomposing the prior covariance matrix of basis coefficients can adjust for strict co-regulation.

\subsection{Joint TV-AR(1) and TV-GARCH(1, 1) Process}
A TV-AR(1) and TV-GARCH(1, 1) process was simulated for 1000 time points from a mean-variance parameterised Gamma model $Y \sim \text{Gamma}(\mu, \sigma^2)$, with
\begin{align*}
a_t &= \text{sin}\left(\frac{t + 10}{75}\right) \left(\frac{50}{t + 100}\right), \\
b_t &= 0.5 \text{exp}\left(-\left(\frac{x - 500}{200} \right)^{10} \right) \\
c_t &= 2.7\times10^{-9}(t - 350)^3 + 5\times10^{-5}t.
\end{align*}
Where,
\[\mu_t = e^{\mu_0 + a_t y_{t-1}}
\]
and
\[\sigma^2 = e^{\tau_0 + b_t\varepsilon^2_{t-1} + c_t\sigma^2_{t-1}},
\]with mean and variance intercepts $\mu_0 = 3$ and $\tau_0 = 2.25$. A selection of different functions for $a_t$, $b_t$ and $c_t$ were selected to assess the capacity of basis coefficients to deal with over-fitting through the decomposing of priors into local and global shrinkage terms with a correlation structure. \\
\\
The LOOIC \citep{vehtari2017practical} was chosen to compare the posterior predictive fit of each model. There was no overwhelming differences in the posterior predictive fit between the models with Inverse-Wishart, Horseshoe or multivariate Horseshoe (Supplementary Figure 6), suggesting good model fits across all functions. There were no large differences between the use of Horseshoe and multivariate Horseshoe priors in the estimation of the TV-ARCH(1) process for multivariate Horseshoe priors (Figure 3). The use of an Inverse-Wishart prior on the covariance matrix of basis coefficients tended to provide wigglier estimates for the TV-ARCH and TV-GARCH functions and a slight lack of smoothing compared to other prior specifications.

\section{Park Grass Experiment}
\subsection{Model}
Herbage yields from 1966 to 2018 of the Park Grass experiment were modelled using a mean-variance parameterised Gamma model specified in Section 4 with $Y \sim \text{Gamma}(\mu, \sigma^2)$. Mean estimates are given as\[
\mu_{jt} = e^{\sum_{j}x_{ij}\beta_j + a_{t} y_{t-1}}
\] with $\beta_j$ a list of seasonally summarised variables presented in Table 1 estimating the effect of climate change on herbage production from 1966 to 2018. The TV-AR(1) process is estimated as a smoothed function $a_t$. To deal with heteroskedastic errors across mean estimates and time-varying generalised autoregressive conditional heteroskedasticity, the model variance was estimated as \[
\sigma^2_{it} = e^{\tau_i + b_t{\varepsilon^s}^2_{it-1} + c_t\sigma^2_{it-1}}.
\] The TV-GARCH(1, 1) process was estimated as a smooth function $b_t$ and $c_t$. The mean-variance relationship was modelled by a smoothed function $\tau_i = \sum_p \tau_pB_p(\mu_i)$, with $\tau_p$ basis coefficients and $B_p$ basis dimensions with multivariate Horseshoe priors on smooth terms. It should be noted that the mean-variance relationship and the TV-GARCH process are modelled from the mean estimate $\mu_{jt}$ given above. The full list of multivariate Horseshoe priors for the non-parametric mean-variance, TV-AR(1) and TV-GARCH(1, 1) functions of the Park Grass herbage yield model are given below.
\begingroup
\allowdisplaybreaks
\begin{align*} 
\boldsymbol{\tau} | \boldsymbol{\lambda_\tau}, \lambda_\tau^G, \Omega_\tau &\sim \text{MVN}(0, {\lambda^G_\tau}^2\boldsymbol{\lambda_\tau}\boldsymbol{\lambda}^T_{\boldsymbol{\tau}} \Omega_\tau) \\
\boldsymbol{a} | \boldsymbol{\lambda_{a}}, \lambda_{a}^G, \Omega_{a} &\sim \text{MVN}(0, {\lambda^G_a}^2\boldsymbol{\lambda_a}\boldsymbol{\lambda}^T_{\boldsymbol{a}} \Omega_a) \\
\boldsymbol{b} | \boldsymbol{\lambda_{b}}, \lambda_{b}^G, \Omega_{b} &\sim \text{MVN}(0, {\lambda^G_b}^2\boldsymbol{\lambda_b}\boldsymbol{\lambda}^T_{\boldsymbol{b}} \Omega_b) \\
\boldsymbol{c} | \boldsymbol{\lambda_{c}}, \lambda_{c}^G, \Omega_{c} &\sim \text{MVN}(0, {\lambda^G_c}^2\boldsymbol{\lambda_c}\boldsymbol{\lambda}^T_{\boldsymbol{c}} \Omega_c)\\
\lambda_{\tau_p} &\sim C^{+} (0, 1)\\
\lambda_{a_p} &\sim C^{+} (0, 1)\\
\lambda_{b_p} &\sim C^{+} (0, 1)\\
\lambda_{c_p} &\sim C^{+} (0, 1)\\
\lambda^G_{\tau} &\sim C^{+} (0, 1)\\
\lambda^G_{a} &\sim C^{+} (0, 1)\\
\lambda^G_{b} &\sim C^{+} (0, 1)\\
\lambda^G_{c} &\sim C^{+} (0, 1)\\
\Omega_{\tau} &\sim \text{LKJ}(\phi_\tau)\\
\Omega_{a} &\sim \text{LKJ}(\phi_a)\\
\Omega_{b} &\sim \text{LKJ}(\phi_b)\\
\Omega_{c} &\sim \text{LKJ}(\phi_c)\\
\phi_\tau &\sim C^{+} (0, 1)\\
\phi_a &\sim C^{+} (0, 1)\\
\phi_b &\sim C^{+} (0, 1)\\
\phi_c &\sim C^{+} (0, 1)
\end{align*}
\endgroup
$\lambda_\tau^G$, $\lambda_{a_t}^G$, $\lambda_{b_t}^G$ and $\lambda_{c_t}^G$ are the corresponding global smoothing parameters for each non-parametric function; $\boldsymbol{\tau}$, $\boldsymbol{a}$, $\boldsymbol{b}$ and $\boldsymbol{c}$ are a list of estimated terms from a non-parametric function; and $\Omega_{\tau}$, $\Omega_{a}$, $\Omega_{b}$, and $\Omega_{c}$ are the estimated correlation matrix sampled from a respective LKJ(${\phi_\tau}$), LKJ($\phi_{a}$), LKJ($\phi_{b}$), and LKJ($\phi_{c}$) prior. Comparisons of the Inverse-Wishart and Horseshoe prior specifications for basis coefficients from smooth functions $\tau_i$, $a_t$, $b_t$ and $c_t$ are also given (Figure 4).

\subsection{Climate Change and Time-Varying Autoregression}
From 1966 to 2018, warmer winter temperatures were negatively associated with lower herbage yields on Park Grass (Table 1). There was a slight positive linear association with total rainfall in the autumn, winter and summer and an increase in herbage yields (Table 1). However, there was an inverse polynomial effect of spring weather (both total rainfall and mean temperature) on herbage yield. Suggesting too much spring rainfall or warmer spring temperatures would result in lower herbage yields. The TV-AR(1) process increased in the 1960s, mid-1980s to late-1990s, and from the late-2000 to 2018, and decreased in the mid-1970s to 1980s, and mid-1990s to late-2000s (Figure 4). Overall there was a declining trend of the AR(1) process over time. The estimation of the TV-AR(1) was consistent across all prior specifications and shows the autoregressive process a lack of stationarity of the autoregressive process.

\subsection{Mean-Variance and Time-Varying Variance}
The mean-variance relationship was successfully modelled by a non-parametric smoothed function along with the TV-GARCH(1, 1) process (Figure 4). As herbage yields increase the variation around herbage yields also increases, therefore suggesting more yield instability (see yield plots in Figure 1).\\
\\
The estimated time-varying ARCH(1) process using Horseshoe priors and multivariate Horseshoe priors declined in the 1980s, and increased from the 1990s to the early 2000s.
For the Inverse-Wishart prior specification of the covariance matrix of basis coefficients the estimated TV-ARCH(1) was wigglier with more drastic changes (Figure 4). Slight changes in the TV-ARCH(1) process between 2010 and 2018 were not identified using Horseshoe or multivariate Horseshoe priors, but were identified using Inverse-Wishart priors on the covariance matrix of basis coefficients. Given the slight overfitting of the TV-ARCH(1) process from the first simulation study (Section 5.1), the regulation from the Horseshoe and multivariate Horseshoe priors of basis coefficients prevented the slight overfitting of the TV-ARCH(1) term compared to the Inverse-Wishart prior. The time-varying GARCH process was similar to the ARCH process with a slight increase in variance between the 1990s and mid-2000s. This was identified across all prior specifications on basis coefficients (Figure 4), with Inverse-Wishart prior estimating a slightly wigglier function. Therefore, with no real differences between the TV-AR(1) process across prior specifications, the use of Horseshoe and multivariate Horseshoe priors for basis coefficients was required for preventing slight overfitting of time-varying variance terms, with no large differences in the posterior predictive fit across all prior specifications (Supplementary Figure 7).\\
\\
Model validation of the two-step fitting procedure for the Park Grass data (TV-AR(1) and then TV-GARCH(1, 1)) follows the Bayesian workflow \citep{gelman2020bayesian, gabry2019visualization}. The Posterior Predictive Distribution (PPD) for mean estimates from the TV-AR(1) model represented the data well for all prior specifications (Supplementary Figure 8). Similarity for standard deviation estimates for the TV-GARCH(1, 1) model, the PPD of the variance estimates in the Park Grass data were well represented (Supplementary Figure 8). Further diagnostics for LOOIC sampling and Pareto-smoothed importance sampling (PSIS) are given in Supplementary Figure 9, with most observations below the Pareto 0.7 threshold \citep{vehtari2017practical}.

\section{Discussion}
We proposed the use of the multivariate Horseshoe prior on non-parametric basis coefficients when modelling time-varying AR(1) and GARCH(1, 1) processes of the herbage yield data from the Park Grass Experiment, UK. The use of the multivariate Horseshoe prior decomposes the covariance matrix of a zero-centred multivariate-Normal prior into a correlation matrix along with local and global shrinkage terms, with a LKJ-prior on the correlation matrix along with half-Cauchy(0, 1) priors on local and global shrinkage terms. The multivariate Horseshoe prior specification on non-parametric basis coefficients successfully modelled the mean-variance relationship, TV-AR(1) and TV-GARCH(1, 1) processes of Park Grass herbage yield from 1966 to 2018. We introduced and assessed the use of the Inverse-Wishart prior \citep{kass2006default} for a non-decomposed covariance matrix on a zero-centred Normal prior on basis coefficients as a multivariate approach to global shrinkage of basis coefficients. Compared to previously defined P-splines \citep{lang2004bayesian} where a user defined penalty matrix $K$ on the covariance structure of zero-centred multivariate-Normal distribution is used, the Inverse-Wishart prior is now proper and not rank deficient. There were no large differences in the parameterisation of the TV-AR process across the three different prior specifications (Inverse-Wishart, Horseshoe and multivariate Horseshoe). However, differences in the prior specifications were more pronounced in the estimation of the TV-GARCH(1, 1) process suggesting the co-regulation of basis coefficients from the Inverse-Wishart prior provided less smoothing of time-varying autocorrelation on variance terms. The decomposition of a variance matrix was suggested by \cite{barnard2000modeling} and in the application to time-varying processes we agree this decomposition provides more flexibility in covariate shrinkage than the Inverse-Wishart prior. Further work could be proposed into the use of the LKJ-prior on the correlation matrix of a regulated Horseshoe priors \citep{piironen2017sparsity}, and how co-regulation could be achieved. Other prior distributions may also be used for the basis coefficients correlation terms such as the Beta distribution \citep{gokhale1982assessment}. However, this would require $\frac{3m!}{2!(m-2)}$ priors and with two shape parameters each which could prove computationally intensive. \\
\\
For the Park Grass data from 1966 to 2018 we used a generalised mean-variance parameterisation of the Gamma function to model the TV-AR(1) and TV-GARCH(1, 1) processes. TV-AR(1) and TV-GARCH(1, 1) processes have been used previously for count GLMs \citep{roy2021time}, however we generalised this process for continuous non-negative data via a Gamma model. From the analysis of the Park Grass herbage yield data we conclude similar results to \citep{addy2021changes, cashen1947influence} that warmer temperatures in spring and winter resulted in lower herbage yields. In both \cite{addy2022heteroskedastic} and this study, we observed a consistently positive autocorrelation between yields, indicating that yield-related ecological processes such as seed production and/or energy storage play a role in determining yields in the following year \citep{ma2021soil}. However, in this study, by modelling a TV-AR(1) process of herbage yield we also found the autoregressive process to be non-stationary, but always positive, alongside the influence of climate change and warmer temperatures (Figure 4). Although the mechanism behind the time-varying pattern in yield autocorrelation is not known, it is of interest especially for forecasting future herbage yields into the mid and late-21st century \citep{addy2022heteroskedastic}.\\
\\
By formalising the modelling procedure to include time-varying stochastic time series processes on Park Grass, the TV-GARCH(1, 1) process showed how the variance around the model prediction of the Park Grass herbage yield data are autoregressive and have slightly increased between the 1990s and early-2000s, confirming previous findings in yield stability changing during the 1990s \citep{macholdt2023long, dodd1994stability}. With the mean-variance relationship, we also found that changes in yield stability over time were more severe for higher-yield sections of the Park Grass Experiment. \cite{macholdt2023long} observed a peak in yield variance in the 1990s, at least partially explained by an increase in summer water stress and temperatures during this decade. In contrast, the identification of the TV-GARCH process suggests that variance increased in the 1990s and remained high from the 2000s onwards (Figure 4), and meteorological trends from the Park Grass site do suggest a sustained increase in temperatures from the 1990s onwards \citep{addy2022heteroskedastic}. \cite{zhang2018climate} observed that higher temperatures could decrease grassland species richness and consequently yield stability in northern China, so higher temperatures may have had a similar effect on Park Grass. Another ecologically significant event that occurred during the 1990s was a drop in nitrogen and sulphur deposition, which was associated with an increase in soil pH and an increased proportion of legume species in the herbage cuts \citep{storkey2015grassland, macholdt2023long}, and this may also have affected yield stability.\\
\\
Our conclusions differ slightly to previous work on identifying the drivers of yield stability in Park Grass, with \cite{macholdt2023long} observing an increase in variance in the 1990s but a subsequent decrease, whilst we identified a sustained increase in variance from the 1990s onwards. We modelled different treatments (only two were consistent), and we also modelled yield variance after estimating mean predictions from weather variables and a TV-AR(1) process. Our finding show that Park Grass yield stability has followed a consistently different pattern since the 1990s and this has important implications for understanding how grasslands respond to climate change. However, it is important to note that the range of values from the time-varying GARCH(1, 1) process was smaller compared to the mean-variance process (Figure 4). Overall, our paper demonstrates that variance functions are highly sensitive and dependent on regulation, and considerations should be given when trying to attribute changes in variance over time to other stochastic processes such as weather.

\bibliographystyle{agsm}
\bibliography{RefList2}

@article{gelman2006prior,
  title={Prior distributions for variance parameters in hierarchical models (comment on article by {B}rowne and {D}raper)},
  author={Gelman, Andrew and others},
  journal={Bayesian {A}nalysis},
  volume={1},
  number={3},
  pages={515--534},
  year={2006},
  publisher={International Society for Bayesian Analysis}
}

@book{GLM,
  title={Generalized Linear Models},
  author={McCullagh, P and Nelder, J, A},
  year={1983},
  publisher={Chapman and Hall}
}

@book{macdonald2018guide,
  title={Guide to the Classical and Other Long-Term Experiments, Datasets and Sample Archive},
  author={Macdonald, Andy},
  year={2018},
  publisher={Rothamsted {R}esearch}
}

@article{eilers1996flexible,
  title={Flexible smoothing with {B}-splines and penalties},
  author={Eilers, Paul HC and Marx, Brian D},
  journal={Statistical Science},
  volume={11},
  number={2},
  pages={89--121},
  year={1996},
  publisher={Institute of Mathematical Statistics}
}

@article{lang2004bayesian,
  title={Bayesian {P}-splines},
  author={Lang, Stefan and Brezger, Andreas},
  journal={Journal of Computational and Graphical Statistics},
  volume={13},
  number={1},
  pages={183--212},
  year={2004},
  publisher={Taylor \& Francis}
}

@book{wood2017generalized,
  title={Generalized {A}dditive {M}odels: an {I}ntroduction with {R}},
  author={Wood, Simon N},
  year={2017},
  publisher={CRC press}
}

@article{vehtari2017practical,
  title={Practical {B}ayesian model evaluation using leave-one-out cross-validation and {WAIC}},
  author={Vehtari, Aki and Gelman, Andrew and Gabry, Jonah},
  journal={Statistics and Computing},
  volume={27},
  number={5},
  pages={1413--1432},
  year={2017},
  publisher={Springer}
}

@article{tibshirani1996regression,
  title={Regression shrinkage and selection via the lasso},
  author={Tibshirani, Robert},
  journal={Journal of the Royal Statistical Society: Series B (Methodological)},
  volume={58},
  number={1},
  pages={267--288},
  year={1996},
  publisher={Wiley Online Library}
}

@book{dobson1990introduction,
  title={An {I}ntroduction to {G}eneralized {L}inear {M}odels},
  author={Dobson, Annette J},
  year={1990},
  publisher={CRC press}
}

@article{rice1991estimating,
  title={Estimating the mean and covariance structure nonparametrically when the data are curves},
  author={Rice, John A and Silverman, Bernard W},
  journal={Journal of the Royal Statistical Society: Series B (Methodological)},
  volume={53},
  number={1},
  pages={233--243},
  year={1991},
  publisher={Wiley Online Library}
}

@article{silverman1985some,
  title={Some aspects of the spline smoothing approach to non-parametric regression curve fitting},
  author={Silverman, Bernhard W},
  journal={Journal of the Royal Statistical Society: Series B (Methodological)},
  volume={47},
  number={1},
  pages={1--21},
  year={1985},
  publisher={Wiley Online Library}
}

@article{faulkner2020horseshoe,
  title={Horseshoe-based Bayesian nonparametric estimation of effective population size trajectories},
  author={Faulkner, James R and Magee, Andrew F and Shapiro, Beth and Minin, Vladimir N},
  journal={Biometrics},
  volume={76},
  number={3},
  pages={677--690},
  year={2020},
  publisher={Wiley Online Library}
}

@article{carvalho2010horseshoe,
  title={The horseshoe estimator for sparse signals},
  author={Carvalho, Carlos M and Polson, Nicholas G and Scott, James G},
  journal={Biometrika},
  volume={97},
  number={2},
  pages={465--480},
  year={2010},
  publisher={Oxford University Press}
}

@article{faulkner2018locally,
  title={Locally adaptive smoothing with Markov random fields and shrinkage priors},
  author={Faulkner, James R and Minin, Vladimir N},
  journal={{B}ayesian {A}nalysis},
  volume={13},
  number={1},
  pages={225},
  year={2018},
  publisher={NIH Public Access}
}

@article{bollerslev1986generalized,
  title={Generalized autoregressive conditional heteroskedasticity},
  author={Bollerslev, Tim},
  journal={{J}ournal of {E}conometrics},
  volume={31},
  number={3},
  pages={307--327},
  year={1986},
  publisher={Elsevier}
}

@Misc{RStan,
  title = {{RStan}: the {R} interface to {Stan}},
  author = {{Stan Development Team}},
  note = {R package version 2.21.3},
  year = {2021},
  url = {https://mc-stan.org/}
}

@article{addy2022heteroskedastic,
  title={A heteroskedastic model of {P}ark {G}rass spring hay yields in response to weather suggests continuing yield decline with climate change in future decades},
  author={Addy, John WG and Ellis, Richard H and MacLaren, Chloe and Macdonald, Andy J and Semenov, Mikhail A and Mead, Andrew},
  journal={Journal of the {R}oyal {S}ociety {I}nterface},
  volume={19},
  number={193},
  year={2022},
  publisher={The Royal Society}
}

@article{lewandowski2009generating,
  title={Generating random correlation matrices based on vines and extended onion method},
  author={Lewandowski, Daniel and Kurowicka, Dorota and Joe, Harry},
  journal={Journal of {M}ultivariate {A}nalysis},
  volume={100},
  number={9},
  pages={1989--2001},
  year={2009},
  publisher={Elsevier}
}

@misc{IPCC2023,
  title = {{S}ynthesis {R}eport of the {IPCC} {S}ixth {A}ssessment {R}eport},
  author = {{IPCC}},
  year = {2023},
  url = {https://www.ipcc.ch/report/ar6/syr/}
}

@article{cashen1947influence,
  title={The influence of rainfall on the yield and botanical composition of permanent grass at {R}othamsted},
  author={Cashen, Rose O},
  journal={The {J}ournal of {A}gricultural {S}cience},
  volume={37},
  number={1},
  pages={1--10},
  year={1947},
  publisher={{C}ambridge {U}niversity {P}ress}
}

@article{addy2021changes,
  title={Changes in agricultural climate in South-Eastern England from 1892 to 2016 and differences in cereal and permanent grassland yield},
  author={Addy, John WG and Ellis, Richard H and Macdonald, Andy J and Semenov, Mikhail A and Mead, Andrew},
  journal={{A}gricultural and {F}orest {M}eteorology},
  volume={308},
  pages={108560},
  year={2021},
  publisher={Elsevier}
}

@article{jenkinson1994trends,
  title={Trends in herbage yields over the last century on the {R}othamsted long-term continuous hay experiment},
  author={Jenkinson, DS and Potts, JM and Perry, JN and Barnett, V and Coleman, K and Johnston, AE},
journal={The {J}ournal of {A}gricultural {S}cience},
  volume={122},
  number={3},
  pages={365--374},
  year={1994},
  publisher={Cambridge {U}niversity {P}ress}
}

@article{hoover1998nonparametric,
  title={Nonparametric smoothing estimates of time-varying coefficient models with longitudinal data},
  author={Hoover, Donald R and Rice, John A and Wu, Colin O and Yang, Li-Ping},
  journal={Biometrika},
  volume={85},
  number={4},
  pages={809--822},
  year={1998},
  publisher={Oxford University Press}
}

@article{zhang2012inference,
  title={Inference of Time-varying Regression Models},
  author={Zhang, Ting and Wu, Wei Biao},
  journal={The {A}nnals of {S}tatistics},
  volume={40},
  number={3},
  pages={1376--1402},
  year={2012}
}

@article{karmakar2021bayesian,
  title={Bayesian modelling of time-varying conditional heteroscedasticity},
  author={Karmakar, Sayar and Roy, Arkaprava},
  journal={Bayesian {A}nalysis},
  volume={16},
  number={4},
  pages={1157--1185},
  year={2021},
  publisher={International Society for Bayesian Analysis}
}

@article{engle1982autoregressive,
  title={Autoregressive conditional heteroscedasticity with estimates of the variance of United Kingdom inflation},
  author={Engle, Robert F},
  journal={Econometrica},
  pages={987--1007},
  year={1982},
  publisher={JSTOR}
}

@article{giudici2023bayesian,
  title={Bayesian time-varying autoregressive models of COVID-19 epidemics},
  author={Giudici, Paolo and Tarantino, Barbara and Roy, Arkaprava},
  journal={{B}iometrical {J}ournal},
  volume={65},
  number={1},
  pages={2200054},
  year={2023},
  publisher={Wiley Online Library}
}

@article{roy2021time,
  title={Time-varying auto-regressive models for count time-series},
  author={Roy, Arkaprava and Karmakar, Sayar},
  journal={Electronic {J}ournal of {S}tatistics},
  volume={15},
  number={1},
  pages={2905--2938},
  year={2021},
  publisher={The Institute of Mathematical Statistics and the Bernoulli Society}
}

@article{piironen2017sparsity,
  title={Sparsity information and regularization in the {H}orseshoe and other shrinkage priors},
  author={Piironen, Juho and Vehtari, Aki},
  journal={{E}lectronic {J}ournal of {S}tatistics},
  volume={11},
  pages={5018--5051},
  year={2017}
}

@article{gokhale1982assessment,
  title={Assessment of a prior distribution for the correlation coefficient in a bivariate {N}ormal distribution},
  author={Gokhale, DV and Press, S James},
  journal={{J}ournal of the {R}oyal {S}tatistical {S}ociety: {S}eries {A} ({G}eneral)},
  volume={145},
  number={2},
  pages={237--249},
  year={1982},
  publisher={Wiley Online Library}
}

@book{little2019statistical,
  title={Statistical analysis with missing data},
  author={Little, Roderick JA and Rubin, Donald B},
  year={1986},
  publisher={John Wiley \& Sons}
}

@article{kass2006default,
  title={A default conjugate prior for variance components in generalized linear mixed models (comment on article by {B}rowne and {D}raper)},
  author={Kass, Robert E and Natarajan, Ranjini},
  journal={Bayesian {A}nalysis},
  volume={1},
  number={3},
  pages={535--542},
  year={2006},
  publisher={International Society for Bayesian Analysis}
}

@article{barnard2000modeling,
  title={Modeling covariance matrices in terms of standard deviations and correlations, with application to shrinkage},
  author={Barnard, John and McCulloch, Robert and Meng, Xiao-Li},
  journal={{S}tatistica {S}inica},
  pages={1281--1311},
  year={2000}
}

@article{dodd1994stability,
  title={Stability in the plant communities of the Park Grass Experiment: the relationships between species richness, soil p{H} and biomass variability},
  author={Dodd, Mike E and Silvertown, Jonathan and McConway, Kevin and Potts, Jacqueline and Crawley, Mick},
  journal={Philosophical {T}ransactions of the {R}oyal {S}ociety of {L}ondon. {S}eries {B}: {B}iological {S}ciences},
  volume={346},
  number={1316},
  pages={185--193},
  year={1994},
  publisher={The {R}oyal {S}ociety {L}ondon}
}

@article{macholdt2023long,
  title={Long-term trends in yield variance of temperate managed grassland},
  author={Macholdt, Janna and Hadasch, Steffen and Macdonald, Andrew and Perryman, Sarah and Piepho, Hans-Peter and Scott, Tony and Styczen, Merete Elisabeth and Storkey, Jonathan},
  journal={{A}gronomy for {S}ustainable {D}evelopment},
  volume={43},
  number={3},
  pages={37},
  year={2023},
  publisher={Springer}
}

@article{piepho1998methods,
  title={Methods for comparing the yield stability of cropping systems},
  author={Piepho, H-P},
  journal={{J}ournal of {A}gronomy and {C}rop {S}cience},
  volume={180},
  number={4},
  pages={193--213},
  year={1998},
  publisher={Wiley Online Library}
}

@article{gabry2019visualization,
  title={Visualization in {B}ayesian workflow},
  author={Gabry, Jonah and Simpson, Daniel and Vehtari, Aki and Betancourt, Michael and Gelman, Andrew},
  journal={Journal of the {R}oyal {S}tatistical {S}ociety {S}eries {A}: {S}tatistics in {S}ociety},
  volume={182},
  number={2},
  pages={389--402},
  year={2019},
  publisher={Oxford University Press}
}

@article{gelman2020bayesian,
  title={Bayesian workflow},
  author={Gelman, Andrew and Vehtari, Aki and Simpson, Daniel and Margossian, Charles C and Carpenter, Bob and Yao, Yuling and Kennedy, Lauren and Gabry, Jonah and B{\"u}rkner, Paul-Christian and Modr{\'a}k, Martin},
  journal={ar{X}iv preprint ar{X}iv:2011.01808},
  year={2020}
}

@article{bova2021seasonal,
  title={Seasonal origin of the thermal maxima at the Holocene and the last interglacial},
  author={Bova, Samantha and Rosenthal, Yair and Liu, Zhengyu and Godad, Shital P and Yan, Mi},
  journal={Nature},
  volume={589},
  number={7843},
  pages={548--553},
  year={2021},
  publisher={Nature Publishing Group UK London}
}

@article{willcock2023earlier,
  title={Earlier collapse of Anthropocene ecosystems driven by multiple faster and noisier drivers},
  author={Willcock, Simon and Cooper, Gregory S and Addy, John and Dearing, John A},
  journal={Nature {S}ustainability},
  pages={1--12},
  year={2023},
  publisher={Nature Publishing Group UK London}
}

@article{bengtsson2019grasslands,
  title={Grasslands—more important for ecosystem services than you might think},
  author={Bengtsson, J and Bullock, JM and Egoh, B and Everson, C and Everson, T and O'connor, T and O'farrell, Pj and Smith, HG and Lindborg, Regina},
  journal={Ecosphere},
  volume={10},
  number={2},
  pages={e02582},
  year={2019},
  publisher={Wiley Online Library}
}

@article{ma2021soil,
  title={Soil seed banks, alternative stable state theory, and ecosystem resilience},
  author={Ma, Miaojun and Collins, Scott L and Ratajczak, Zak and Du, Guozhen},
  journal={Bio{S}cience},
  volume={71},
  number={7},
  pages={697--707},
  year={2021},
  publisher={Oxford University Press}
}

@article{zhang2022intensification,
  title={Intensification of disturbance destabilizes productivity through effects on dominant species},
  author={Zhang, Feng and Bennett, Jonathan A and Zhang, Bin and Zhao, Mengli and Han, Guodong},
  journal={Ecological {I}ndicators},
  volume={143},
  pages={109383},
  year={2022},
  publisher={Elsevier}
}

@article{zhang2018climate,
  title={Climate variability decreases species richness and community stability in a temperate grassland},
  author={Zhang, Yunhai and Loreau, Michel and He, Nianpeng and Wang, Junbang and Pan, Qingmin and Bai, Yongfei and Han, Xingguo},
  journal={Oecologia},
  volume={188},
  pages={183--192},
  year={2018},
  publisher={Springer}
}

@article{storkey2015grassland,
  title={Grassland biodiversity bounces back from long-term nitrogen addition},
  author={Storkey, J and Macdonald, AJ and Poulton, PR and Scott, T and K{\"o}hler, IH and Schnyder, H and Goulding, KWT and Crawley, MJ},
  journal={Nature},
  volume={528},
  number={7582},
  pages={401--404},
  year={2015},
  publisher={Nature Publishing Group UK London}
}

\newpage
\section*{Figures and Tables}

\begin{figure}[h]
\begin{center}
\caption{Herbage yields (t dm ha$^{-1}$) from all Park Grass limed sections plots 14.2, 16, 7, 2.2 and 3 from 1966 to 2018. Limed sections are staggered from no liming (light grey) to pH maintained at 7 (black).}
\includegraphics[width = 6.5cm]{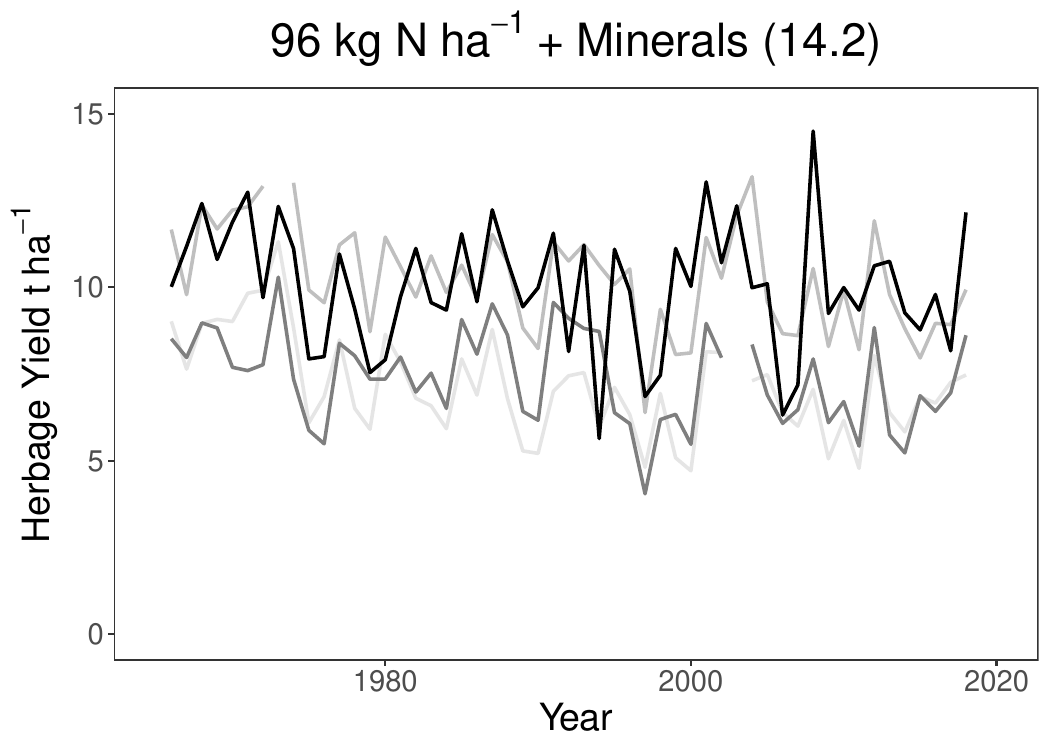}
\includegraphics[width = 6.5cm]{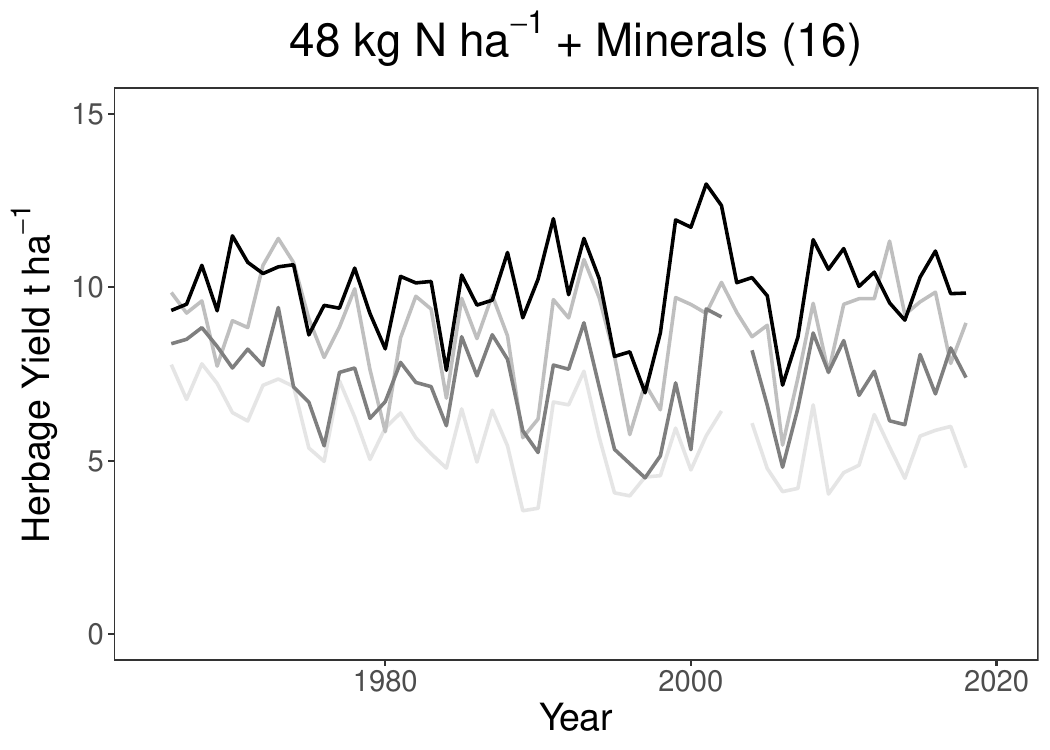}\\
\includegraphics[width = 6.5cm]{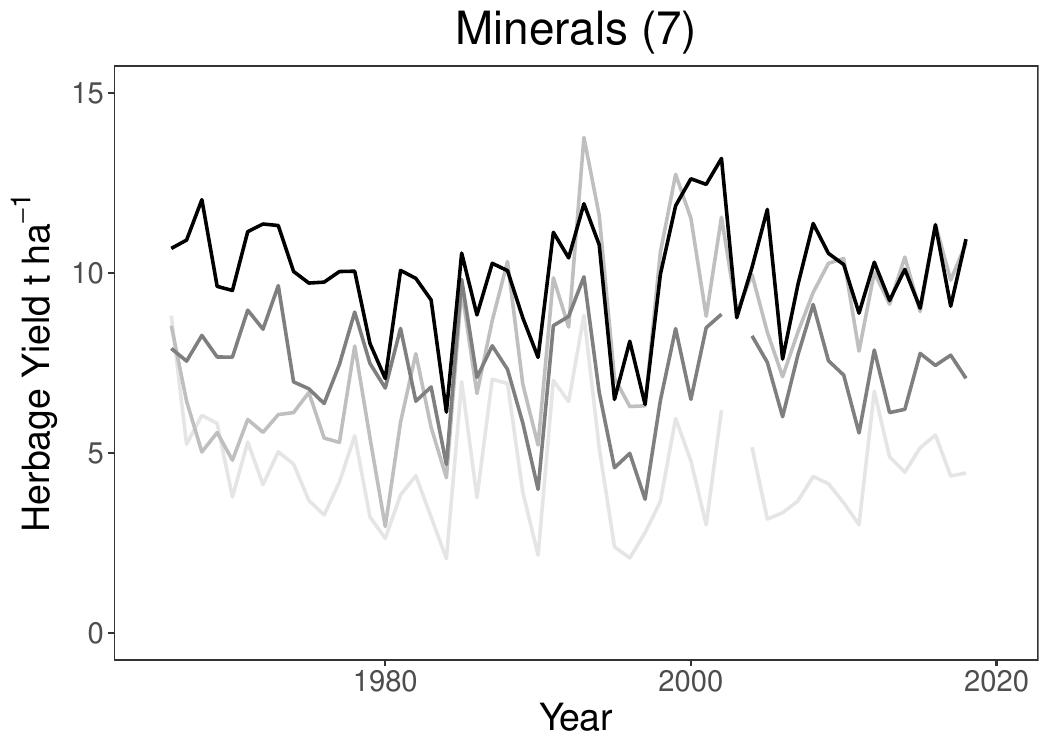}
\includegraphics[width = 6.5cm]{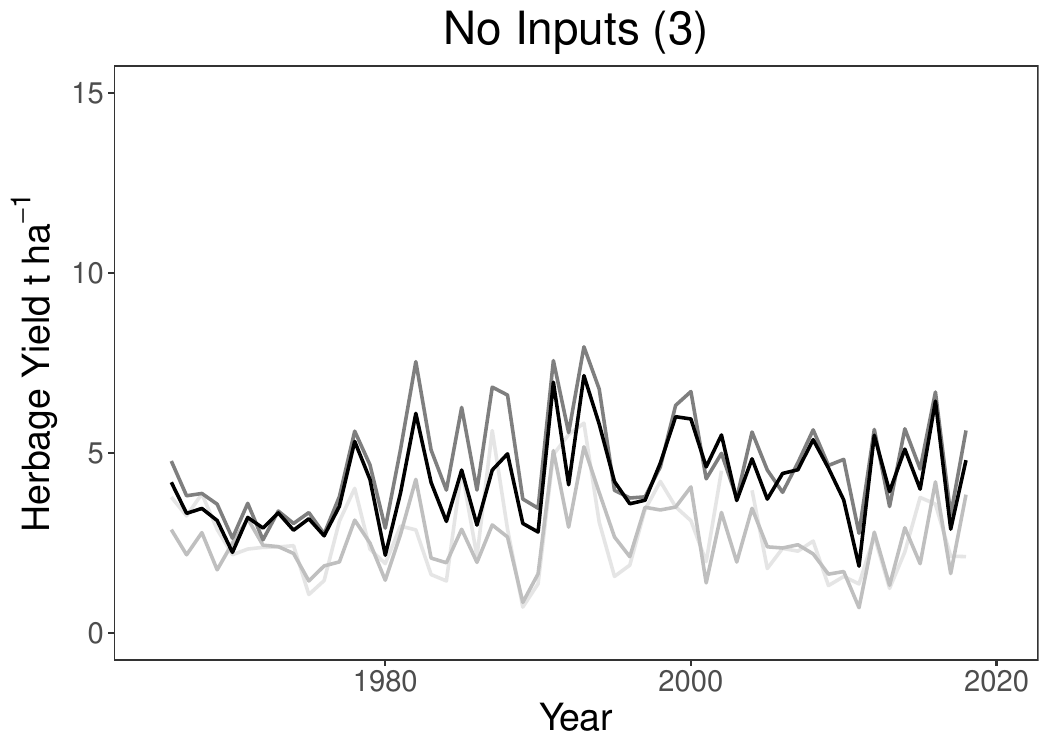}
\end{center}
\end{figure}

\begin{figure}[h]
\begin{center}
\caption{The distribution of correlation terms from 10000 simulations of the LKJ-distribution for different $\phi$ values (top row). The bivariate distribution of basis coefficients from a multivariate Horseshoe distribution with correlation terms sampled from a prior LKJ-distribution for different $\phi$ values (middle row). The influence of covariate shrinkage between basis coefficients with varying correlations sampled from 10000 simulations (bottom row).}
\includegraphics{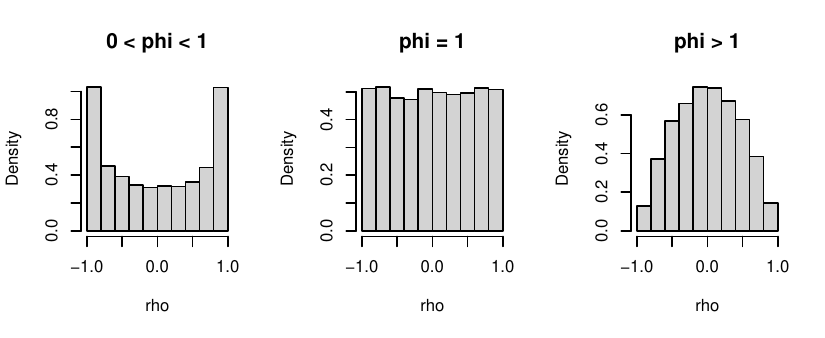}\\
\includegraphics{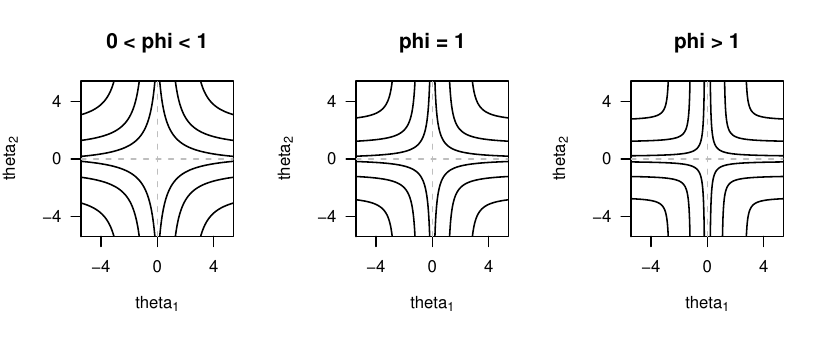}\\
\includegraphics{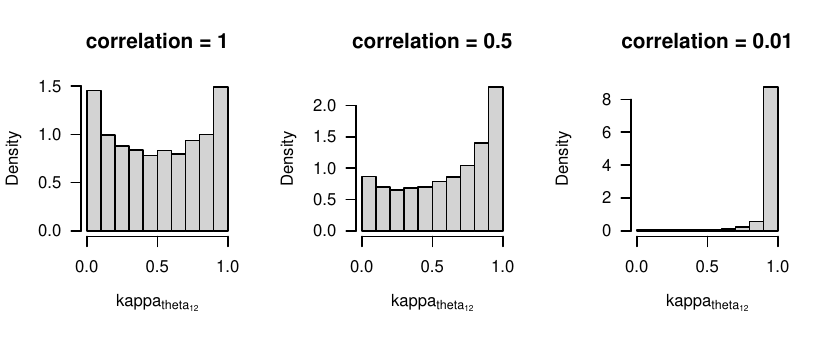}
\end{center}
\end{figure}

\newpage
\begin{figure}[h]
\begin{center}
\caption{The simulated TV-AR(1) and TV-GARCH(1, 1) processes for a mean-variance parameterised Gamma model (top row). The fitted TV-AR(1) (second row) and TV-GARCH(1, 1) (rows three and four) with 80\% credible intervals for each type of basis coefficient prior; Inverse-Wishart($20$) prior on covariance matrix (green), Horseshoe (blue) and multivariate Horseshoe (red). All simulated data and functions are given in grey.}
\includegraphics[width = 4.25cm]{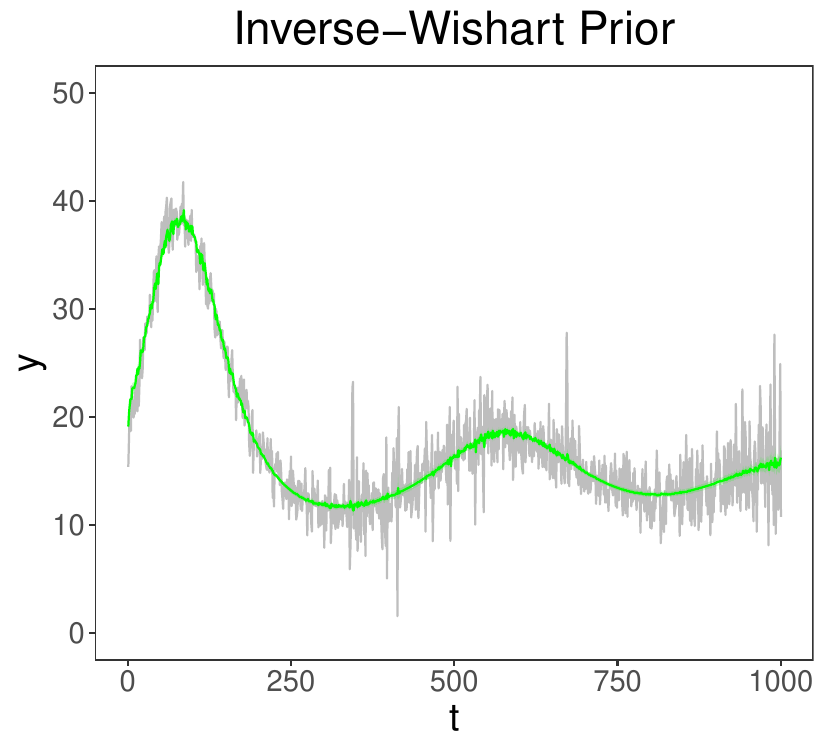}
\includegraphics[width = 4.25cm]{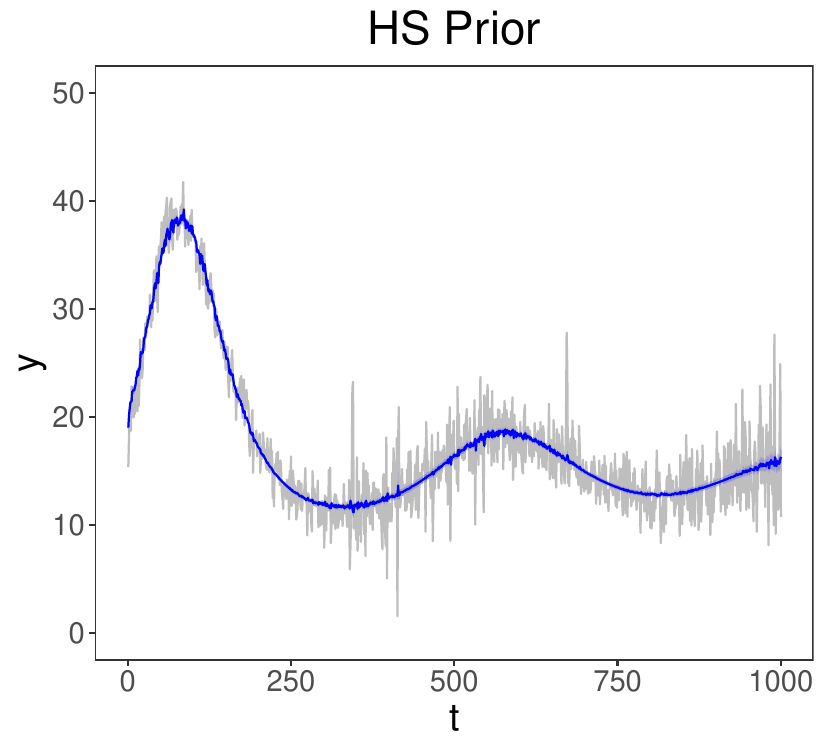}
\includegraphics[width = 4.25cm]{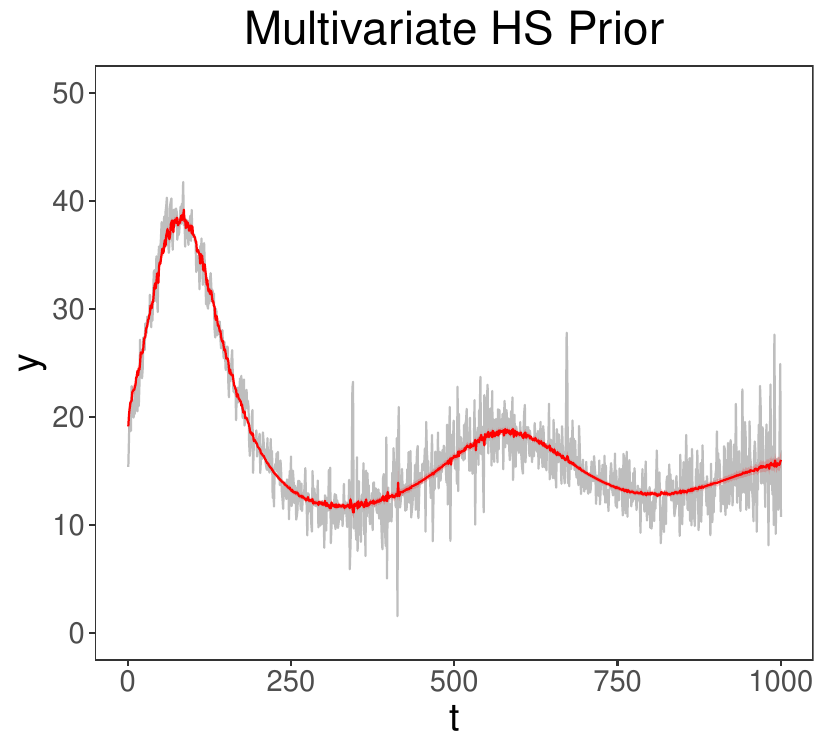}\\
\includegraphics[width = 4.25cm]{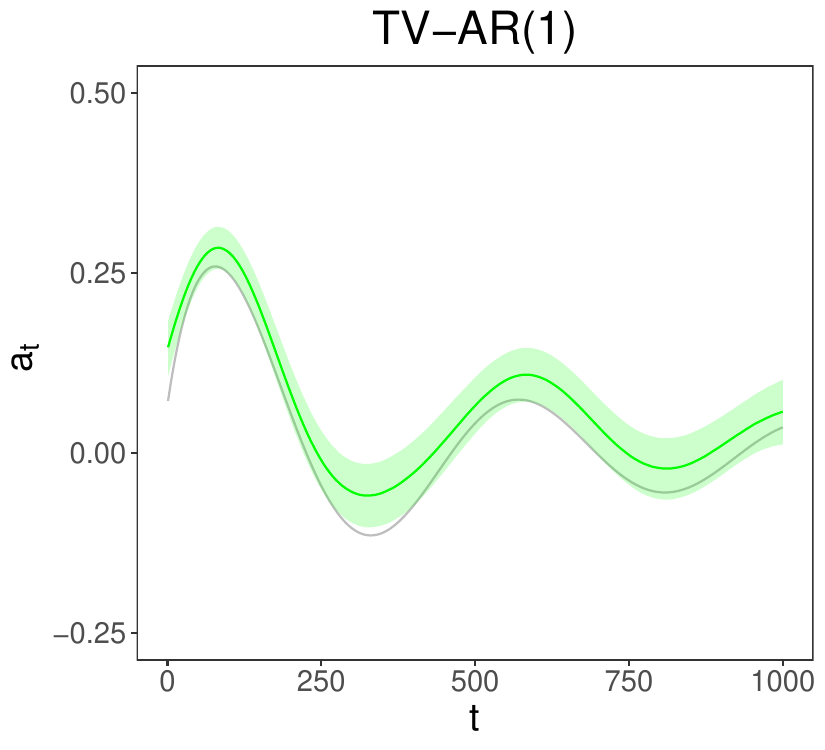}
\includegraphics[width = 4.25cm]{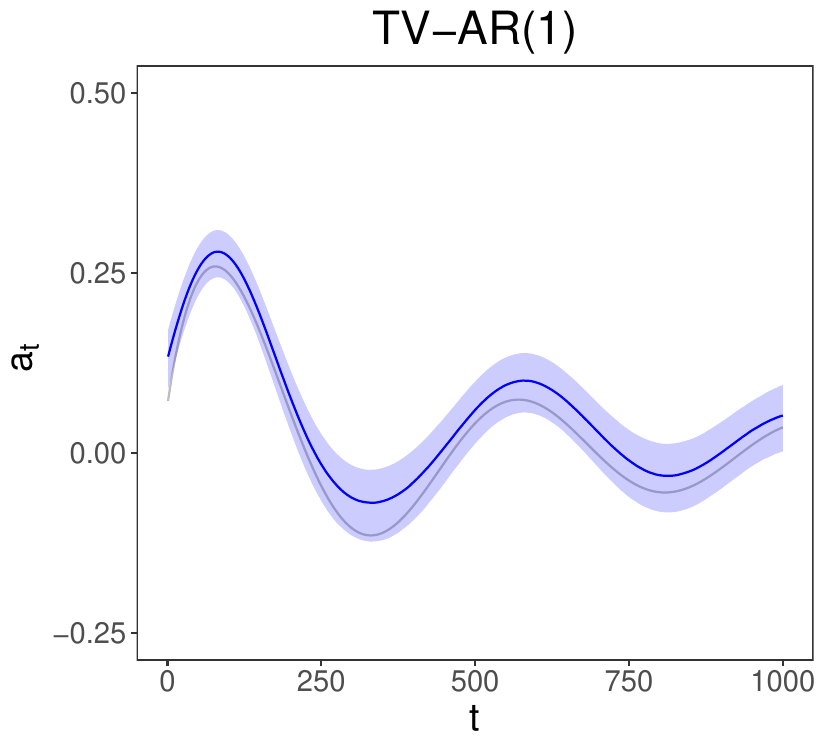}
\includegraphics[width = 4.25cm]{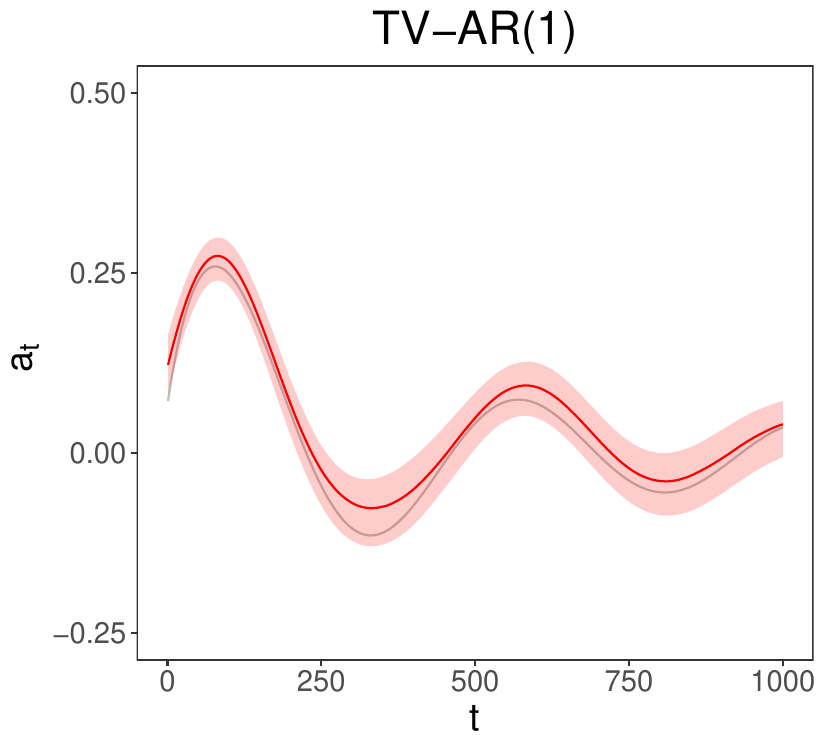}\\
\includegraphics[width = 4.25cm]{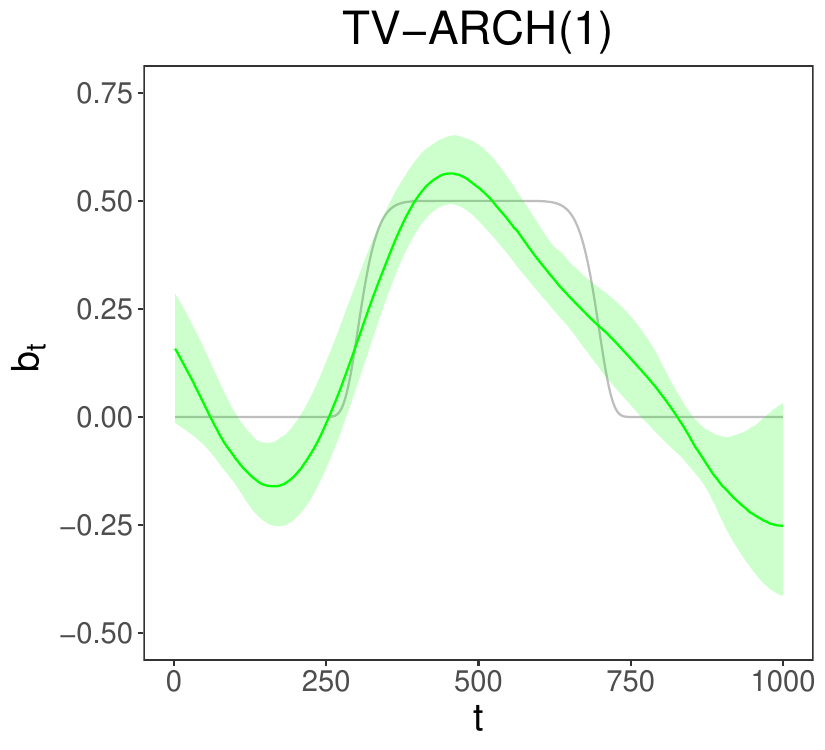}
\includegraphics[width = 4.25cm]{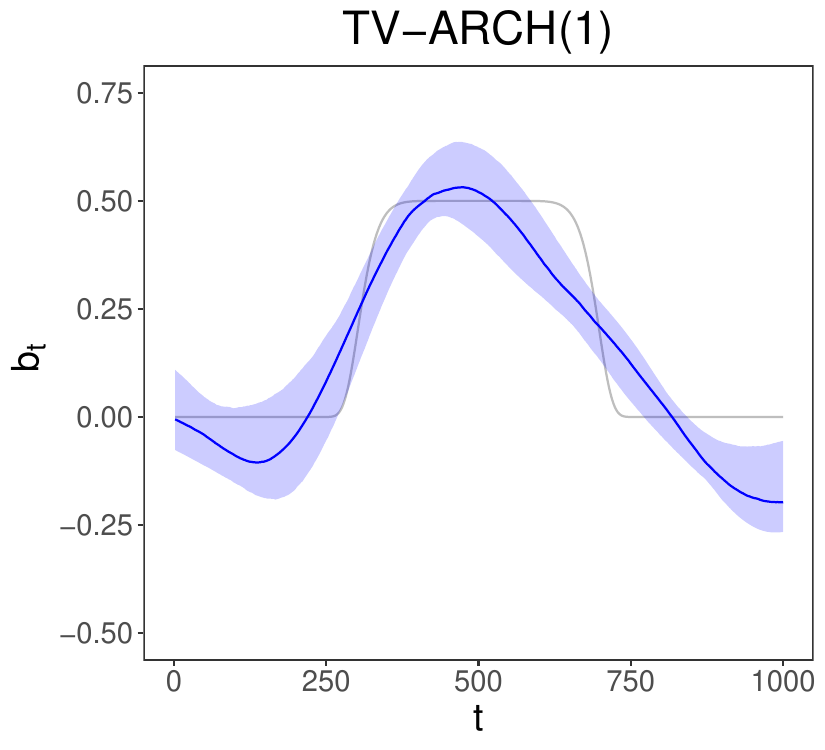}
\includegraphics[width = 4.25cm]{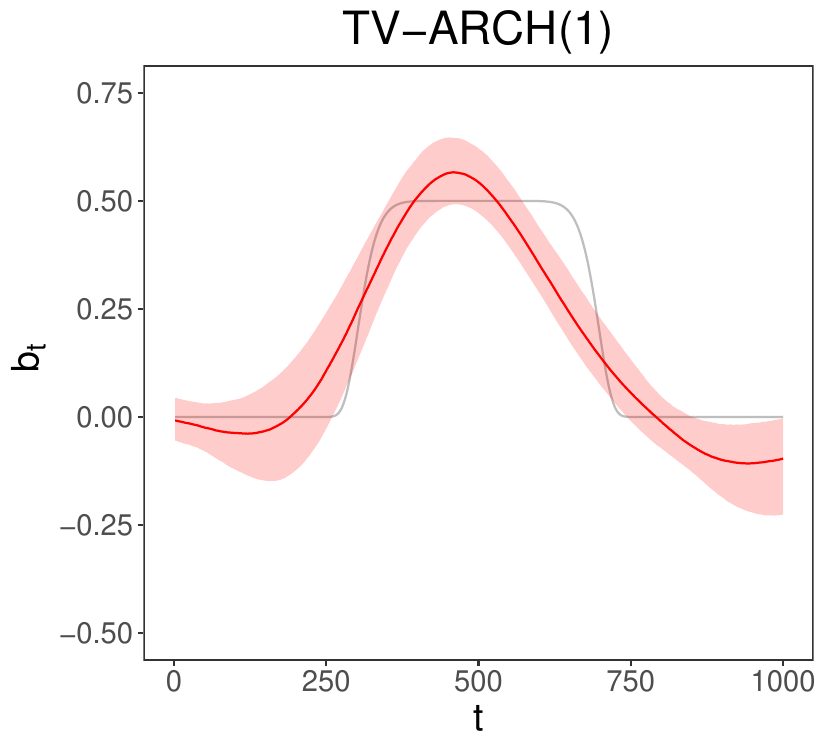}\\
\includegraphics[width = 4.25cm]{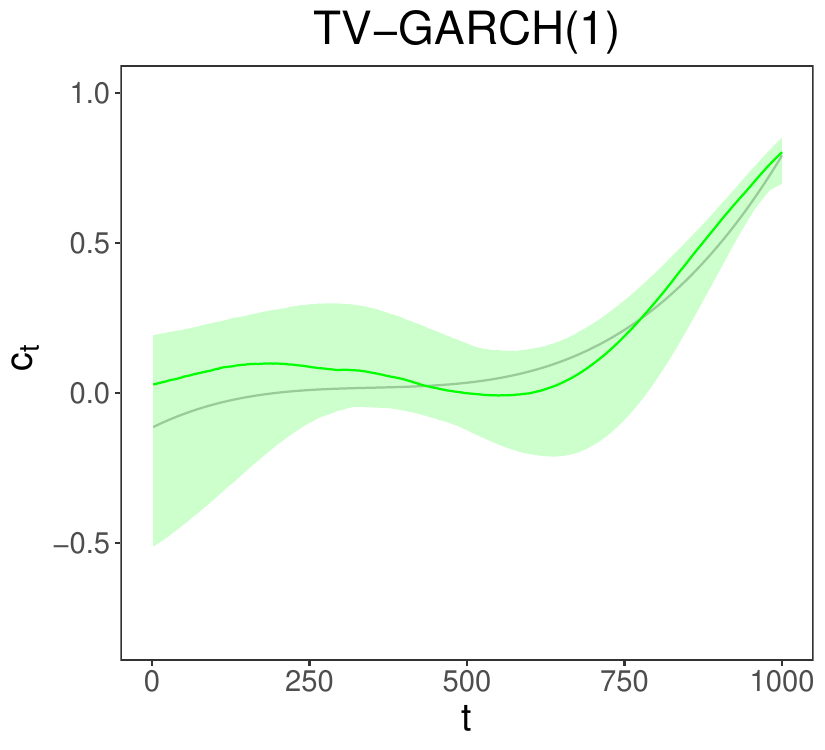}
\includegraphics[width = 4.25cm]{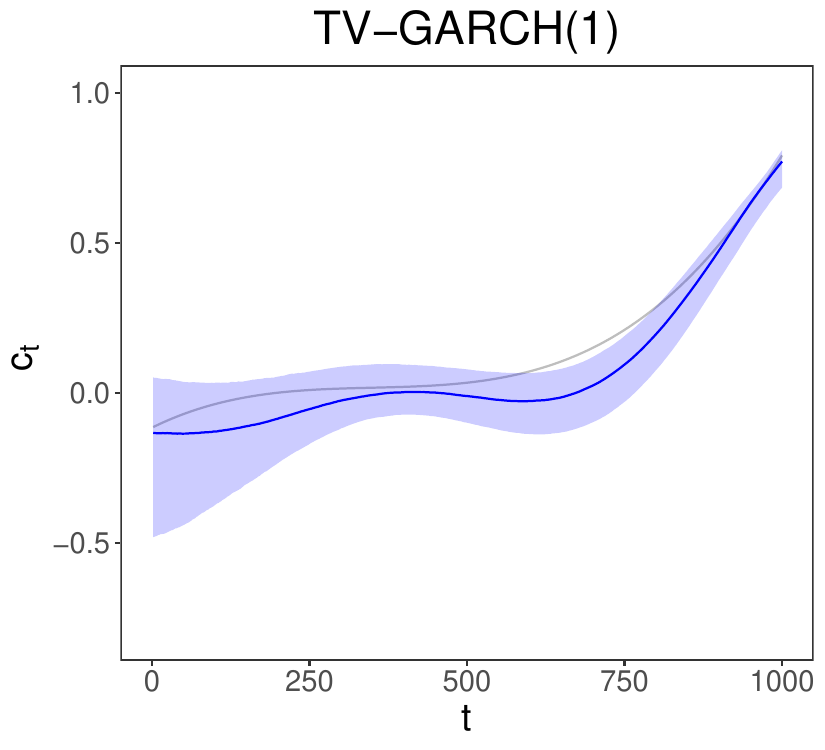}
\includegraphics[width = 4.25cm]{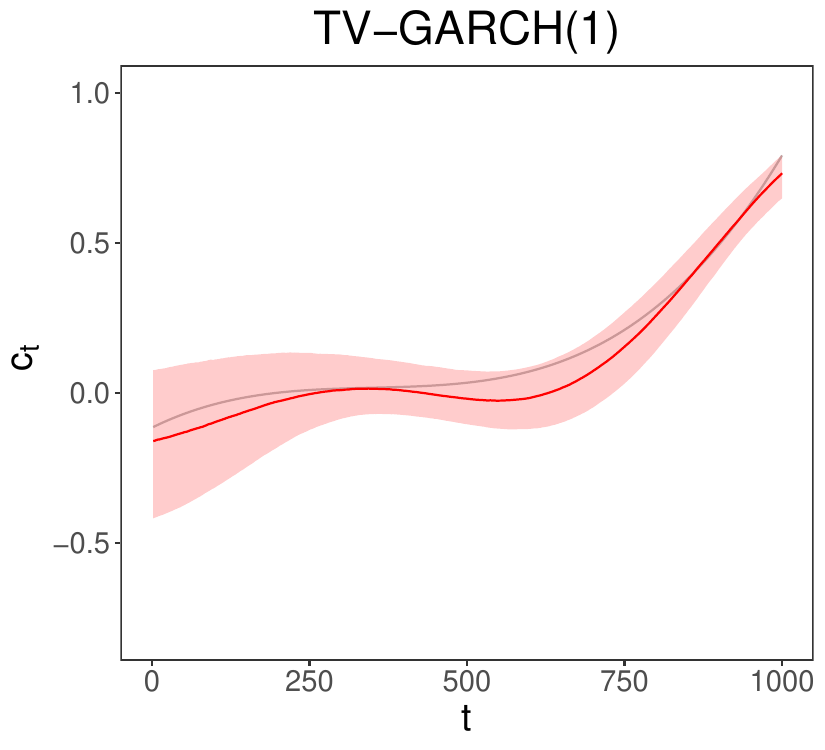}
\end{center}
\end{figure}

\begin{figure}[h]
\begin{center}
\caption{The mean-variance function ($\tau_i$), TV-AR(1) and TV-GARCH(1, 1) processes of the mean-variance parameterised Gamma model for the Park Grass data with 80\% credible intervals for each type of basis coefficient prior; Inverse-Wishart($20$) prior on covariance matrix (green), Horseshoe (blue) and multivariate Horseshoe (red).}
\includegraphics[width = 4.25cm]{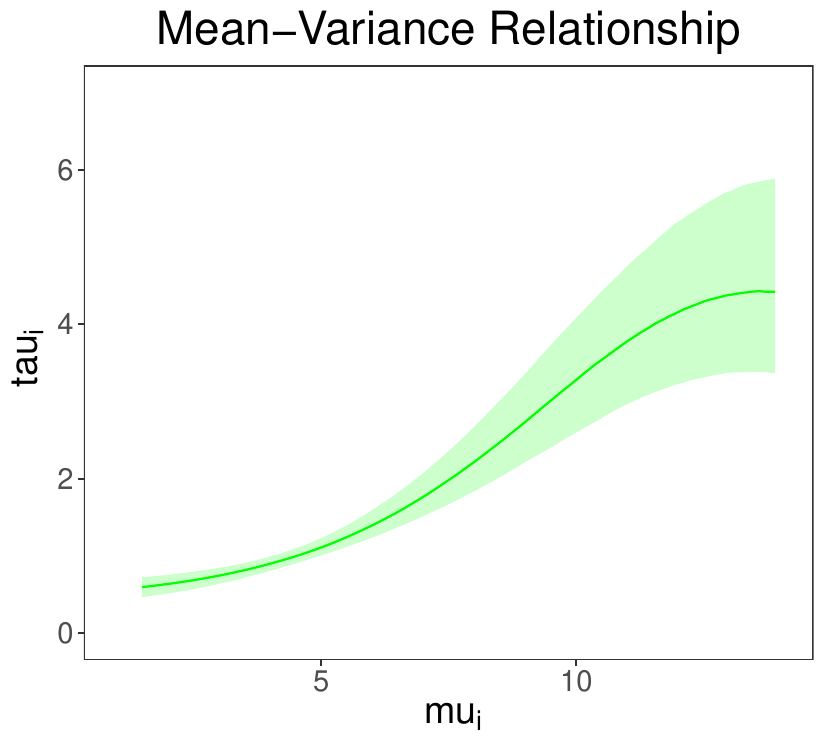}
\includegraphics[width = 4.25cm]{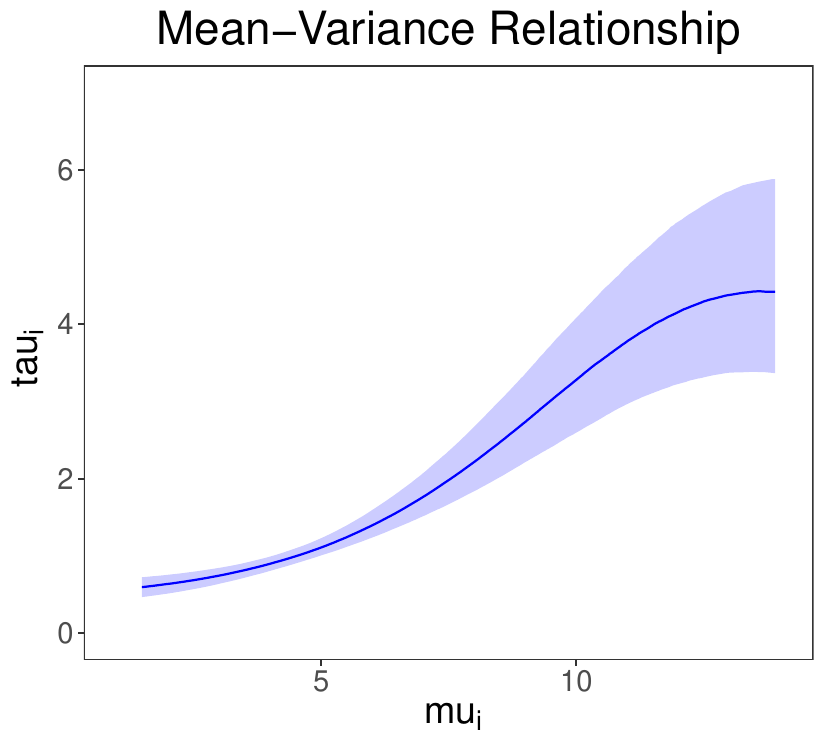}
\includegraphics[width = 4.25cm]{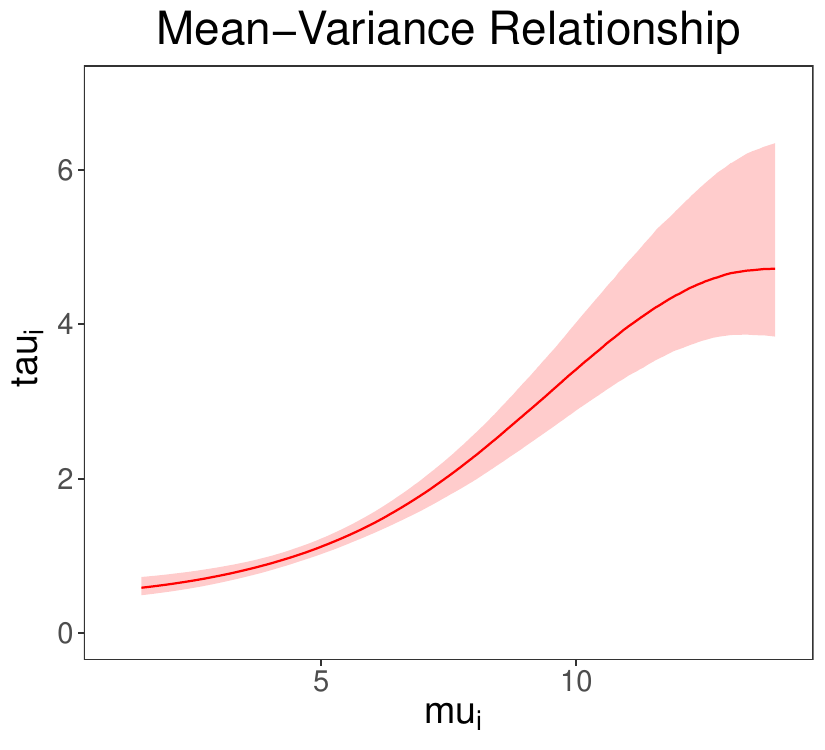}\\
\includegraphics[width = 4.25cm]{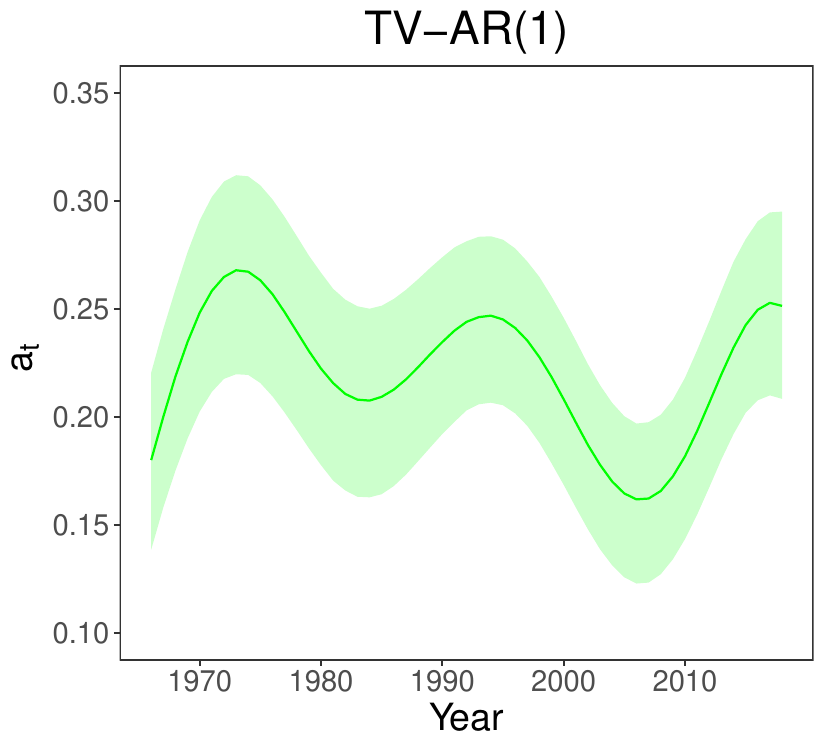}
\includegraphics[width = 4.25cm]{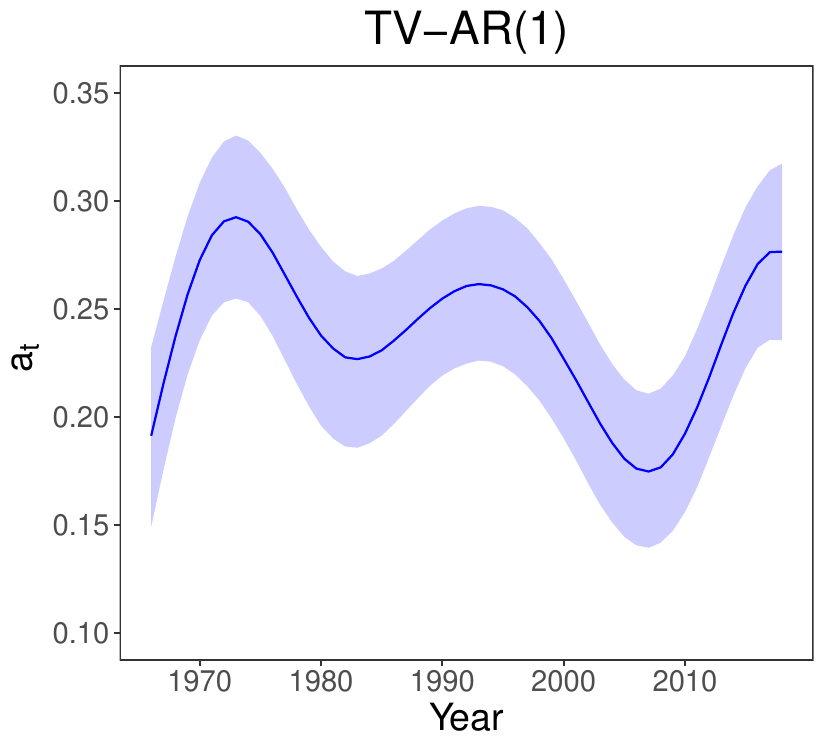}
\includegraphics[width = 4.25cm]{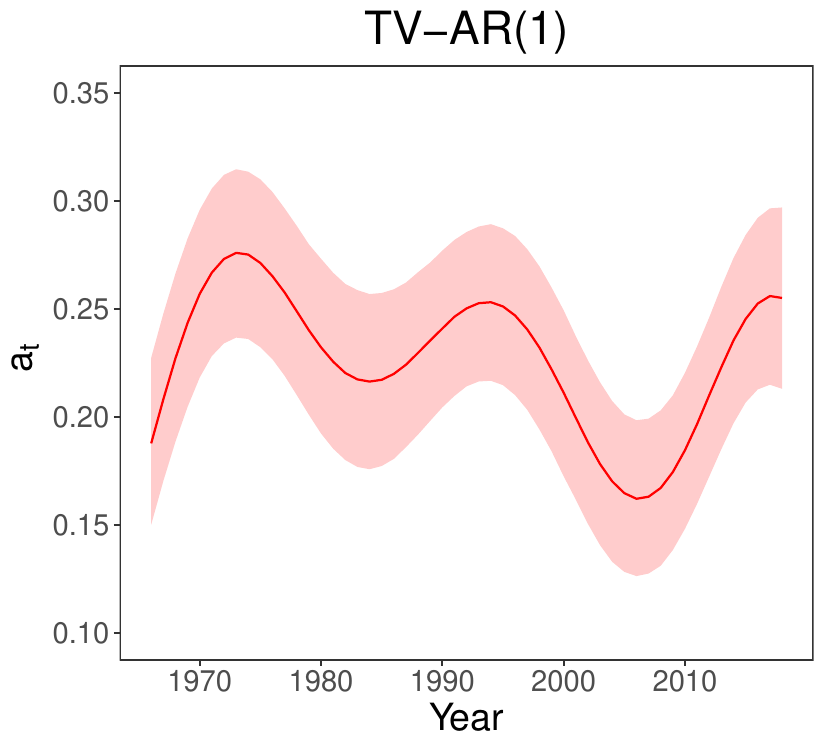}\\
\includegraphics[width = 4.25cm]{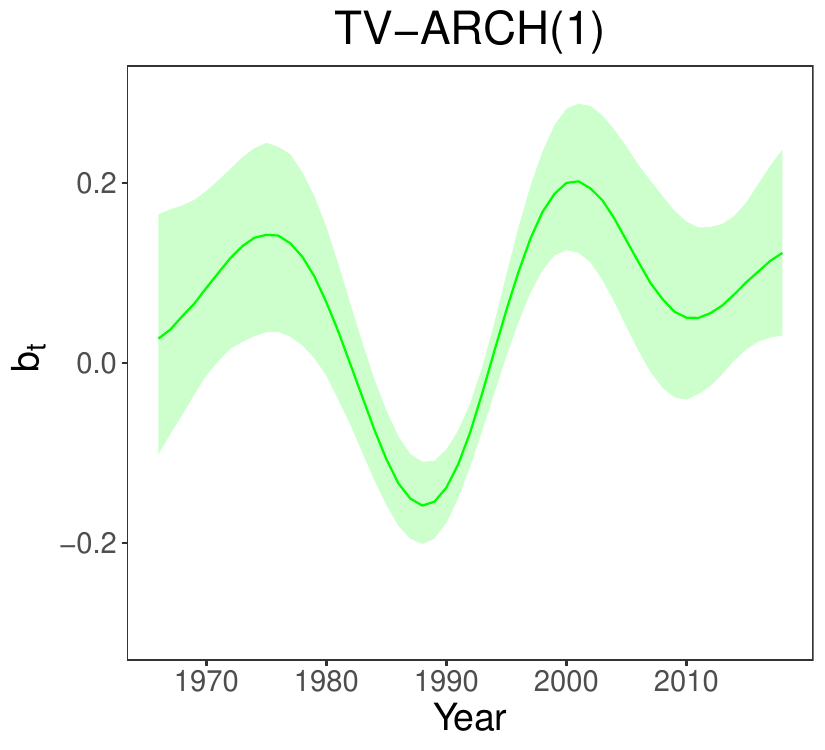}
\includegraphics[width = 4.25cm]{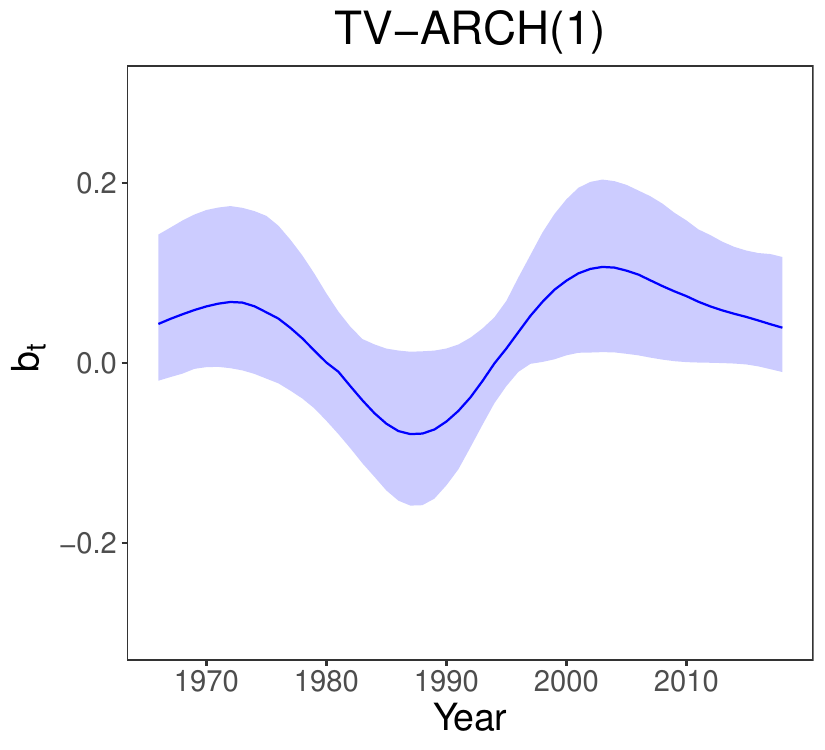}
\includegraphics[width = 4.25cm]{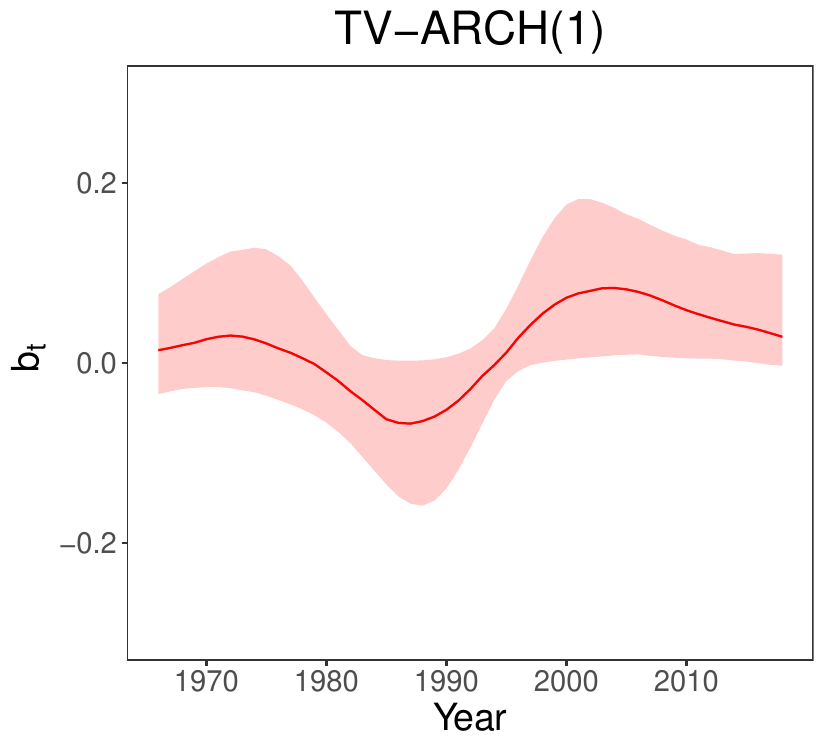}\\
\includegraphics[width = 4.25cm]{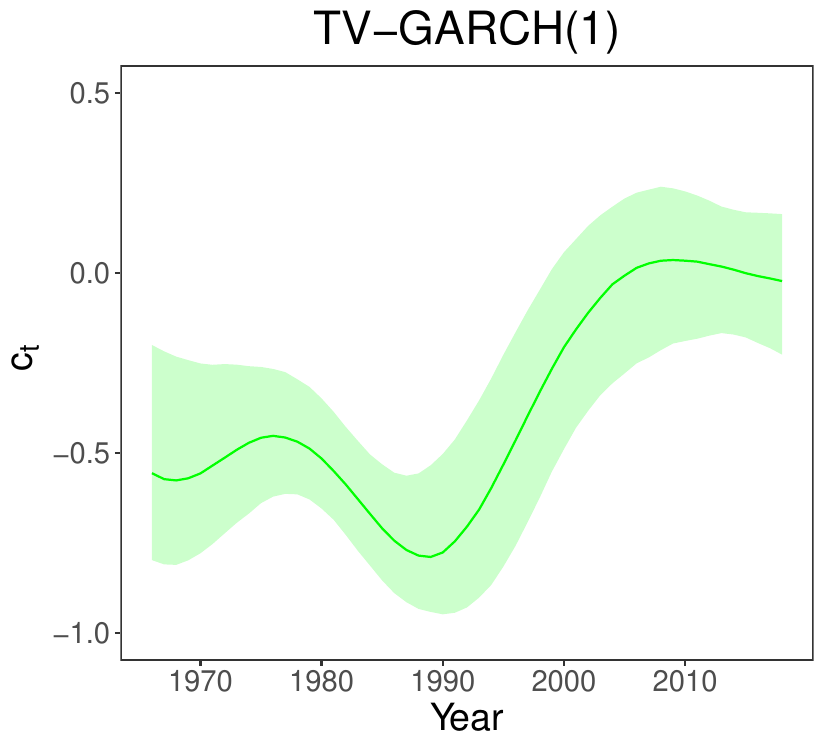}
\includegraphics[width = 4.25cm]{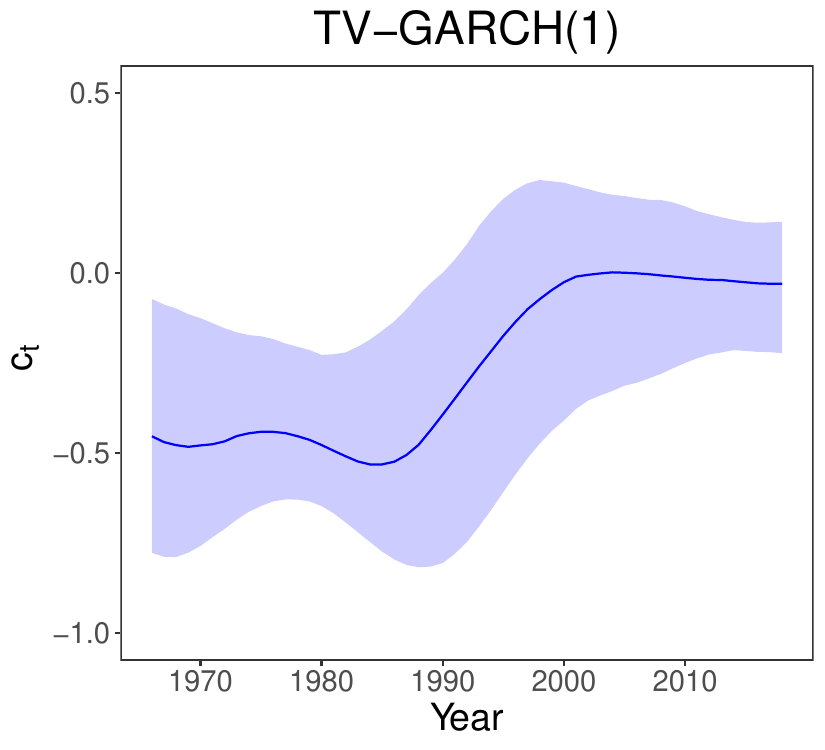}
\includegraphics[width = 4.25cm]{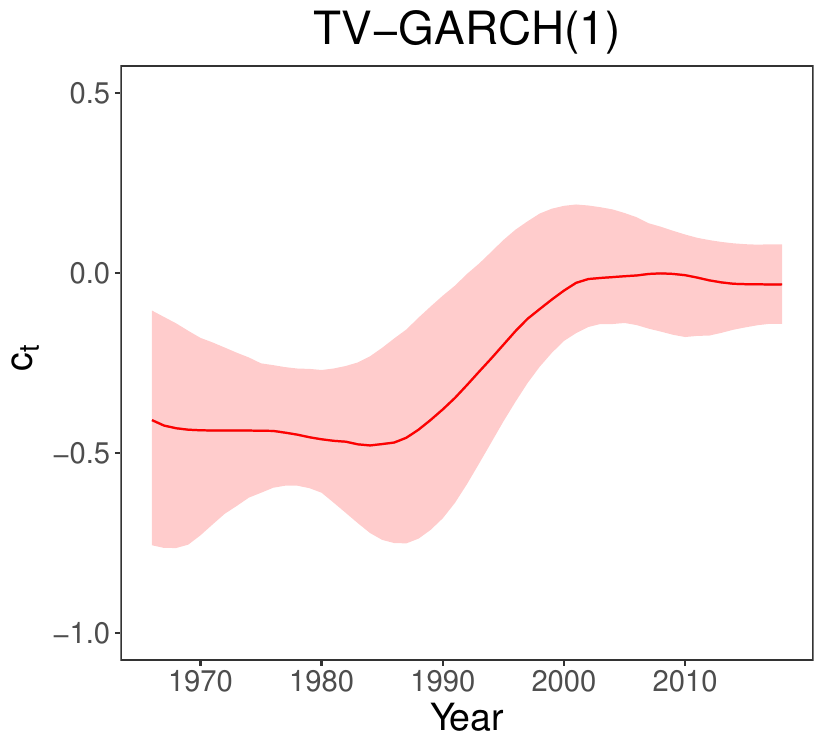}\\
\end{center}
\end{figure}

\begin{figure}[h]
\begin{center}
\caption{Variance estimates from mean-variance and TV-GARCH(1, 1) process with multivariate Horseshoe priors of Park Grass limed sections plots 14.2, 16, 7, 2.2 and 3 from 1966 to 2018. Limed sections are staggered from no liming (light grey) to pH maintained at 7 (black).}
\includegraphics[width = 6.5cm]{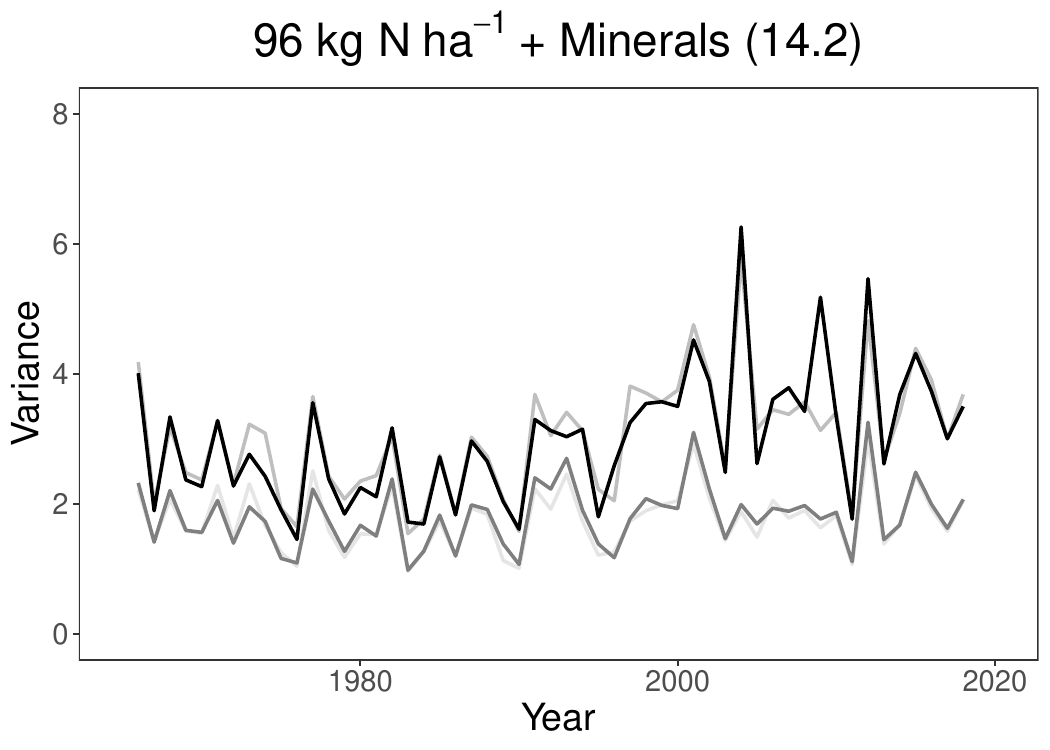}
\includegraphics[width = 6.5cm]{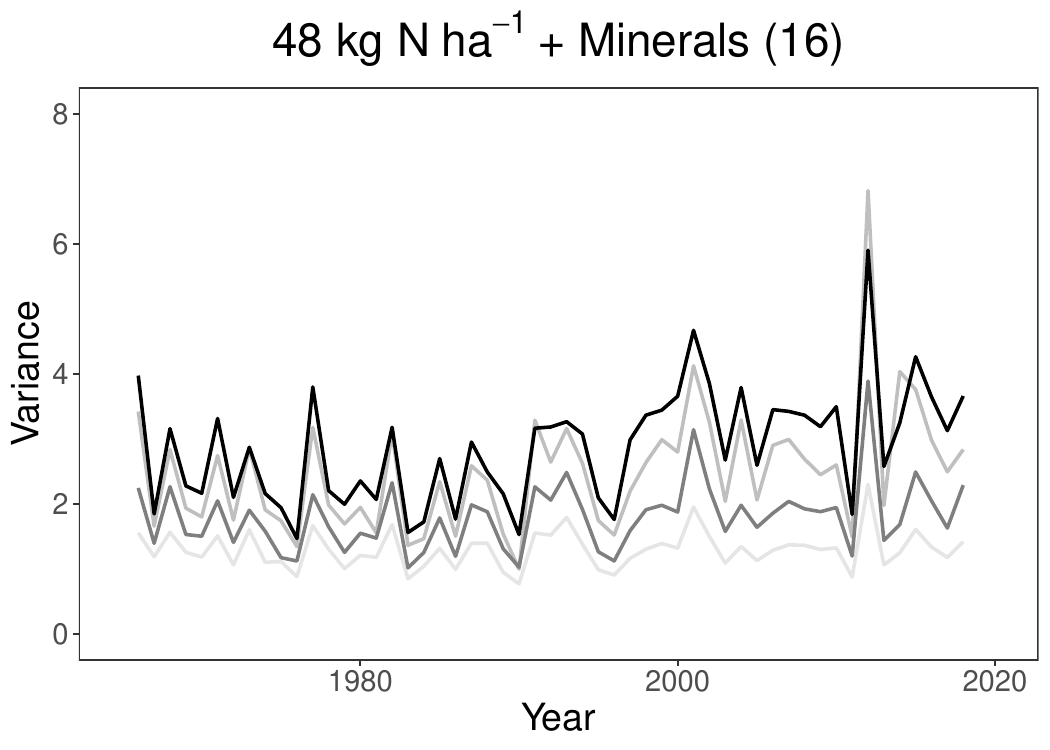}\\
\includegraphics[width = 6.5cm]{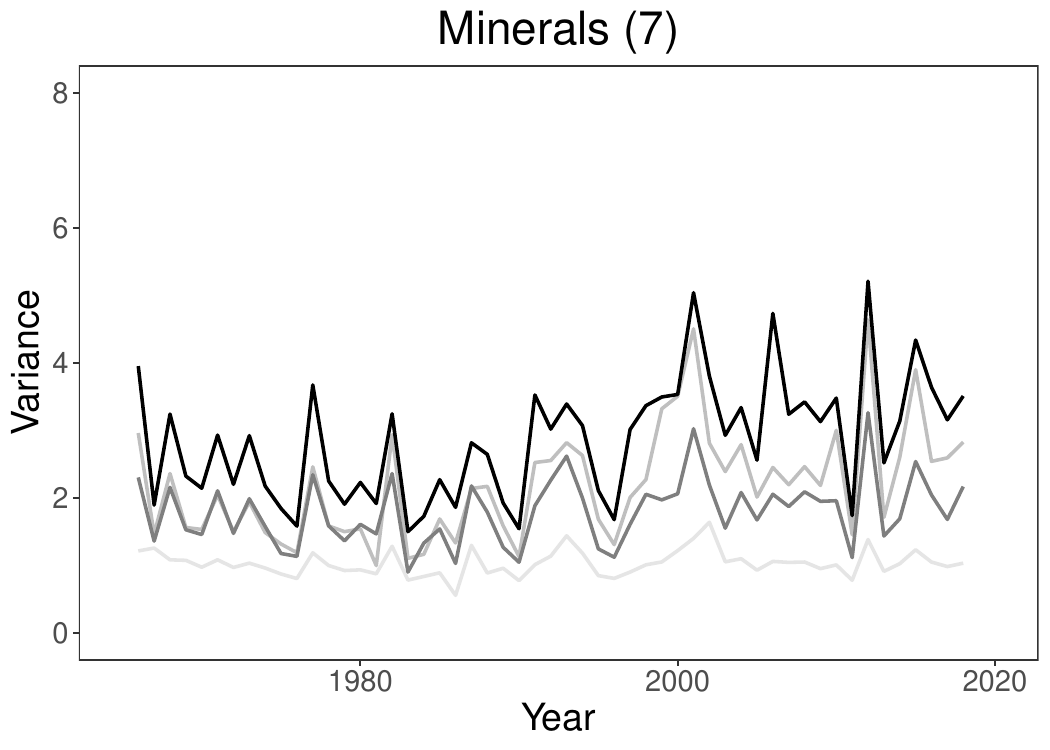}
\includegraphics[width = 6.5cm]{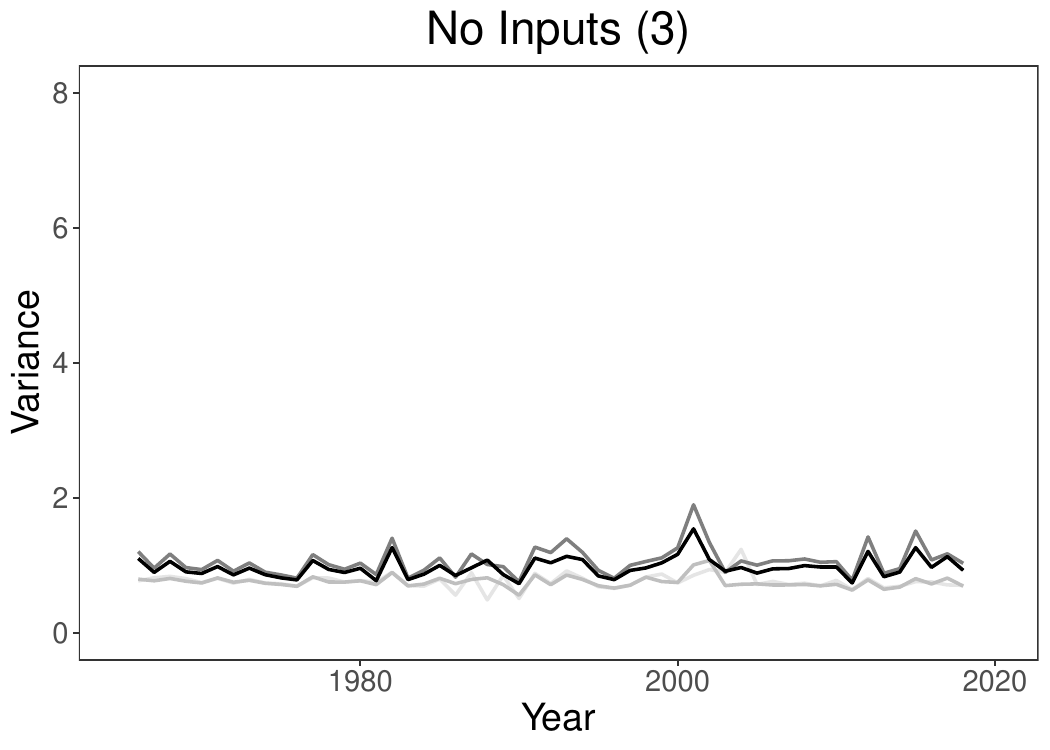}
\end{center}
\end{figure}

\begin{table}
\caption{Weather parameter estimates with standard deviation (SD), 80\% credible intervals from the Park Grass regression model. TR and MT refer to Total Rainfall and Mean Temperature with subscript $_1$ and $_2$ correspond to linear and polynomial terms.}
\begin{center}
\begin{tabular}{ c | c c c c}
 Model Terms & Mean & SD & 10\% & 90\% \\
 \hline
$\beta_\text{TR Aut (mm)}$ & 3.40$\times10^{-4}$ & 1.120$\times10^{-4}$ & 1.97$\times10^{-4}$ & 4.84$\times10^{-4}$ \\
$\beta_\text{TR Win (mm)}$ & 2.56$\times10^{-4}$ & 1.124$\times10^{-4}$ & 1.13$\times10^{-4}$ & 3.98$\times10^{-4}$ \\
$\beta_{\text{TR Spr}_1 \text{(mm)}}$ & 1.99 & 0.266 & 1.64 & 2.33 \\
 $\beta_{\text{TR Spr}_2 \text{(mm)}}$ & -1.74 & 0.308 & -2.12 & -1.34 \\
$\beta_\text{TR Sum (mm)}$ & 1.74$\times10^{-3}$ & 0.186$\times10^{-3}$ & 1.50$\times10^{-3}$ & 1.99$\times10^{-3}$ \\
$\beta_\text{MT Aut (deg C)}$ & 2.23$\times10^{-2}$ & 1.056$\times10^{-2}$ & 0.84$\times10^{-2}$ & 3.58$\times10^{-2}$ \\
$\beta_\text{MT Win (deg C)}$ & -3.83$\times10^{-2}$ & 0.815$\times10^{-2}$ & -4.86$\times10^{-2}$ & -2.78$\times10^{-2}$ \\
 $\beta_{\text{MTSpr}_1 \text{(deg C)}}$ & 1.19 & 0.340 & 0.72 & 1.65 \\
 $\beta_{\text{MTSpr}_2 \text{(deg C)}}$ & -1.89 & 0.259 & -2.22 & -1.57 \\
$\beta_\text{MT Sum (deg C)}$ & 2.96$\times10^{-2}$ & 1.162$\times10^{-2}$ & 1.45$\times10^{-2}$ & 4.49$\times10^{-2}$ \\
\end{tabular}
\end{center}
\end{table}
\newpage
\clearpage

\renewcommand{\figurename}{Supplementary Figure}
\setcounter{table}{0}
\setcounter{figure}{0}

\section*{Supplementary Materials}
\begin{figure}[h]
\begin{center}
\caption{10000 samples of the shrinkage coefficient from an Inverse-Wishart($\psi = 20$) and Horseshoe prior.}
\includegraphics{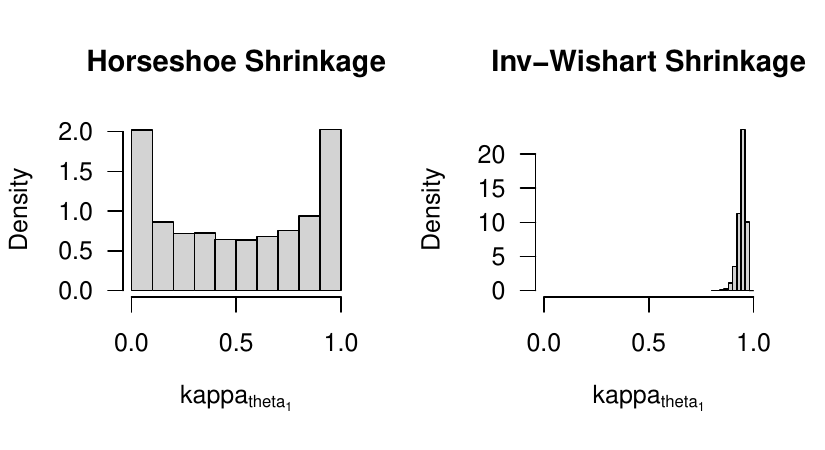}
\end{center}
\end{figure}

\begin{figure}[h]
\begin{center}
\caption{10000 samples of a bivariate Inverse-Wishart prior with $\psi = 3, 10$ and $20$ (top row). The covariate shrinkage of 10000 samples from an Inverse-Wishart prior with $\psi = 3, 10$ and $20$ (bottom row).}
\includegraphics{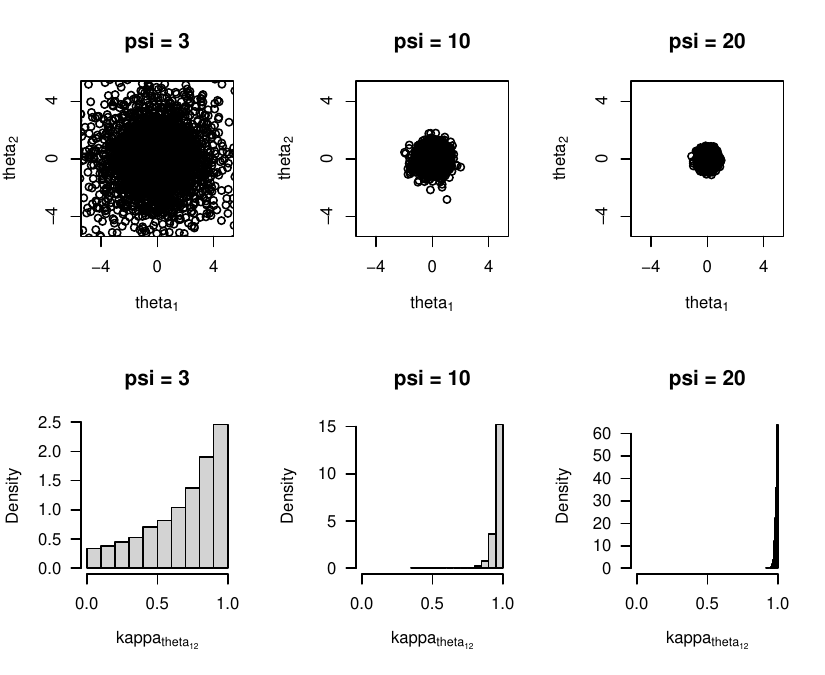}
\end{center}
\end{figure}

\newpage
\begin{figure}[h]
\begin{center}
\caption{The simulated TV-AR(1) processes from simulation example 5.1 for a mean-variance parameterised Gamma model (top row). The fitted TV-AR(1) (second row) and TV-GARCH(1, 1) (rows three and four) with 80\% credible intervals for each type of basis coefficient prior; Inverse-Wishart($20$) prior on covariance matrix (green), Horseshoe (blue) and multivariate Horseshoe (red). All simulated data and functions are given in grey.}
\includegraphics[width = 4.25cm]{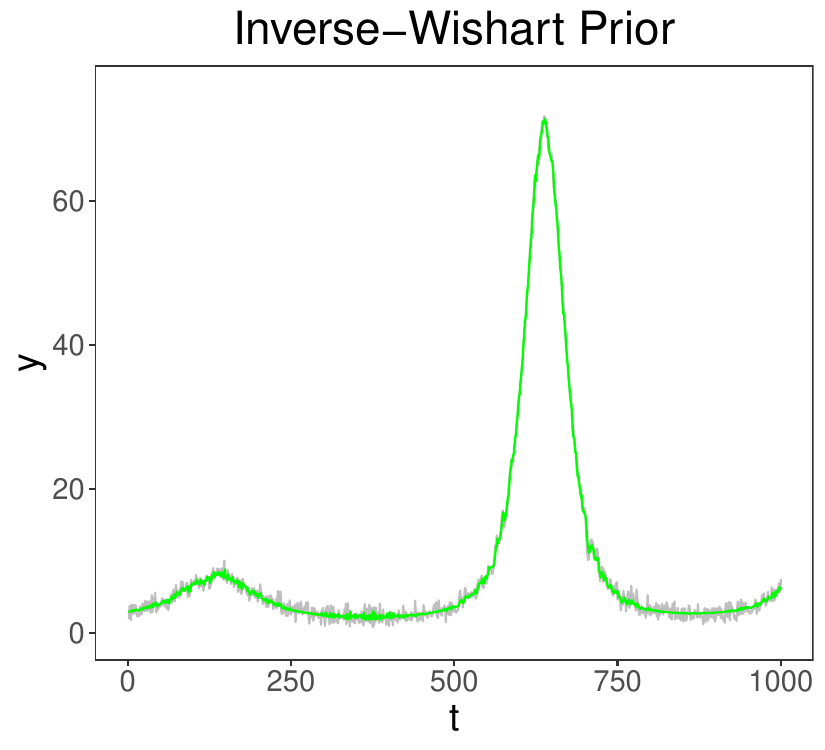}
\includegraphics[width = 4.25cm]{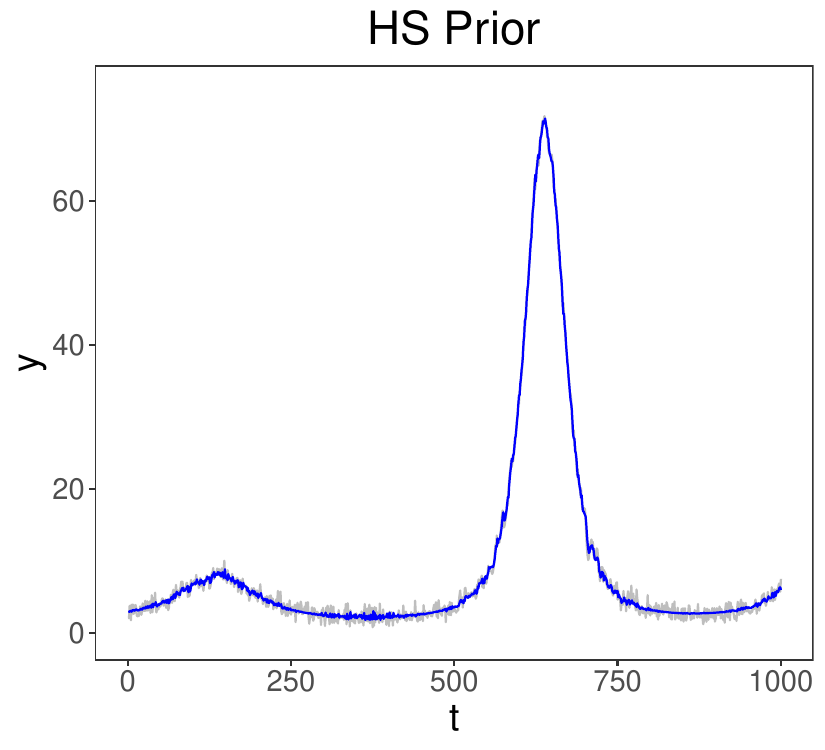}
\includegraphics[width = 4.25cm]{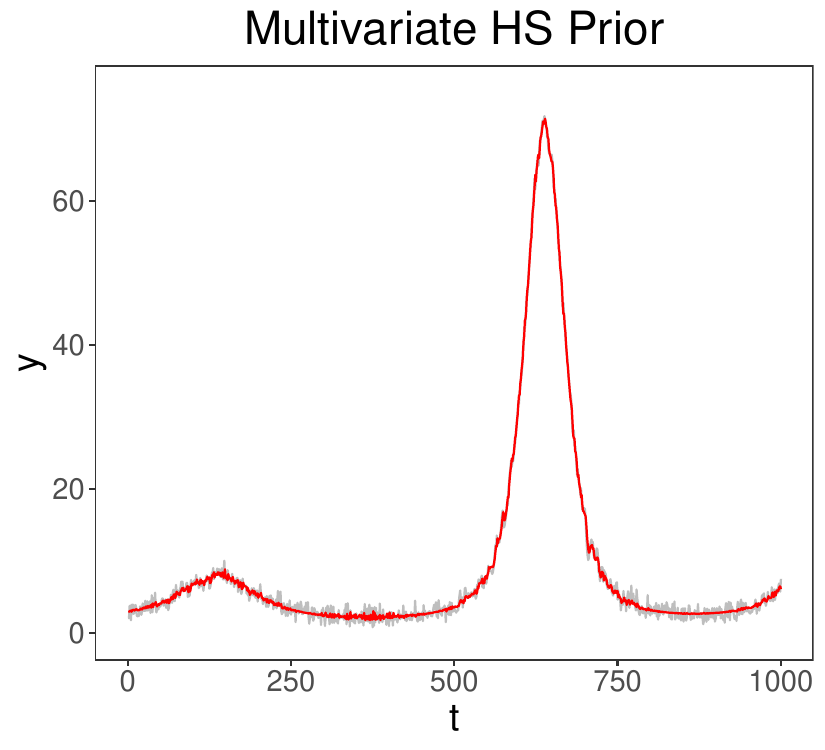}\\
\includegraphics[width = 4.25cm]{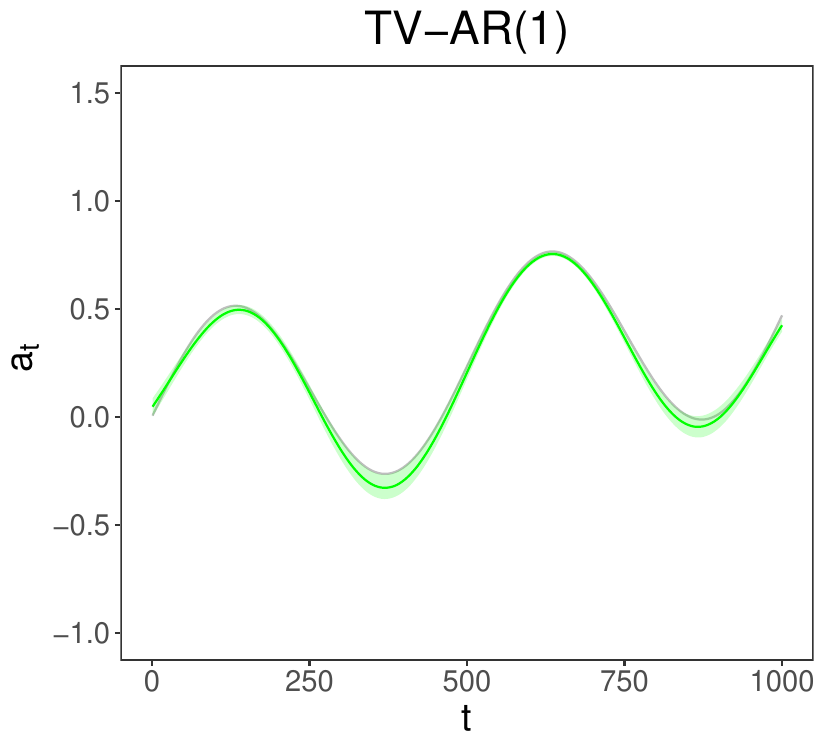}
\includegraphics[width = 4.25cm]{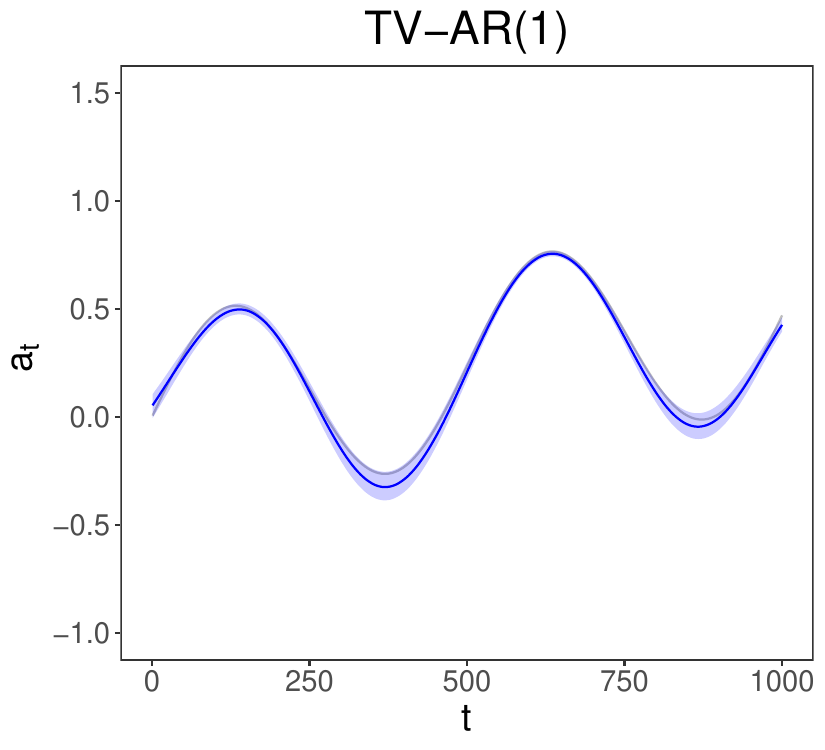}
\includegraphics[width = 4.25cm]{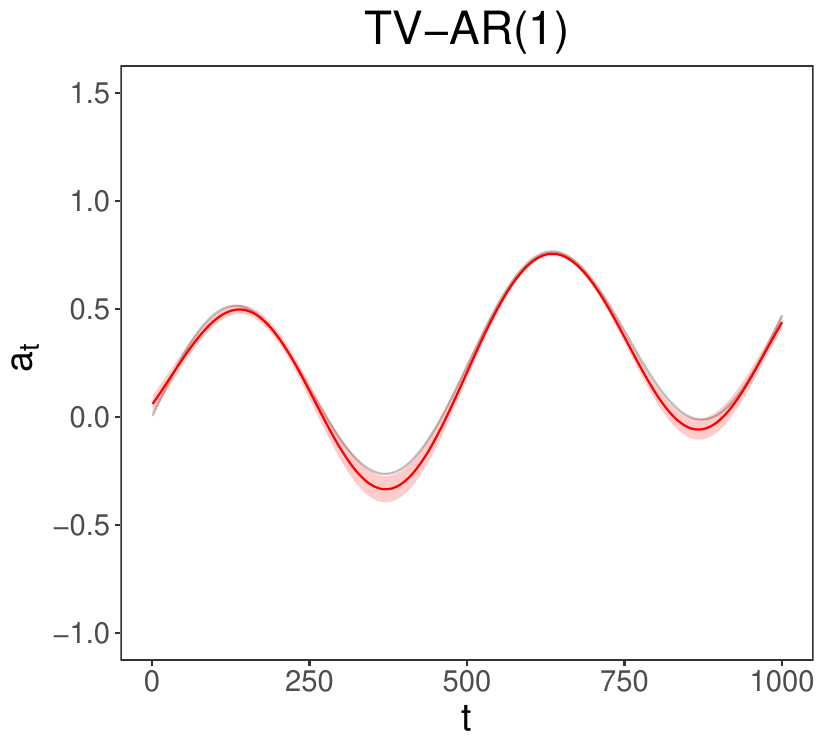}\\
\includegraphics[width = 4.25cm]{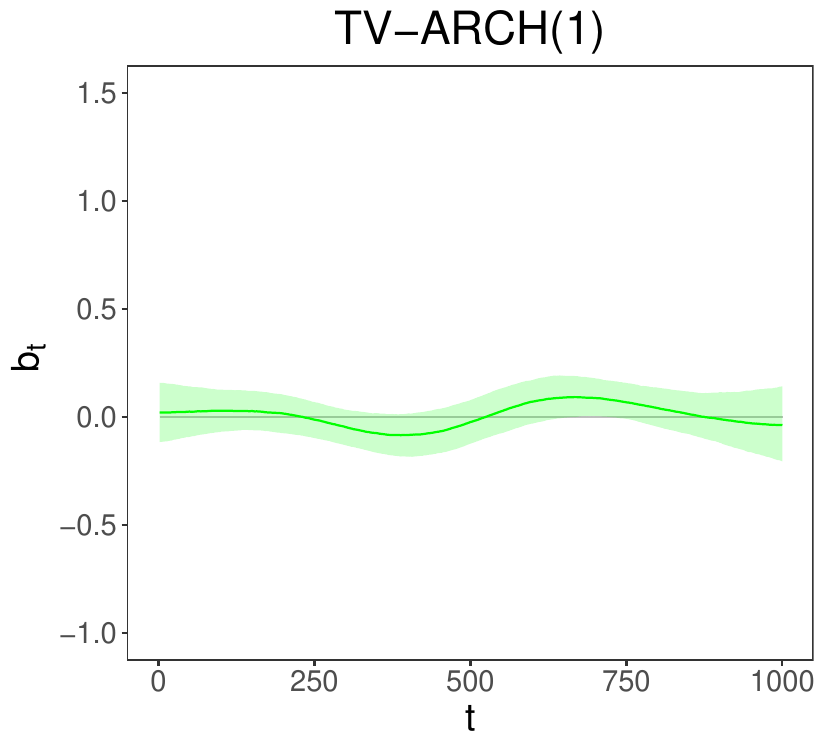}
\includegraphics[width = 4.25cm]{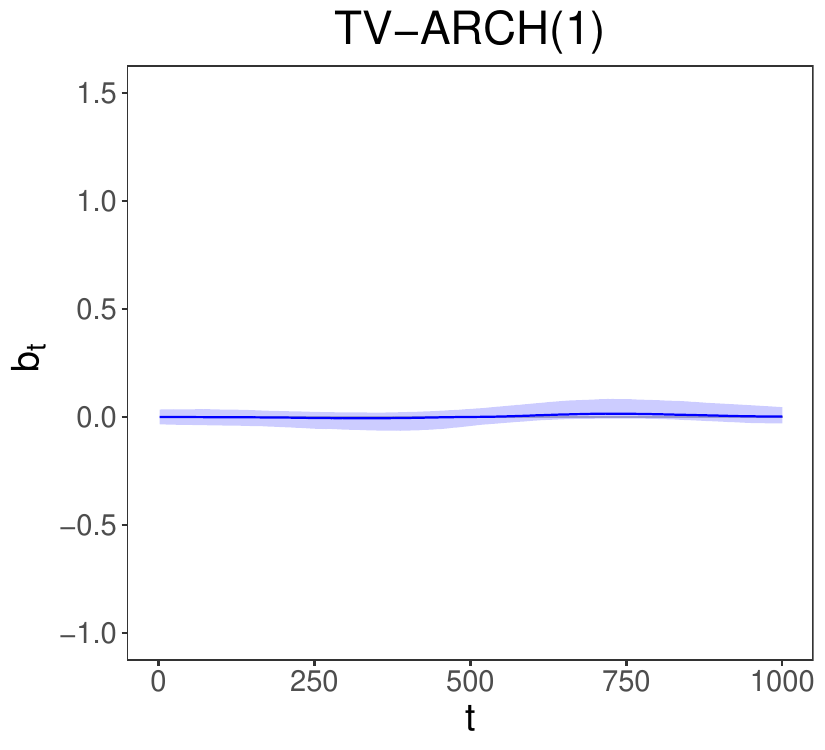}
\includegraphics[width = 4.25cm]{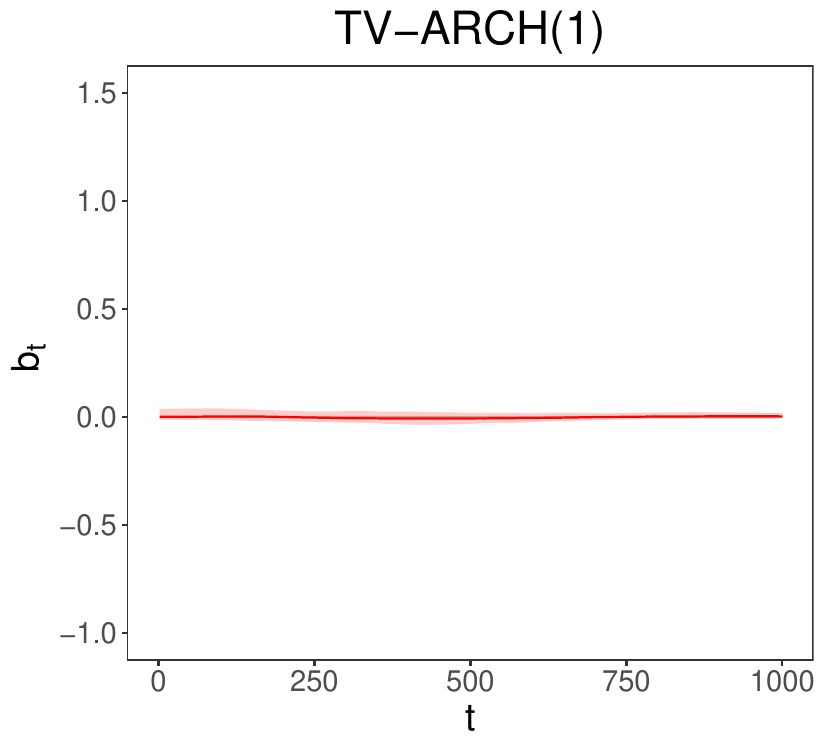}\\
\includegraphics[width = 4.25cm]{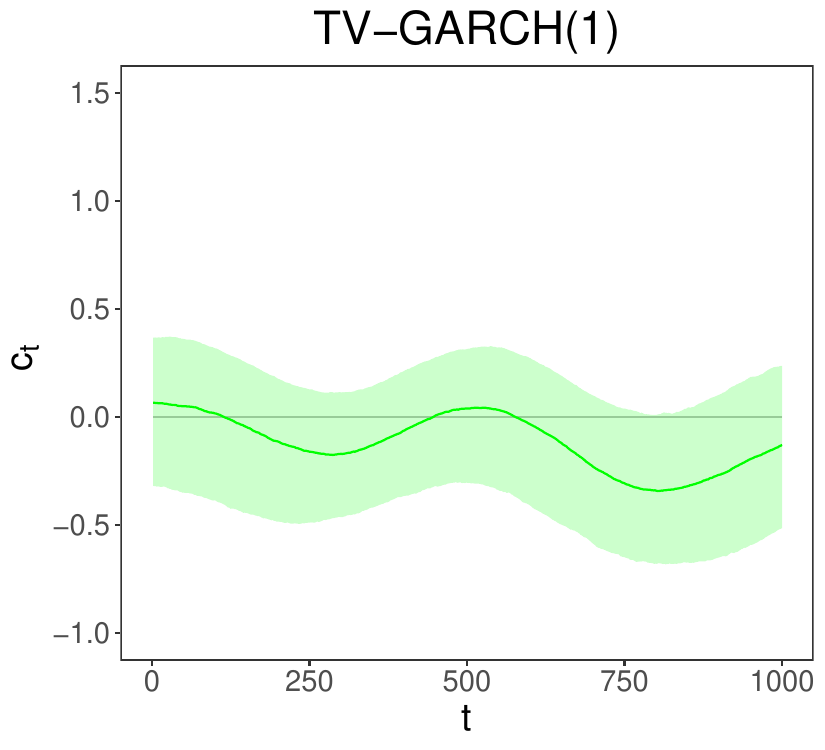}
\includegraphics[width = 4.25cm]{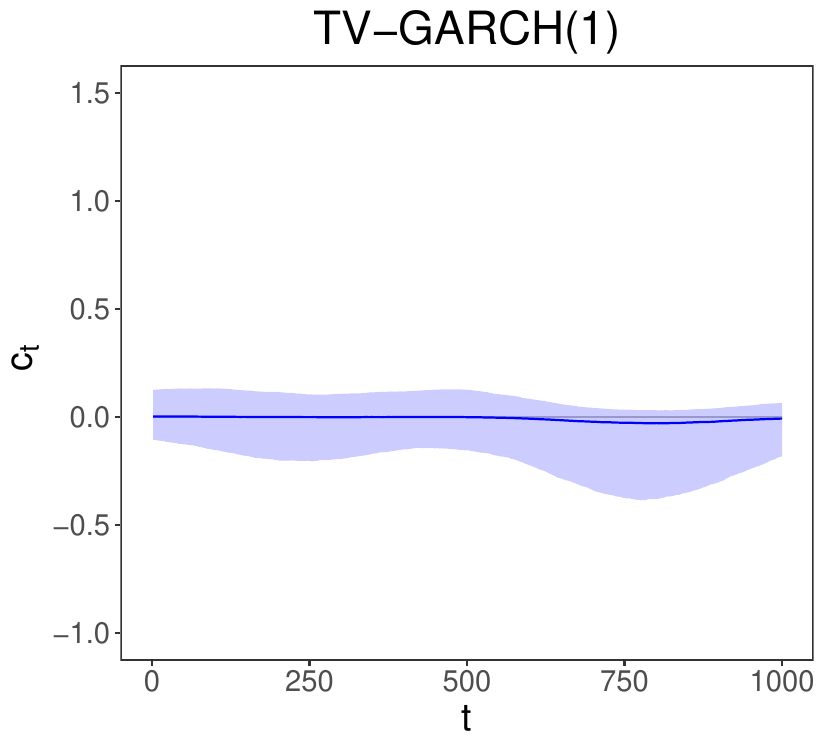}
\includegraphics[width = 4.25cm]{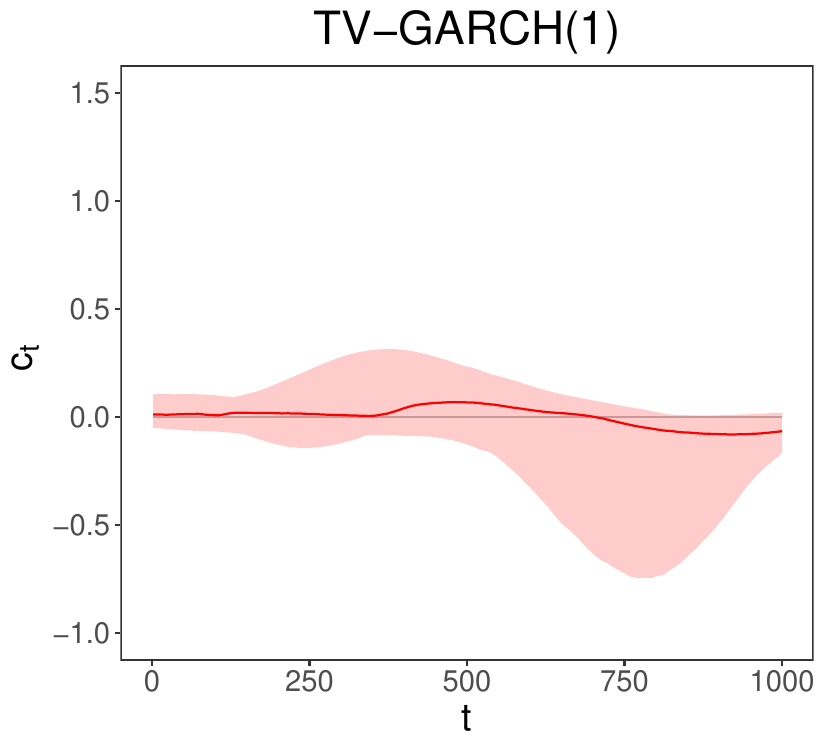}
\end{center}
\end{figure}

\newpage
\begin{figure}[h]
\begin{center}
\caption{The simulated TV-ARCH(1) process from simulation example 5.1 for a mean-variance parameterised Gamma model (top row). The fitted TV-AR(1) (second row) and TV-GARCH(1, 1) (rows three and four) with 80\% credible intervals for each type of basis coefficient prior; Inverse-Wishart($20$) prior on covariance matrix (green), Horseshoe (blue) and multivariate Horseshoe (red). All simulated data and functions are given in grey.}
\includegraphics[width = 4.25cm]{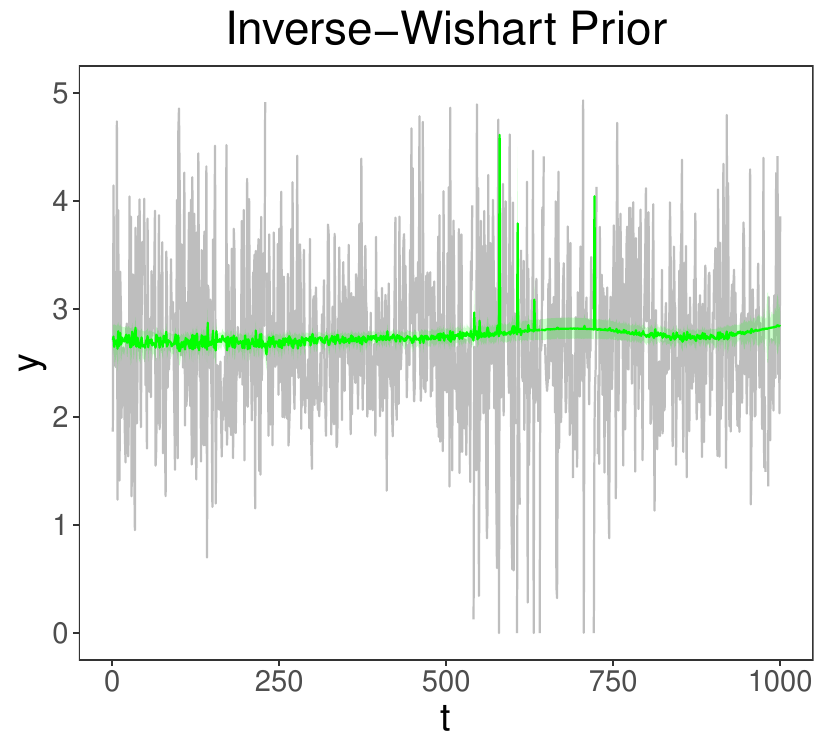}
\includegraphics[width = 4.25cm]{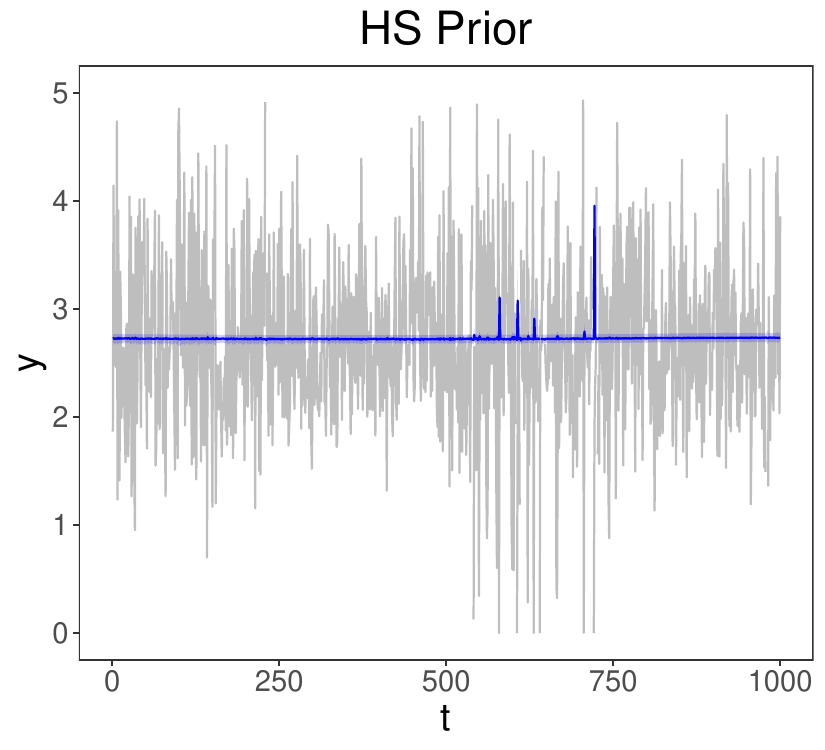}
\includegraphics[width = 4.25cm]{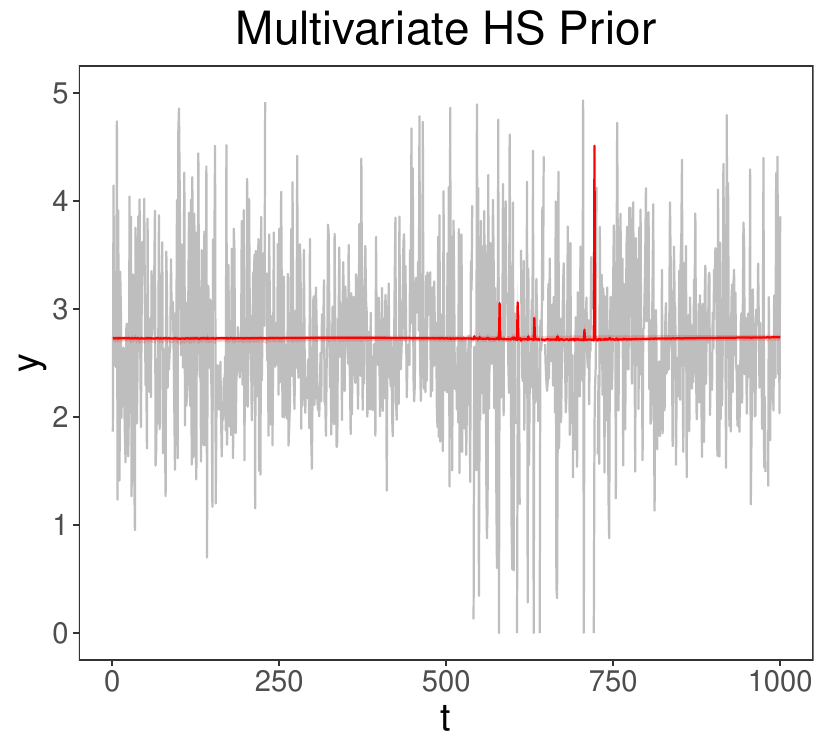}\\
\includegraphics[width = 4.25cm]{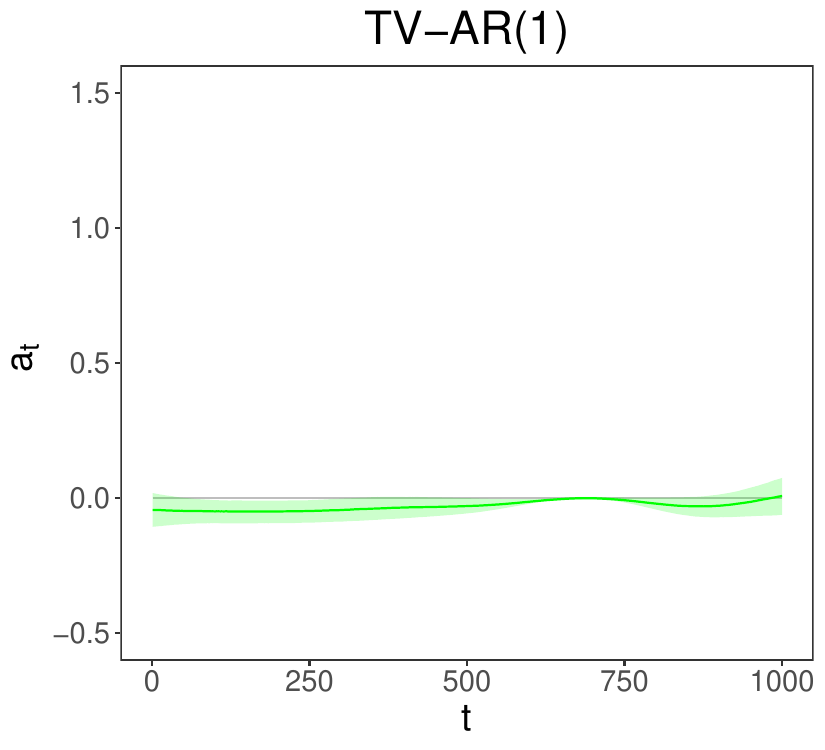}
\includegraphics[width = 4.25cm]{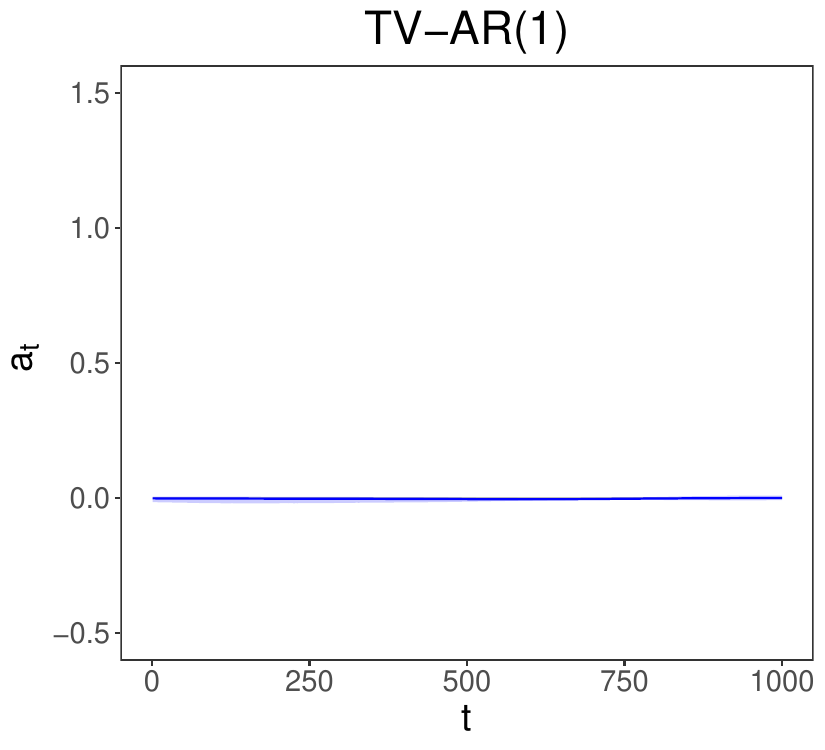}
\includegraphics[width = 4.25cm]{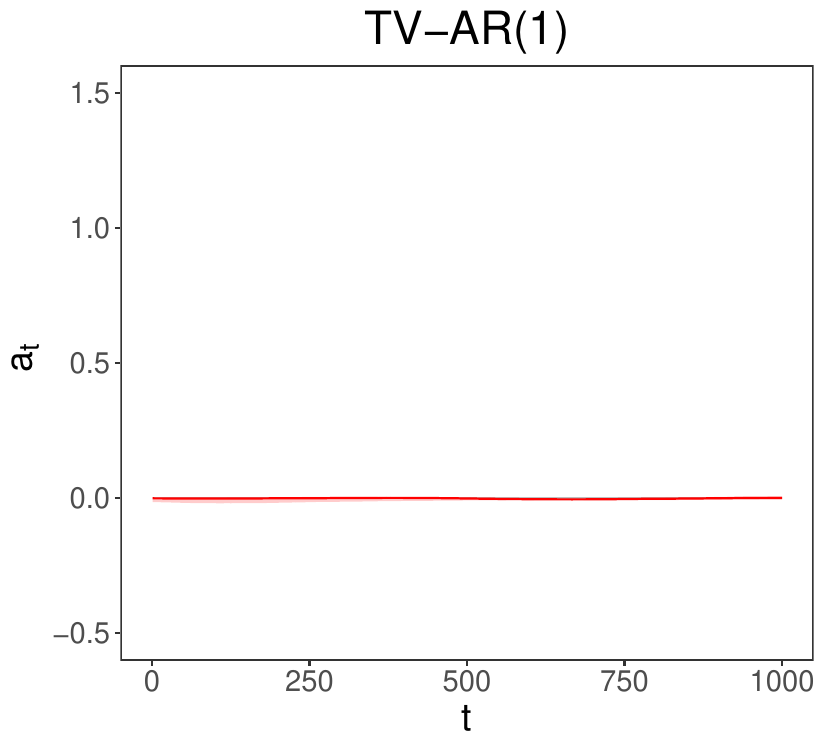}\\
\includegraphics[width = 4.25cm]{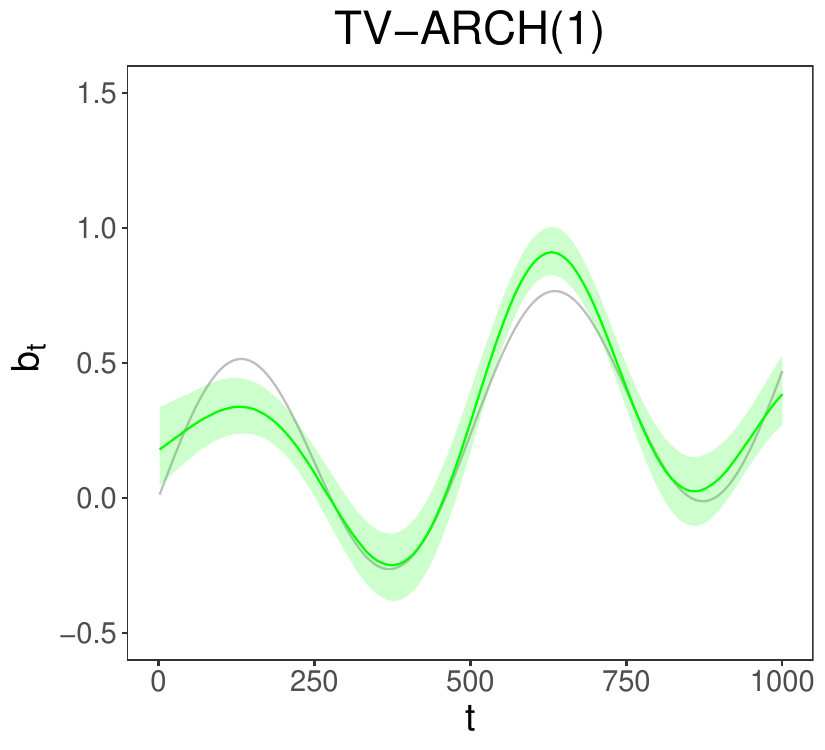}
\includegraphics[width = 4.25cm]{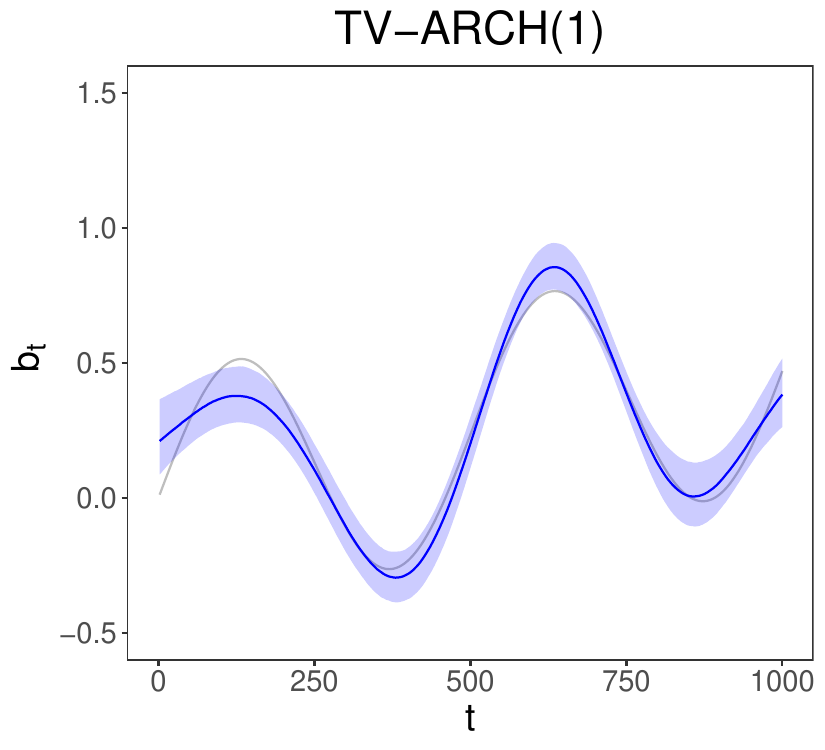}
\includegraphics[width = 4.25cm]{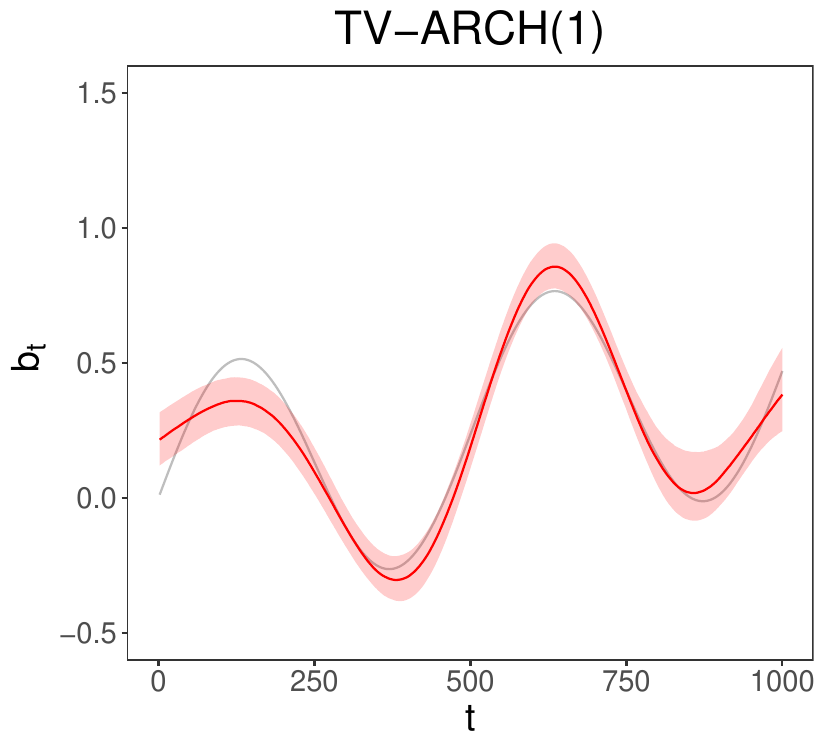}\\
\includegraphics[width = 4.25cm]{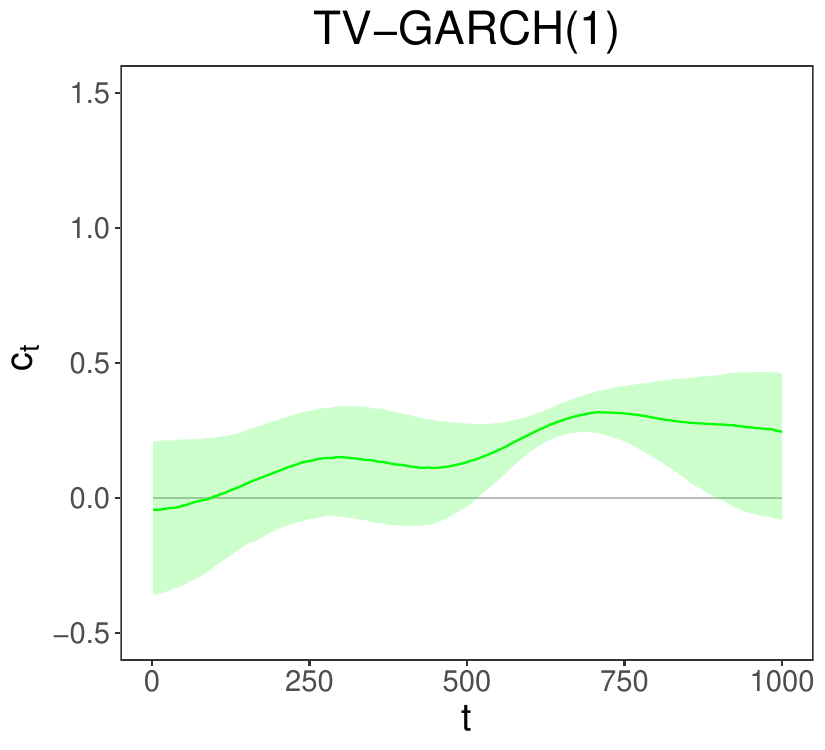}
\includegraphics[width = 4.25cm]{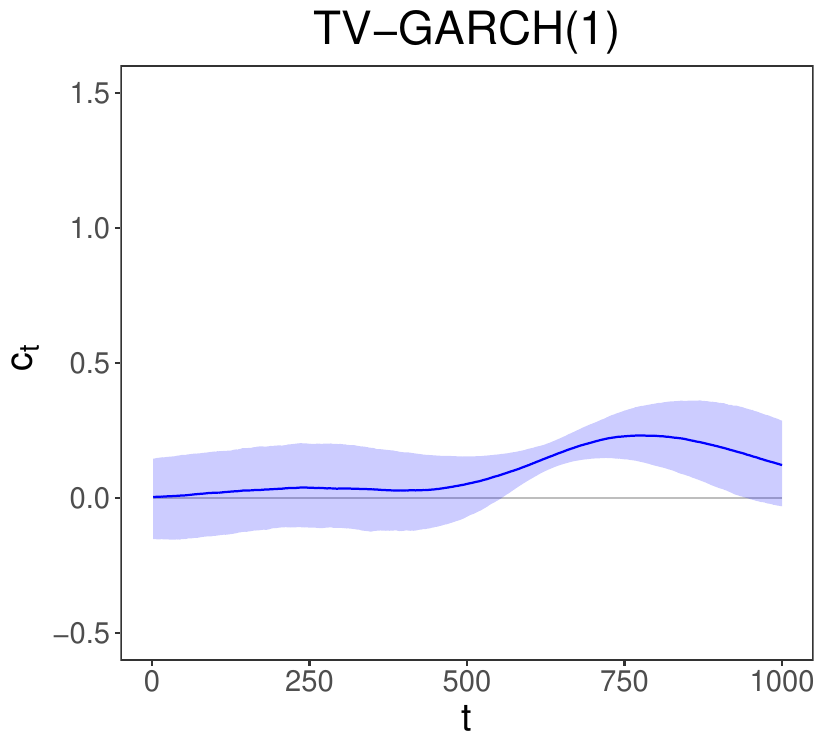}
\includegraphics[width = 4.25cm]{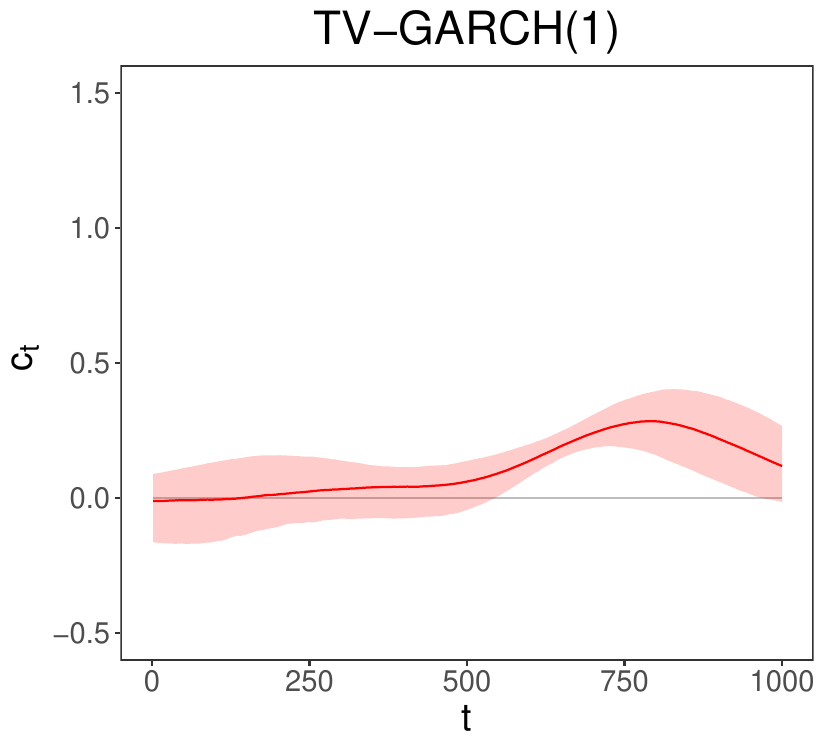}
\end{center}
\end{figure}

\newpage
\begin{figure}[h]
\begin{center}
\caption{The simulated TV-GARCH(1) process from simulation example 5.1 for a mean-variance parameterised Gamma model (top row). The fitted TV-AR(1) (second row) and TV-GARCH(1, 1) (rows three and four) with 80\% credible intervals for each type of basis coefficient prior; Inverse-Wishart($20$) prior on covariance matrix (green), Horseshoe (blue) and multivariate Horseshoe (red). All simulated data and functions are given in grey.}
\includegraphics[width = 4.25cm]{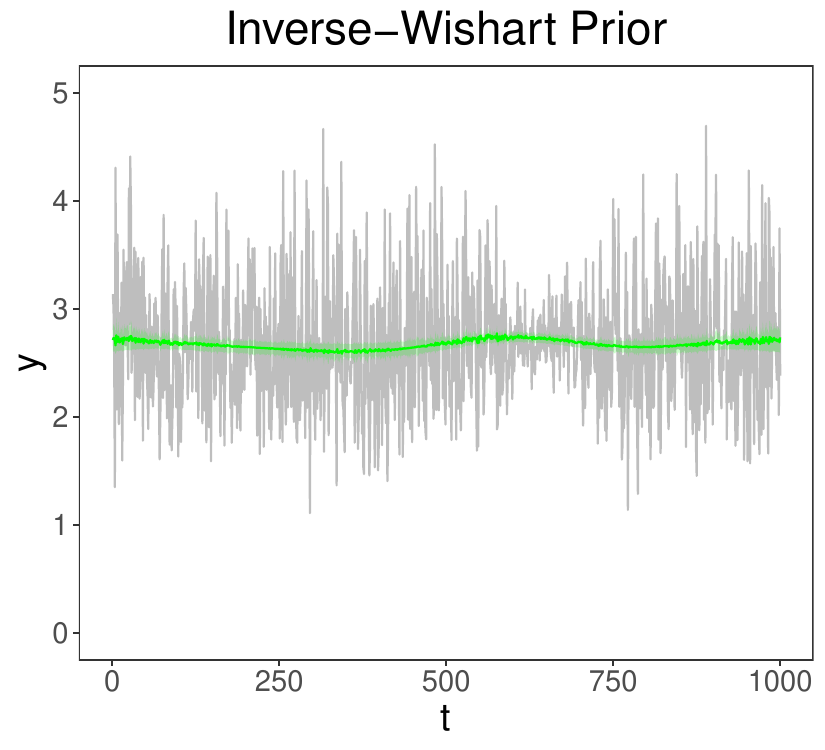}
\includegraphics[width = 4.25cm]{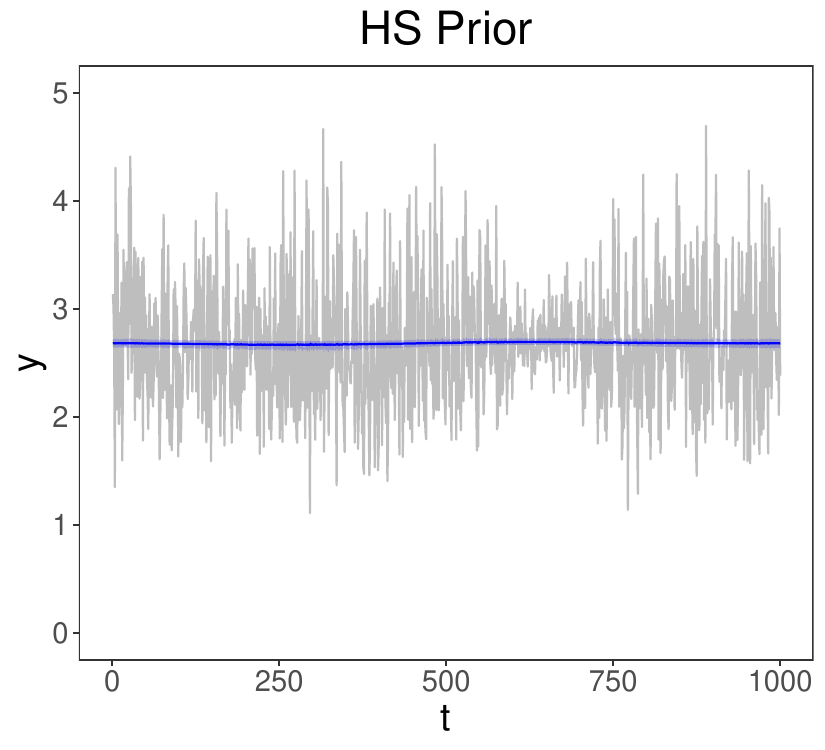}
\includegraphics[width = 4.25cm]{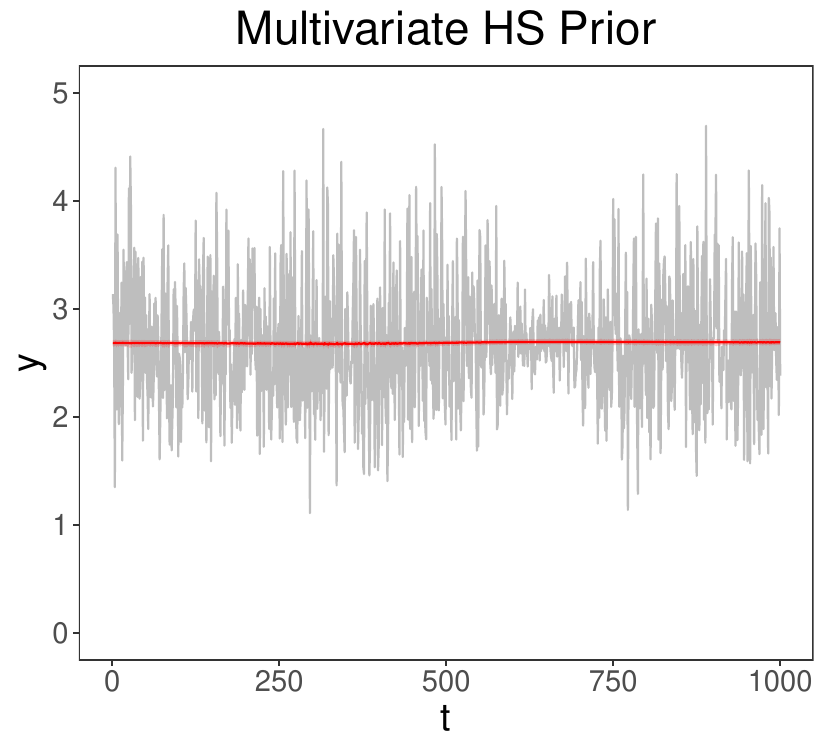}\\
\includegraphics[width = 4.25cm]{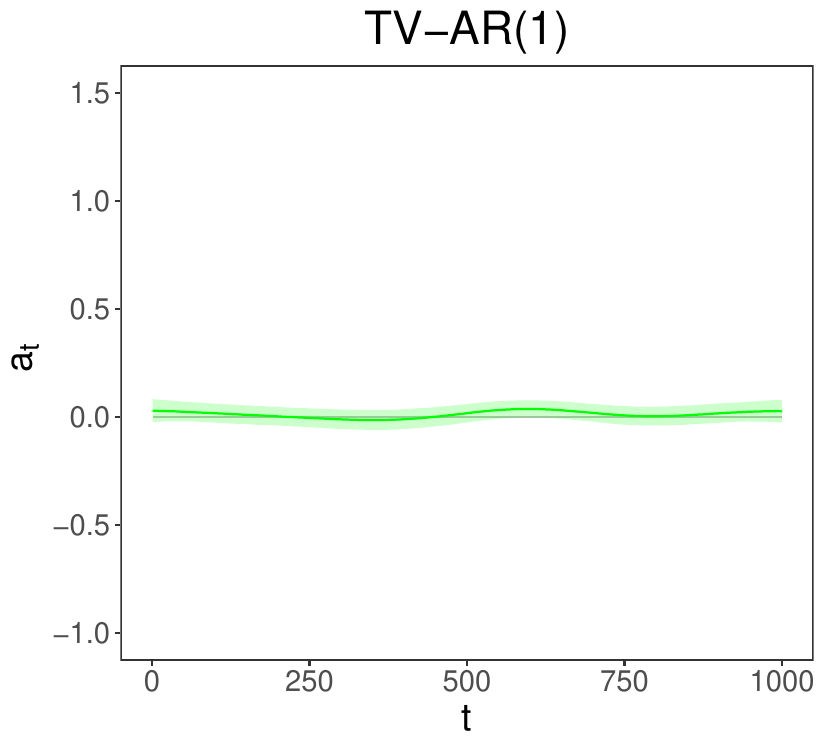}
\includegraphics[width = 4.25cm]{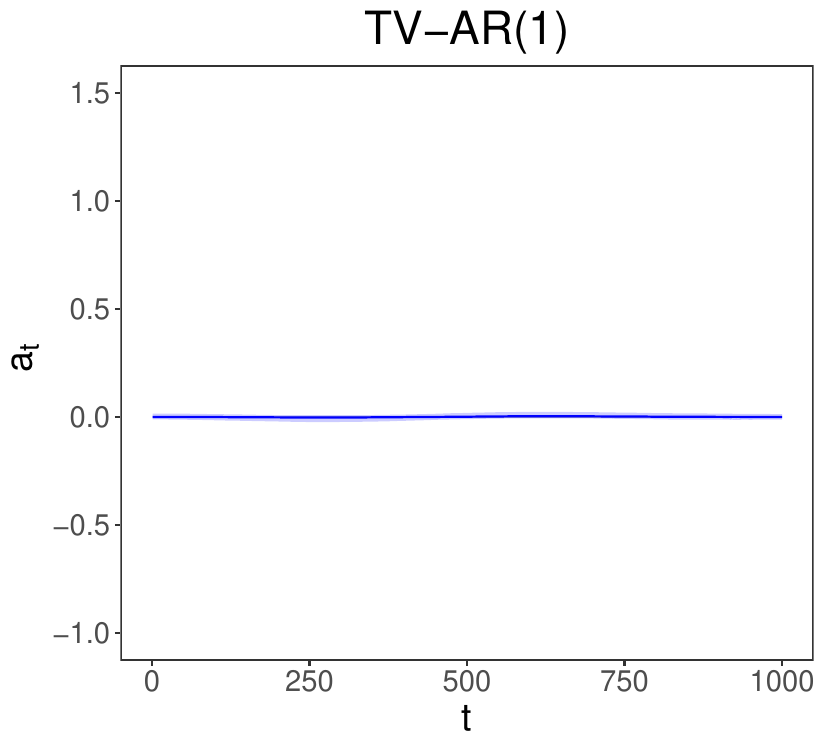}
\includegraphics[width = 4.25cm]{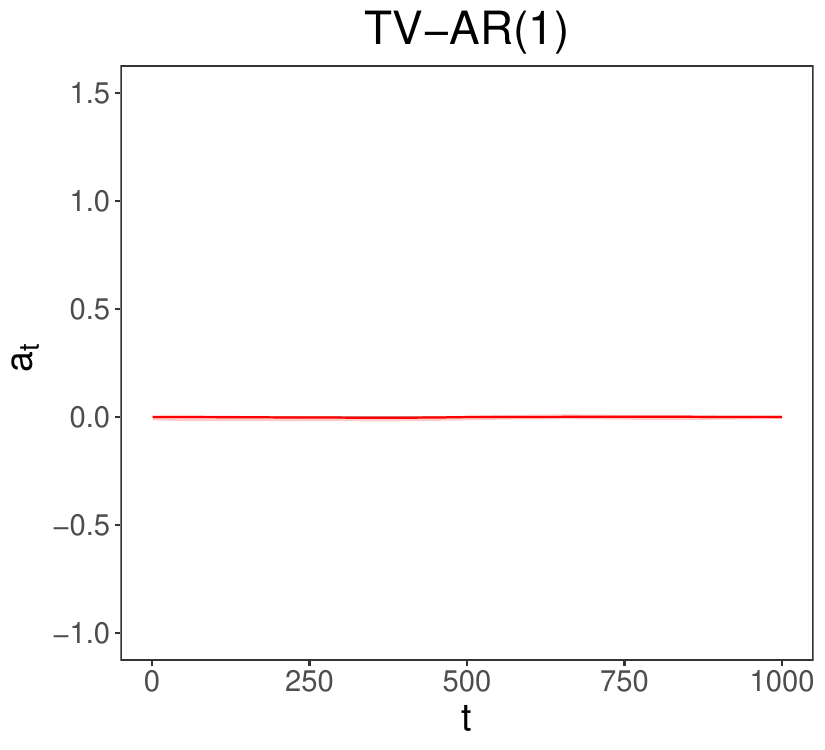}\\
\includegraphics[width = 4.25cm]{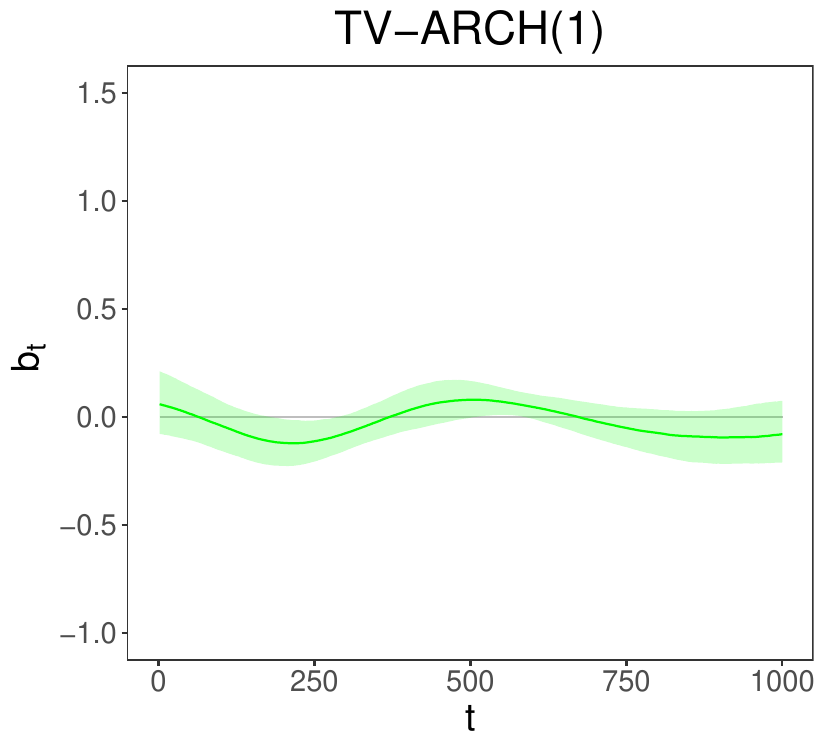}
\includegraphics[width = 4.25cm]{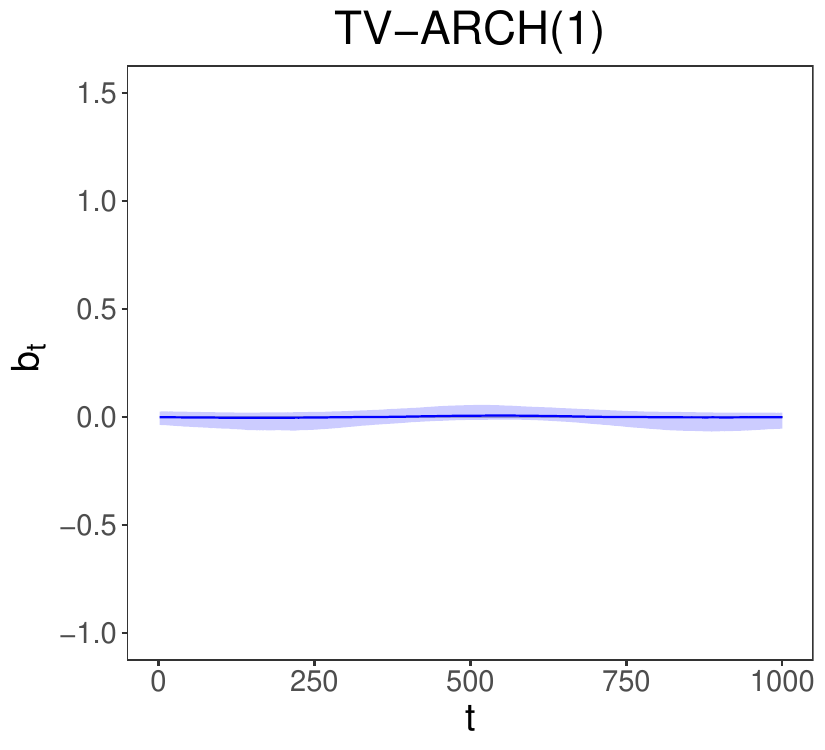}
\includegraphics[width = 4.25cm]{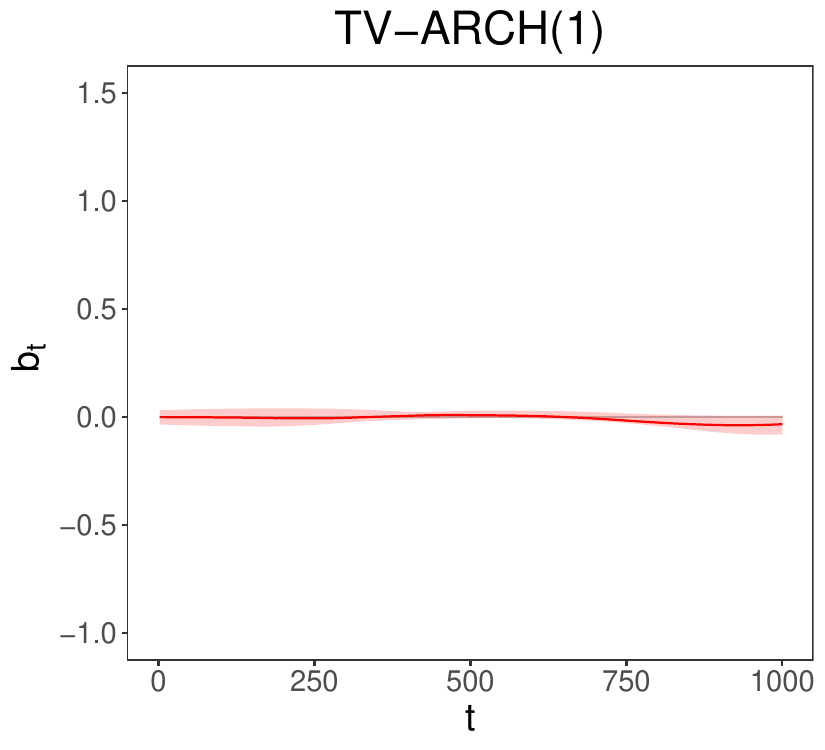}\\
\includegraphics[width = 4.25cm]{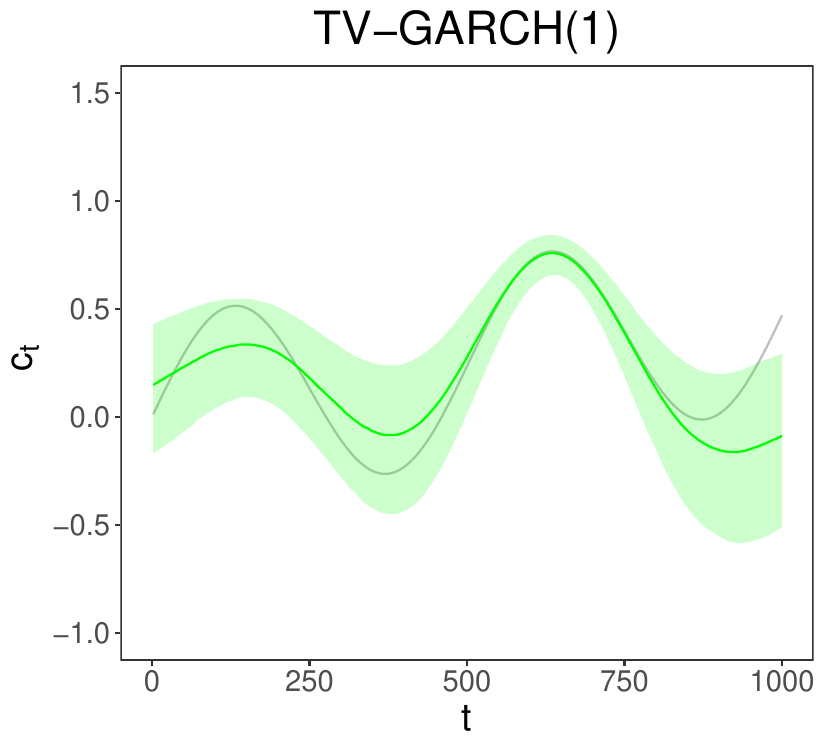}
\includegraphics[width = 4.25cm]{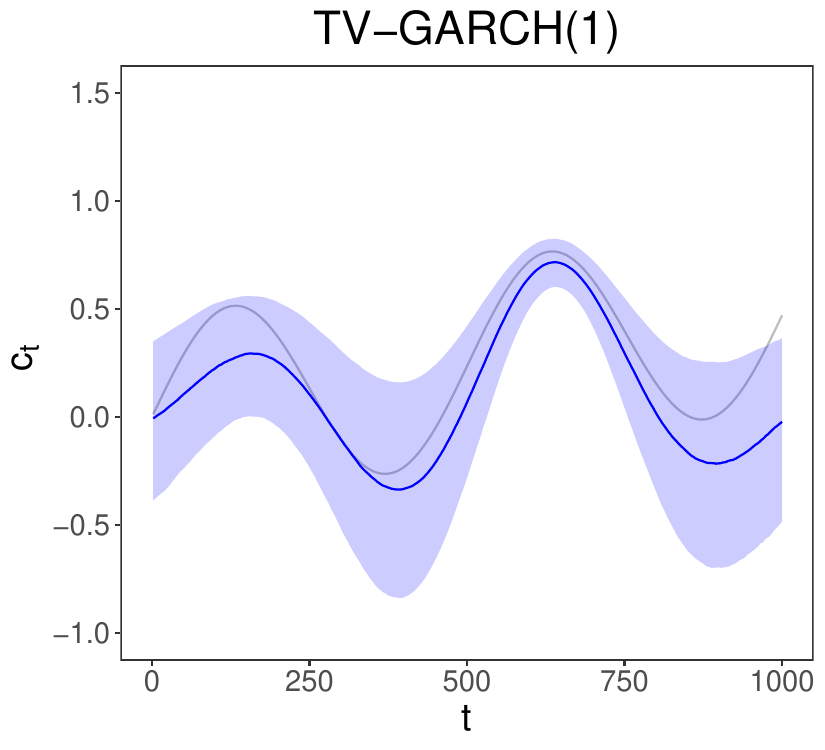}
\includegraphics[width = 4.25cm]{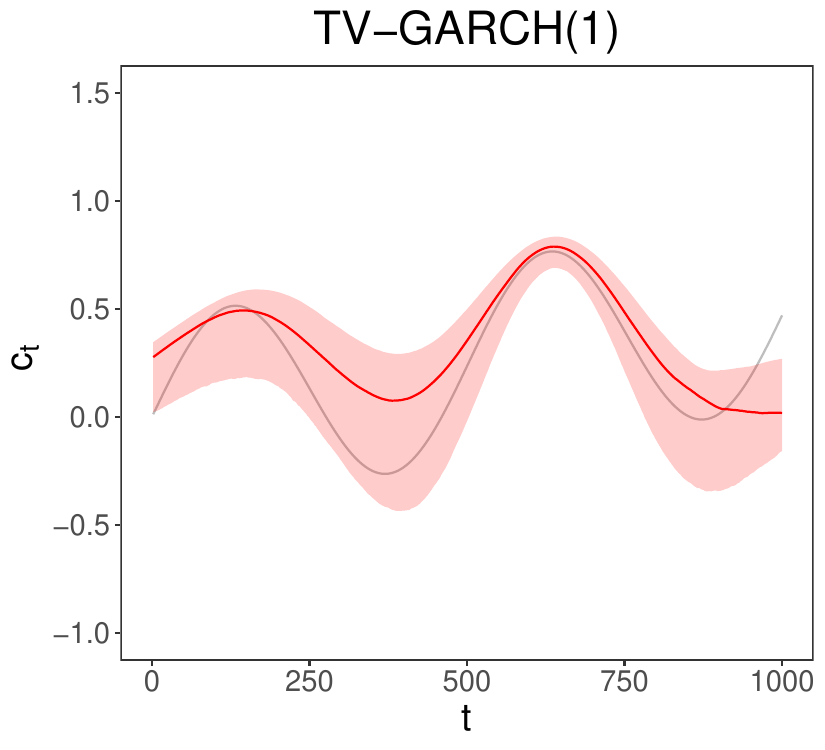}
\end{center}
\end{figure}

\begin{figure}[h]
\begin{center}
\caption{The LOOIC ($+-$standard error) for the simulation study for each prior selection on basis coefficients.}
\includegraphics[width = 6.5cm]{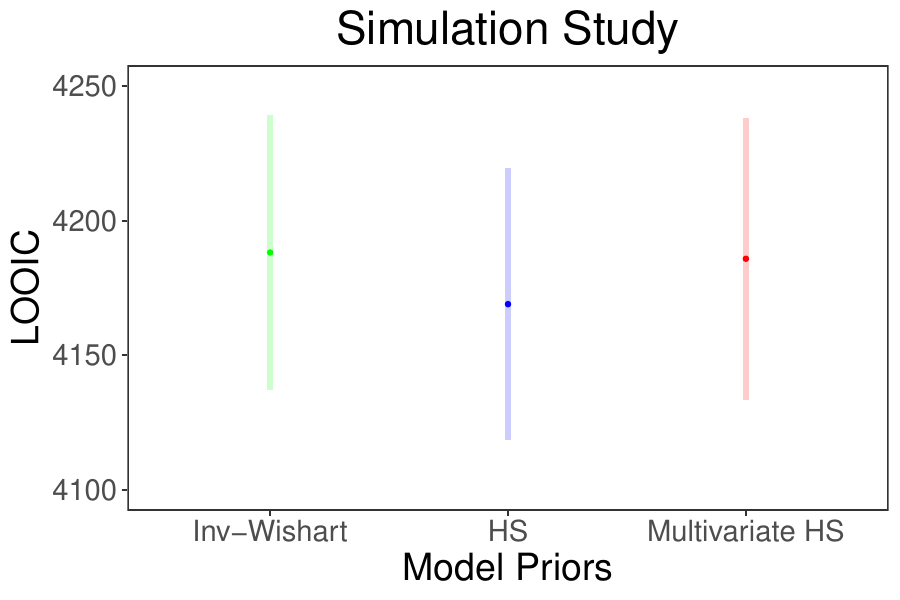}
\end{center}
\end{figure}

\begin{figure}[h]
\begin{center}
\caption{The LOOIC ($+-$standard error) for the Park Grass example for each prior selection on basis coefficients.}
\includegraphics[width = 6.5cm]{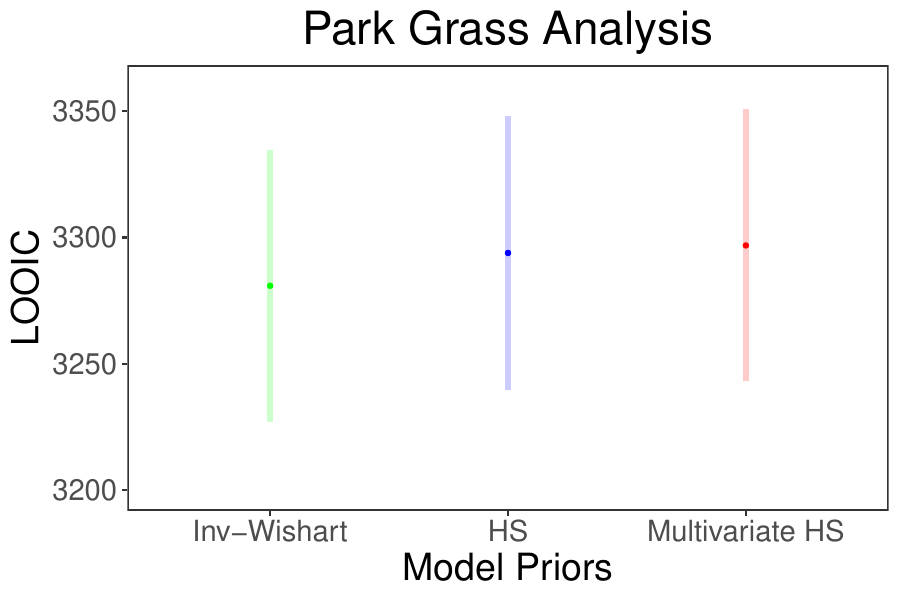}
\end{center}
\end{figure}

\newpage
\begin{figure}[h]
\begin{center}
\caption{The Posterior Predictive Distribution (PPD) of the mean estimates of the TV-AR(1) model for Park Grass for all prior specifications (left). PPD of the standard deviation estimates for the TV-GARCH(1, 1) model for Park Grass for all prior specifications (right). Both mean and SD's were derived from the log-scale of the data.}
\includegraphics[width = 6.5cm]{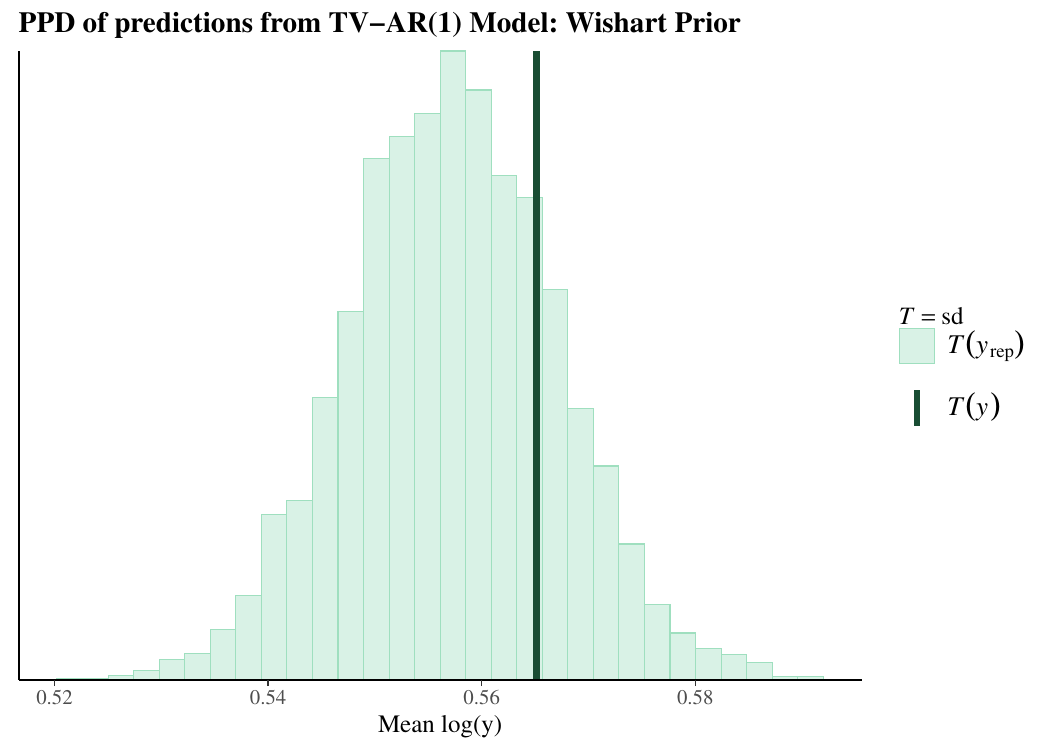}
\includegraphics[width = 6.5cm]{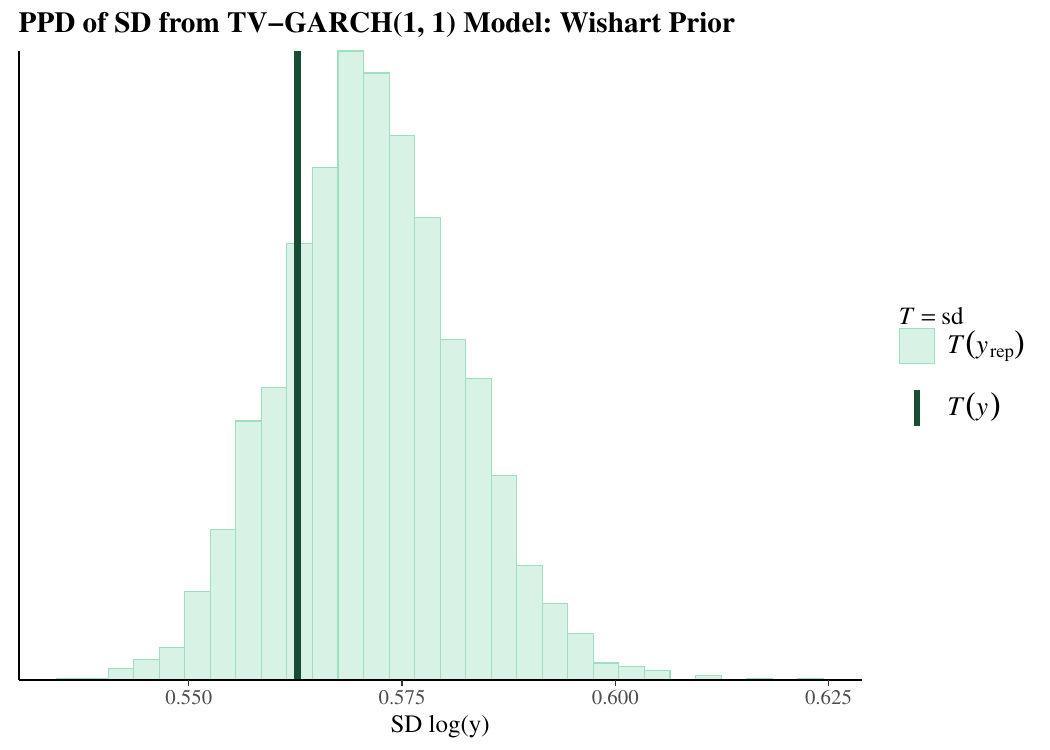}\\
\includegraphics[width = 6.5cm]{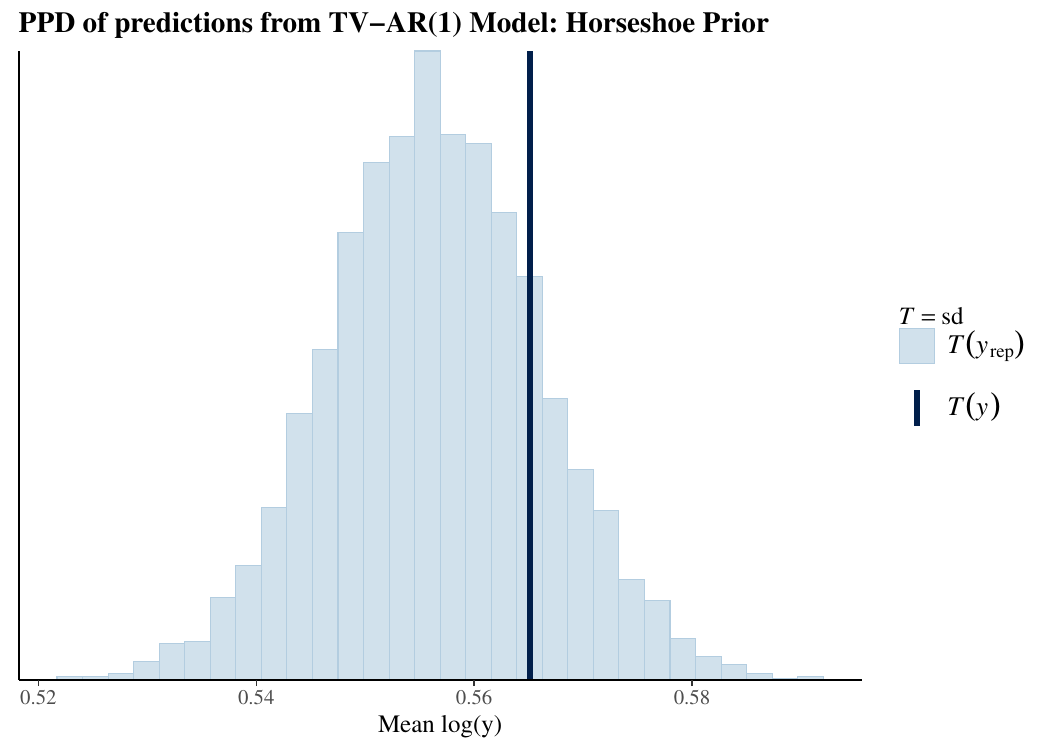}
\includegraphics[width = 6.5cm]{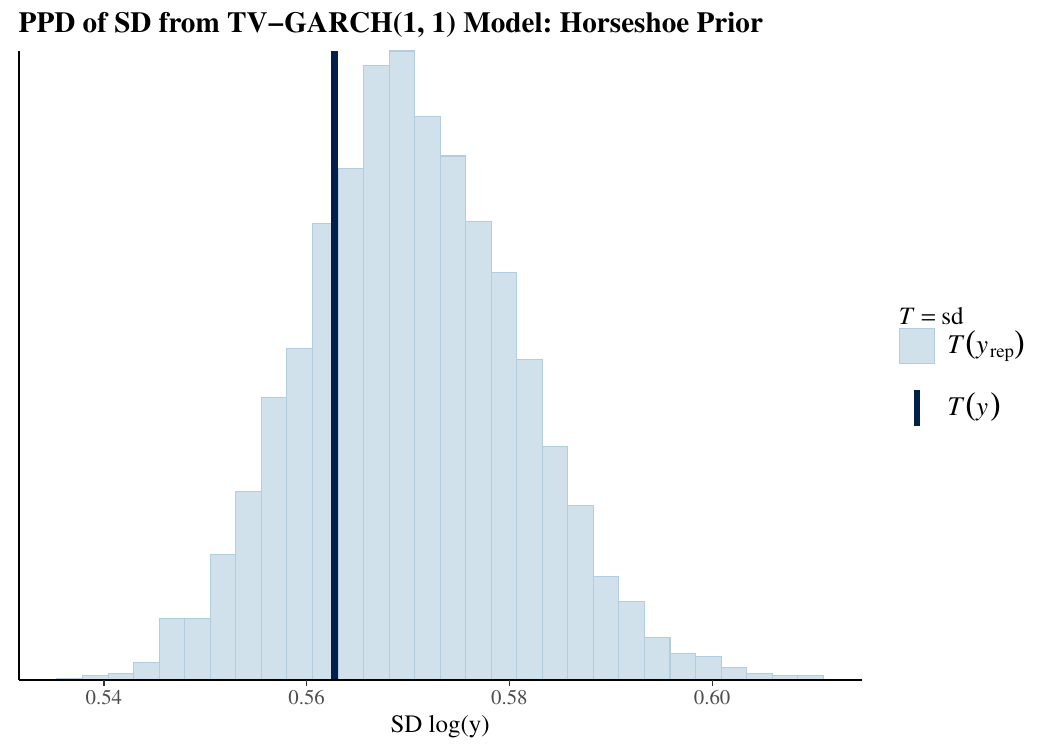}\\
\includegraphics[width = 6.5cm]{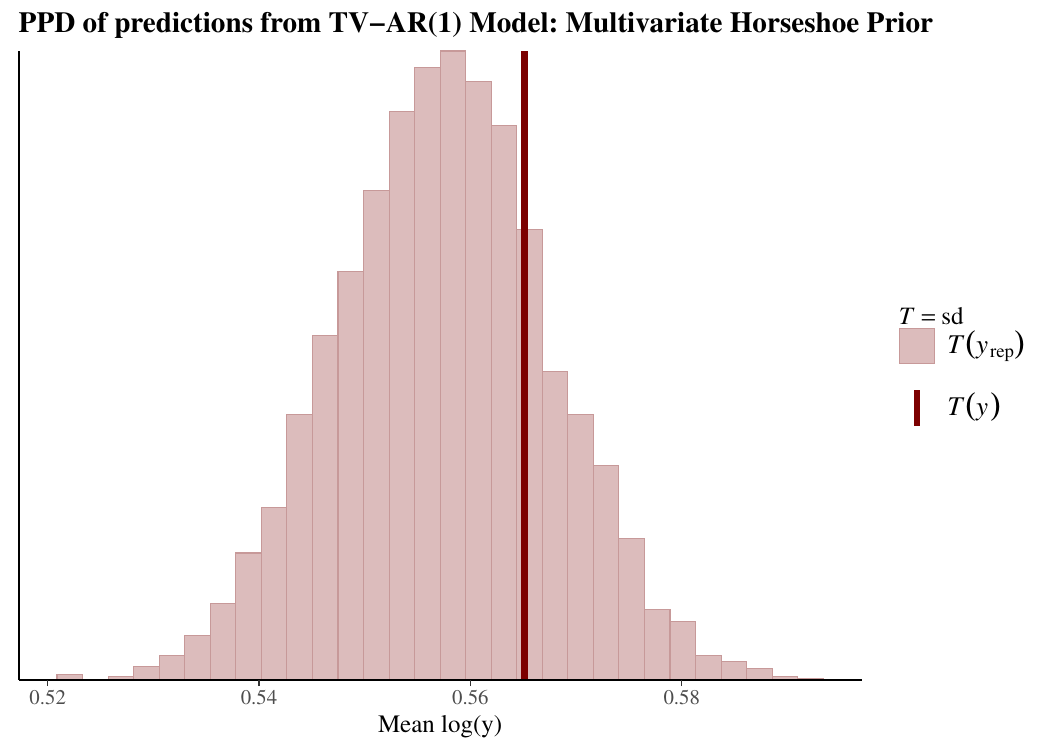}
\includegraphics[width = 6.5cm]{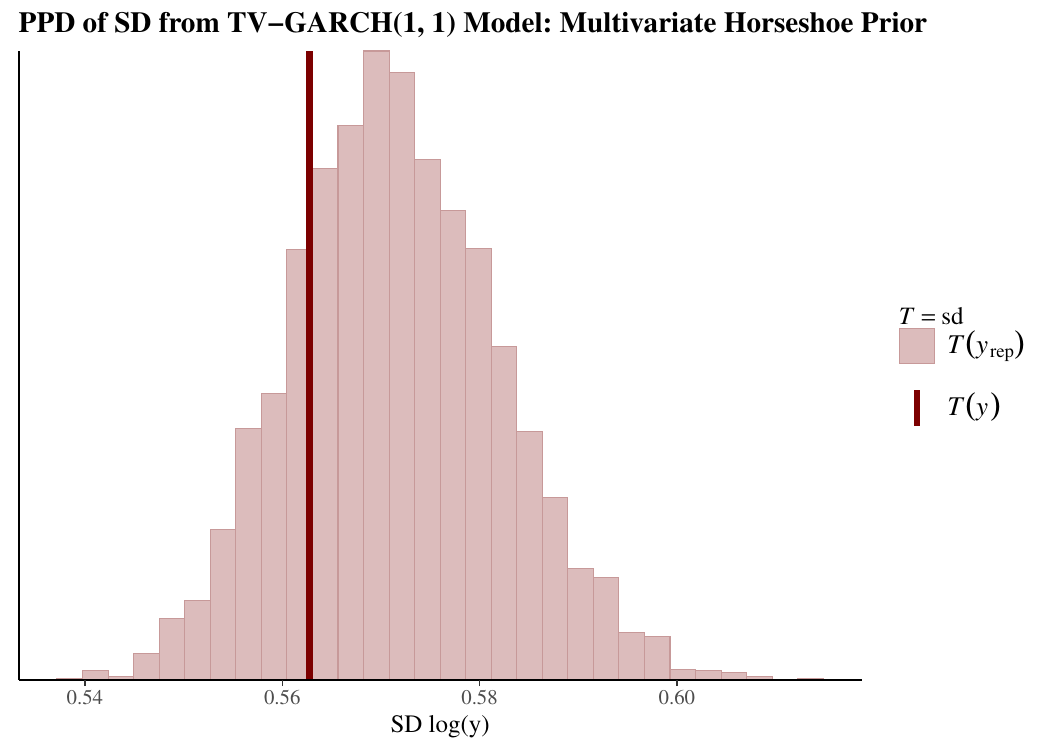}
\end{center}
\end{figure}

\newpage
\begin{figure}[h]
\begin{center}
\caption{The Pareto-smoothed importance sampling (PSIS) output from the TV-AR(1) and TV-GARCH(1, 1) models of the Park Grass dataset across all prior specifications.}
\includegraphics[width = 6.5cm]{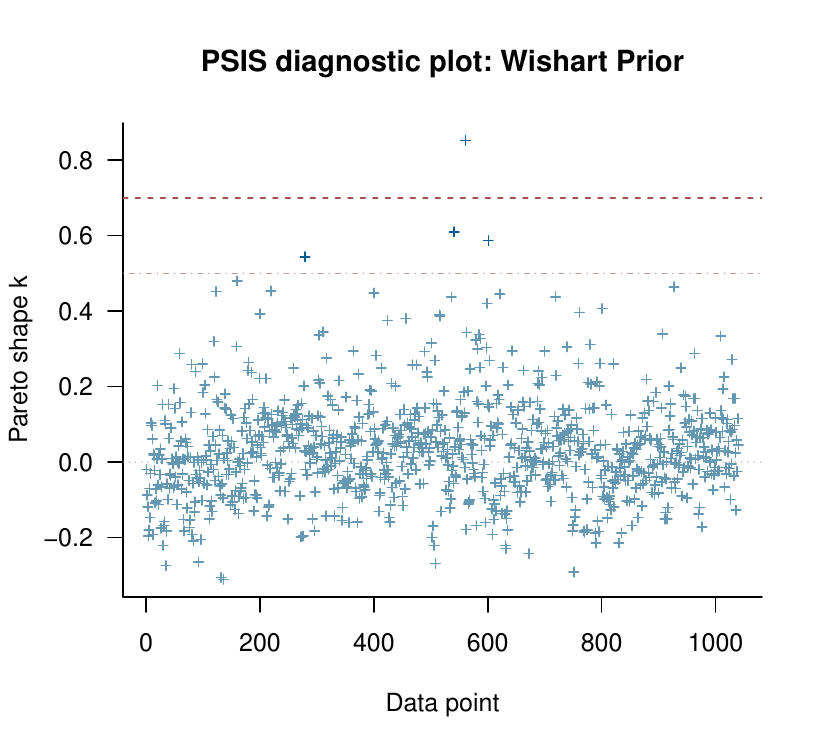}\\
\includegraphics[width = 6.5cm]{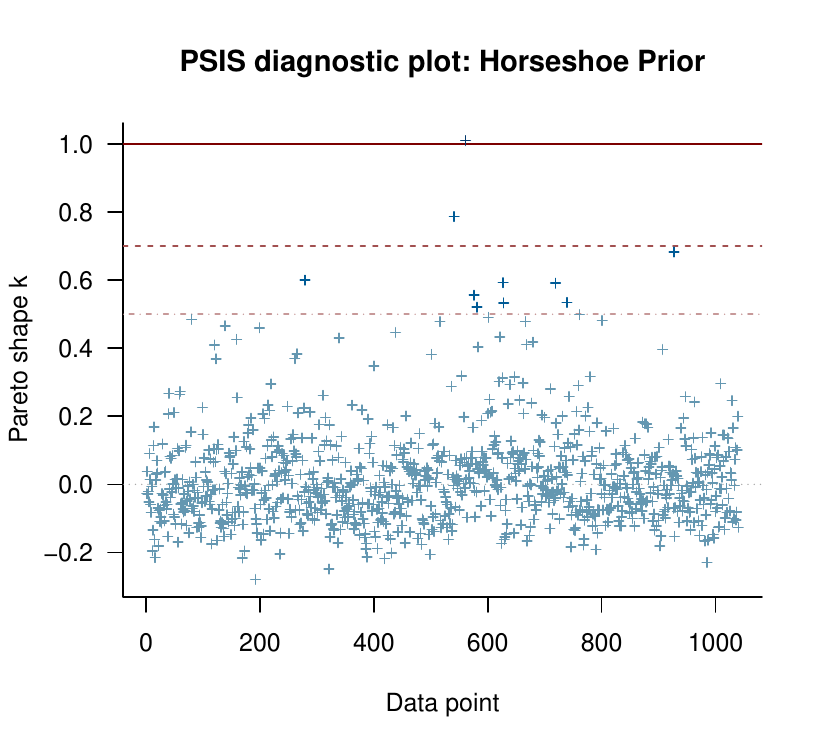}\\
\includegraphics[width = 6.5cm]{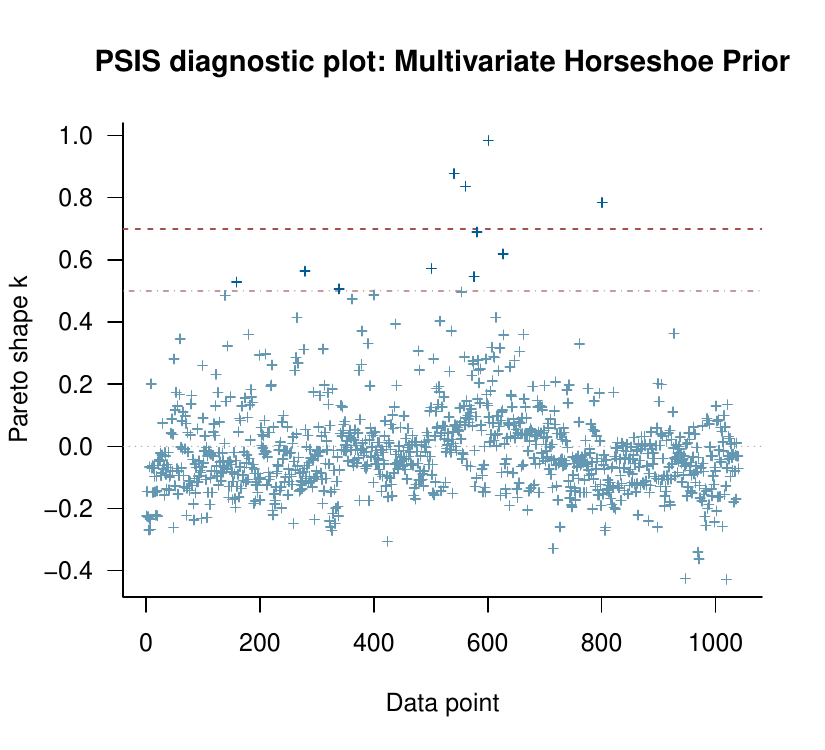}
\end{center}
\end{figure}

\end{document}